\documentclass[prl,twocolumn,superscriptaddress,floatfix]{revtex4}
\usepackage[dvipdfmx,hiresbb]{graphicx}
\usepackage{color}
\usepackage{enumerate}
\usepackage{epsfig}
\usepackage{amsmath,amssymb,latexsym}
\usepackage{ascmac}
\usepackage{bm}
\usepackage{natbib}
\def\U#1{{\rm #1}} 

\newtheorem{theorem}{{\bf Theorem}}
\newtheorem{lemma}[theorem]{{\bf Lemma}}

\newtheorem{pro}[theorem]{{\bf Proposition}}
\newtheorem{coro}[theorem]{{\bf Corollary}}
\newtheorem{define}[theorem]{{\bf Definition}}
\newcommand{\bra}[1]{\langle #1 |}
\newcommand{\ket}[1]{| #1 \rangle}

\newcommand{\expect}[1]{\left\langle #1 \right\rangle} 

\newcommand{\sq}{\qquad $\blacksquare$}
\newcommand{\wt}{\U{wt}}
\newcommand{\test}{\U{Test}}
\newcommand{\negl}{\U{negl}}
\newcommand{\Pos}{\U{Pos}}

\newcommand{\no}{\nonumber}

\def\Pr{\U{Pr}}
\def\tr{\U{tr}}

\def\tiltila{\tilde{\tilde{\alpha}}}

\newcommand{\Acal}{\mathcal{A}}

\newcommand{\Dcal}{\mathcal{D}}
\newcommand{\Fcal}{\mathcal{F}}
\newcommand{\Gcal}{\mathcal{G}}

\newcommand{\Kcal}{\mathcal{K}}
\newcommand{\Rcal}{\mathcal{R}}
\newcommand{\Xcal}{\mathcal{X}}
\newcommand{\Ycal}{\mathcal{Y}}

\newcommand{\chosen}{\leftarrow}
\newcommand{\bit}{\{0, 1\}}

\newcommand{\GEN}{\mathsf{GEN}}
\newcommand{\INV}{\mathsf{INV}}
\newcommand{\CHK}{\mathsf{CHK}}
\newcommand{\SAMP}{\mathsf{SAMP}}
\newcommand{\Supp}{\mathsf{Supp}}

\newcommand{\thetabf}{{\boldsymbol \theta}}

\newcommand{\Nbb}{\mathbb{N}}

\newcommand{\Erm}{\mathrm{E}}

\begin{document}
\setlength{\abovedisplayskip}{6pt} 
\setlength{\belowdisplayskip}{6pt} 
\title{
Computational self-testing for entangled magic states
}
\author{Akihiro Mizutani$^\ast$}
\affiliation{Mitsubishi Electric Corporation, Information Technology R\&D Center,
5-1-1 Ofuna, Kamakura-shi, Kanagawa, 247-8501 Japan}
\author{Yuki Takeuchi}
\affiliation{NTT Communication Science Laboratories, NTT Corporation, 3-1
Morinosato Wakamiya, Atsugi, Kanagawa 243-0198, Japan}
\author{Ryo~Hiromasa}
\affiliation{Mitsubishi Electric Corporation, Information Technology R\&D Center,
5-1-1 Ofuna, Kamakura-shi, Kanagawa, 247-8501 Japan}
\author{Yusuke~Aikawa}
\affiliation{Mitsubishi Electric Corporation, Information Technology R\&D Center,
5-1-1 Ofuna, Kamakura-shi, Kanagawa, 247-8501 Japan}
\author{Seiichiro~Tani}
\affiliation{NTT Communication Science Laboratories, NTT Corporation, 3-1
Morinosato Wakamiya, Atsugi, Kanagawa 243-0198, Japan}

\begin{abstract}
Can classical systems grasp quantum dynamics executed in an untrusted quantum device? 
Metger and Vidick answered this question affirmatively by proposing 
a computational self-testing protocol for Bell states that certifies generation of Bell states and measurements on them. 
Since their protocol relies on the fact that the target states are stabilizer states, 
it is highly 
non-trivial to reveal whether the other class of quantum states, {\it non-stabilizer states}, can be self-tested.  
Among non-stabilizer states, magic states are indispensable resources for universal quantum computation. 
Here, we show that a magic state for the $CCZ$ gate can be self-tested while that for the $T$ gate cannot. 
Our result is applicable to a proof of quantumness, 
where we can classically verify whether a quantum device generates a quantum state having non-zero magic. 
\end{abstract}
\maketitle

{\it Introduction.}
In device-independent quantum information processing, we treat a quantum device as a black box and can only access it classically. 
By using classical input-output statistics obtained through interacting with the 
device, our goal is to make statements about the inner workings of the quantum device. 
A scheme for characterizing a quantum device provides an approach to achieve device-independent quantum key 
distribution~\cite{PhysRevLett.67.661,MY98,PhysRevX.3.031006,PhysRevLett.113.140501,Ekert2014,Arnon-Friedman2018,MeQKD20} 
and delegated quantum computation~\cite{RUV13,HPF15}. 

A stringent form of device-independent certification for quantum devices is self-testing, 
which was introduced by Mayers and Yao~\cite{MY04}. 
In traditional self-testing protocols~(see e.g.,~\cite{McKague_2012,CGS17,selftestreview}), a classical verifier certifies that 
computationally unbounded devices, which are also called provers, have prepared the target state up to some isometry 
(i.e., a change of basis) and measured qubits with the observable as required by the verifier. 
Their crucial assumption 
is that there are multiple provers, and each prover is allowed to be entangled but cannot classically communicate with others. 
In practice, however, this non-communication assumption is difficult to enforce.

Recently, a different type of self-testing called computational self-testing (C-ST) 
was proposed~\cite{MV20}, which replaces the non-communicating multiple provers with a 
single computationally bounded quantum prover who only performs efficient quantum computation. 
To remove the non-communication assumption, their protocol relies on a standard assumption in post-quantum cryptography where 
the Learning with Errors (LWE) problem
\footnote{
The LWE problem is to solve a noisy system of linear equations, and so far 
there exists no efficient quantum algorithm to solve this problem.
}
cannot be solved by quantum computers in polynomial time~\cite{Regev}. 
Since the prover is assumed to be computationally bounded, 
the probability of solving the LWE problem is negligibly small, which 
we call the {\it LWE assumption}. 
Here, it is important to note that unlike in classical public-key cryptography, this LWE assumption must hold {\it only during} execution 
of the self-testing protocol
\footnote{
Note that encrypted messages using classical public-key cryptography are decrypted 
once it becomes technologically feasible to break the underlying computational assumption. 
On the other hand, the LWE assumption supposed in~\cite{MV20} is only 
exploited to prevent the malicious prover from tricking the verifier into accepting the prover as honest. 
Hence, as long as the LWE assumption holds during the self-testing protocol, 
if this assumption is broken after the protocol, the results already obtained never be compromised. 
}. 
The C-ST~\cite{MV20} has been applied to device-independent quantum key distribution~\cite{MeQKD20} 
and oblivious transfer~\cite{BY21}. 

The self-testing protocol~\cite{MV20} consists of interactions between the classical verifier and the prover, and after the interactions, 
the verifier decides to either ``accept" or ``reject" the prover. 
In general, a C-ST protocol must satisfy 
two properties. One is completeness where the honest prover (i.e., the ideal device) is accepted by the verifier with high probability. 
The other is soundness where 
if the verifier accepts the prover with high probability, the device's functionality is close to the ideal one, i.e., 
the device generates the target state and executes measurements on it with high precision as required by the verifier. 
So far, the C-ST protocol has been constructed only for Bell states 
$(\sigma_X^a\otimes\sigma_X^b)(\ket{0}\ket{+}+\ket{1}\ket{-})/\sqrt{2}$ with $a,b\in\{0,1\}$~\cite{MV20}, which are stabilizer states, and 
their protocol measures the stabilizers $\sigma_Z\otimes\sigma_X$ and $\sigma_X\otimes\sigma_Z$ to self-test them. 
Here, $\ket{\pm}:=(\ket{0}\pm\ket{1})/\sqrt{2}$ with $\{\ket{0},\ket{1}\}$ being the computational basis, and 
$\sigma_Z$ and $\sigma_X$ are the Pauli-$Z$ and $X$ operators, respectively. 
The underlying primitives of their protocol 
are the {\it extended noisy trapdoor claw-free function} (ENTCF) families introduced in~\cite{BCM+,mahadev} that are constructed from the 
LWE problem. 
The ENTCF families consist of two families of function pairs, one used 
to check the Pauli-$Z$ operator, and the other used for checking the Pauli-$X$ operator. 
Hence, it should be straightforward to extend the result in~\cite{MV20} to all the stabilizer states whose stabilizers are tensor products 
of the Pauli-$Z$ and $X$ operators. 
However, for other states, such as non-stabilizer states, constructing C-ST protocols is non-trivial.

Among non-stabilizer states, hypergraph states~\cite{RHBM13}, generated by applying controlled-controlled-$Z$ $(CCZ)$ 
gates on graph states~\cite{rausbri}, are useful in various quantum information processing tasks, 
such as preparing a magic state~\cite{magic} for quantum computation, 
decreasing the number of bases for measurement-based quantum computation~\cite{takesci,Miller}, 
enhancing the amount of violation of Bell's inequality~\cite{Ebell}, and demonstrating quantum supremacy~\cite{supre}. 
Experimentally, generating hypergraph states with high fidelity is generally hard since it requires $CCZ$ gates. 
Hence, it is important to certify whether a generated state is the target hypergraph state. 
Indeed, several certification methods have been invented~\cite{pra2017,PRX2018,prap2019}, 
where the measurements are assumed to be {\it trusted}.

In this Letter, 
we construct a C-ST protocol for the entangled magic state $CCZ\ket{+}^{\otimes3}$. 
This hypergraph state is useful for use as a magic state or a building block of Union Jack states~\cite{Miller}, 
and for realizing the violation of Bell's inequality~\cite{Ebell}. 
As for magic states, $T\ket{+}$ with $T:=\ket{0}\bra{0}+e^{\U{i}\pi/4}\ket{1}\bra{1}$ is a major one, but 
we show that no C-ST protocol can be constructed for it within the framework of~\cite{MV20}. 

We explain an intuitive idea of how to construct the C-ST protocol for $CCZ\ket{+}^{\otimes3}$. This state 
is a simultaneous $+1$ eigenstate of $\sigma_{X,1}CZ_{23}$, $\sigma_{X,2}CZ_{13}$, and $\sigma_{X,3}CZ_{12}$, which 
we call {\it generalized stabilizers}. 
Here, $\sigma_{X,i}$ and $CZ_{jk}$ denote the Pauli-$X$ operator acting on the $i^{\U{th}}$ qubit and the controlled-$Z$ ($CZ$) gate 
acting on the $j^{\U{th}}$ and $k^{\U{th}}$ qubits, respectively.
Since these three operators are not the tensor products of Pauli-$Z$ and $X$, the arguments in~\cite{MV20} cannot be directly applied. 
To overcome this problem, we generalize the idea in~\cite{PRX2018}. 
This shows that expected values of the generalized stabilizers for a state $\rho$ can be estimated
by measuring the individual qubits of $\rho$ with the ideal Pauli-$Z$ and $X$ measurements followed by classical processing. 
Since the ideality of the measurements is not assumed in the self-testing scenario, we 
generalize the result in~\cite{PRX2018} so that it works even if the measurements are untrusted.

In constructing C-ST protocols for $n$-qubit states, there are two obstacles that must be overcome. 
Our construction would overcome one of them, and we will discuss that in Discussion section.

Recently, by exploiting the ENTCF families, various protocols have been 
invented for the proof of quantumness~\cite{BCM+,simpler,compBell,LH21,Alex21}, 
verification of quantum computations~\cite{mahadev,alagic,yamakawa,kaimin}, 
remote state preparation~\cite{andru,qfactory}, and zero-knowledge arguments for quantum computations~\cite{moriyama,Coladangelo,tina}. 
We show that our self-testing protocol for the entangled magic state is applicable to another type of proof of 
quantumness where the classical verifier can certify whether the device generates a state having non-zero magic. 
The magic represents the non-stabilizerness, and it is regarded as quantumness in the sense that implementing non-Clifford 
gates via injection of non-stabilizer states upgrades classically simulatable Clifford circuits to universal quantum circuits. 

{\it Computational self-testing of magic states.}
First, we show that it is impossible to construct a C-ST protocol for the magic state 
$T\ket{+}$ with the same usage of ENTCF families in~\cite{MV20}.
More specifically, with the current usage of these families, 
the classical verifier can only check Pauli-$Z$ and $X$ measurements, 
but the statistics of the outcomes of these two measurements are the 
same for $T\ket{+}$ and $T^\dag\ket{+}$
\footnote{
When $T\ket{+}$ is measured in the Pauli-$Z$ basis, the outcomes $0$ and $1$ are obtained with equal probability. 
On the other hand, if it is measured in the Pauli-$X$ basis, they are obtained with probabilities 
$(2+\sqrt{2})/4$ and $(2-\sqrt{2})/4$, respectively.
These statistics are the same for $T^\dag\ket{+}$.}. 
Therefore, the classical verifier 
accepts the prover even when the prover generates $T^\dag\ket{+}$, which violates the aforementioned soundness. 

Next, we turn to the C-ST protocol for the entangled magic state. 
Before we describe it, we briefly introduce the main properties of the ENTCF families~\cite{BCM+,mahadev}, 
where the formal definitions are given in 
Sec.~I of the Supplemental Material~\cite{supple}.

Let $\Xcal$ and $\Ycal$ be finite sets specified by a security 
parameter (i.e., the value that determines the concrete hardness of solving the underlying LWE problem). 
ENTCF families consist of two families, $\Fcal$ and $\Gcal$, of function pairs such that each of the functions
injectively maps an element of $\Xcal$ to the one of $\Ycal$
\footnote{
Note that we assume for simplicity that the outputs of the functions are elements of set $\mathcal{Y}$, 
but precisely, the outputs are probability distributions over $\mathcal{Y}$. The rigorous definitions of ENTCF families are given in 
Sec.~I of the Supplemental Material~\cite{supple}.}.
A function $f$ in these families is injective, namely $f(x)\neq f(x')$ if $x\neq x'\in\mathcal{X}$. 
A function pair $(f_{k, 0}, f_{k, 1})$ in $\Fcal=\{(f_{k, 0}, f_{k, 1})\}_{k}$ is indexed by 
a key $k$, which is public information specifying parameters in the LWE problem, 
and $f_{k, 0}$ and $f_{k, 1}$ have the same image over $\mathcal{X}$. 
Hence, given $y\in\mathcal{Y}$, 
there exists a claw $(x_0(k,y),x_1(k,y))$ in $\mathcal{X}$ satisfying $y=f_{k,0}(x_0(k,y))=f_{k,1}(x_1(k,y))$. 
The function pair is called {\it claw-free} if it is hard to find a claw in quantum polynomial time. 
For a claw $(x_0(k, y), x_1(k, y))$ and $d\in \Xcal$, we define bit $u(k,y,d):=d\cdot(x_0(k,y)\oplus x_1(k,y))$. 
A function pair $(f_{k,0}, f_{k,1})$ in the other family of function pairs $\Gcal=\{(f_{k,0}, f_{k,1})\}_{k}$ is also indexed by a key $k$, 
but $f_{k, 0}$ and $f_{k, 1}$ have disjoint images over $\mathcal{X}$. Because of its disjointness, bit $b(k,y)$ is uniquely
determined such that given $k$ and $y$, there exists an element $x$ satisfying $y=f_{k,b(k,y)}(x)$.

Depending on the family of function pairs, the verifier generates a key $k$ and trapdoor information
$t_k$. The trapdoor is a piece of secret information that enables the 
verifier to efficiently
compute an element $x$ from $y=f_{k, b}(x)$ for any $b\in\{0,1\}$.

\begin{figure}[t]
\includegraphics[width=8.7cm]{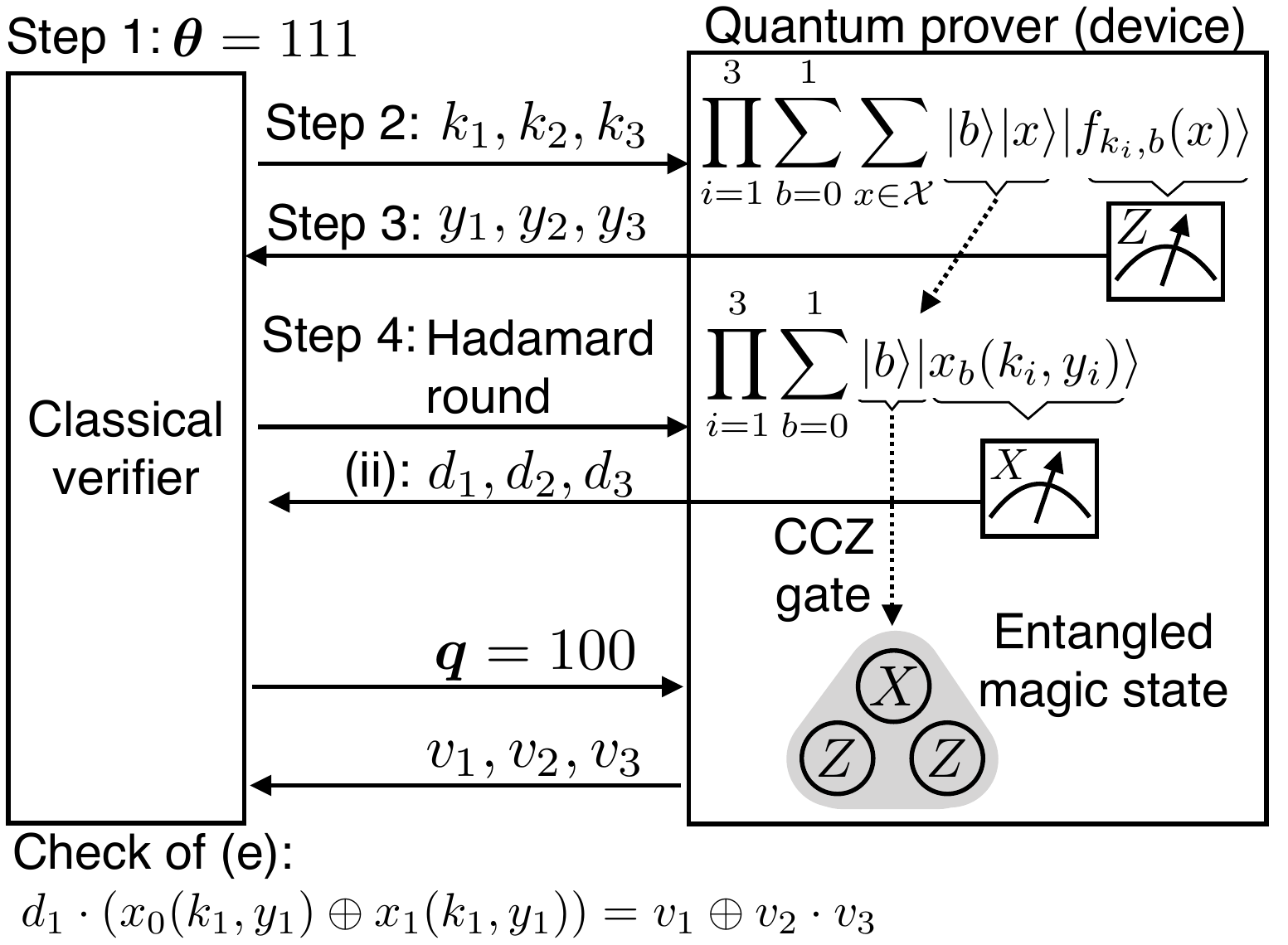}
\caption{
This figure shows the procedures for the honest device that passes step~(e). 
If the device executes the displayed state preparation, measurements, and $CCZ$ gate operation, where the register $\ket{f_{k_i,b}(x)}$
(register $\ket{x_b(k_i,y_i)}$) is measured in the computational (Hadamard) basis, 
the entangled magic state is prepared. 
The measurement with $\bm{q}=100$, which requests Pauli-$X$ ($Z$) measurement 
on the 1$^\U{st}$ qubit ($2^{\U{nd}}$ and 3$^\U{rd}$ qubits), corresponds to measuring the generalized stabilizer of the entangled magic state. 
Therefore, the outcomes $v_1, v_2, v_3$ of this honest device pass the check at step (e).
}
\label{Fig:contri}   
\end{figure}

Below, we describe Protocol~1, which consists of a three-round interaction 
between the classical verifier and the computationally bounded quantum prover (see Fig.~\ref{Fig:contri}). 
The target state of our C-ST protocol is the $Z$-rotated entangled magic state, which is defined for 
$s_1,s_2,s_3\in\{0,1\}$ by
\begin{align}
\ket{\phi_{\U{H}}^{(s_1,s_2,s_3)}}:=(\sigma_Z^{s_1}\otimes\sigma_Z^{s_2}\otimes\sigma_Z^{s_3})CCZ\ket{+}^{\otimes3}.
\label{eq:hyper}
\end{align}
In the protocol description, $x\in_R\mathcal{T}$ means that the variable $x$ is chosen from set $\mathcal{T}$ uniformly at random. 
\\
{\bf Protocol~1}
\begin{enumerate}
\item
The verifier chooses 
bases $\bm{\theta}:=\theta_1\theta_2\theta_3\in_R\mathcal{B}:=\{000,001,010,100,111\}$. 
The basis choices 0 and 1 correspond to the computational and the Hadamard basis, respectively. 
We call the basis choice 
$\bm{\theta}\in\{000,001,010,100\}$ the {\it test case} and $\bm{\theta}=111$ the {\it hypergraph case}. 
\item
For each $i\in\{1,2,3\}$, the verifier chooses 
the function family $\mathcal{G}$ ($\mathcal{F}$) if $\theta_i=0$ ($\theta_i=1$). 
Depending on the chosen families, 
the verifier generates keys $k_1,k_2,k_3$ and trapdoors $t_{k_1},t_{k_2},t_{k_3}$. 
Then, the verifier sends keys $k_1,k_2,k_3$ to the prover but keeps trapdoors $t_{k_1},t_{k_2},t_{k_3}$ secret from the prover. 
\item
The verifier receives $y_1,y_2,y_3\in\mathcal{Y}$ from the prover. 
\item
The verifier chooses a round type from $\{\U{preimage~round,~Hadamard~round}\}$ uniformly at random and sends 
it to the prover. \\
(i)
For a preimage~round: the verifier receives 
preimages $(b_1,x_1;b_2,x_2;b_3,x_3)$ from the prover with $b_i\in\{0,1\}$ and $x_i\in\mathcal{X}$. 
The verifier rejects the prover and sets a flag $flag\leftarrow fail_{\U{Pre}}$ unless all the preimages are correct 
(namely, $f_{k_i,b_i}(x_i)=y_i$ holds for $i=1,2,3$). 
\\\\
(ii)
For an Hadamard~round: 
the verifier receives $d_1,d_2,d_3\in\mathcal{X}$ from the prover. Then, the verifier sends measurement bases 
$q_1,q_2,q_3\in_R\{0,1\}$ to the prover, and 
the prover returns measurement outcomes $v_1,v_2,v_3\in\{0,1\}$. 
Depending on the bases $\bm{\theta}$, the verifier executes the following checks. 
If the flag is set, the verifier rejects the prover. 
\begin{enumerate}
\item  $\bm{\theta}=$000: set $flag\leftarrow fail_{\test}$ if for $i\in_R\{1,2,3\}$, $q_i=0$ and $b(k_i,y_i)\neq v_i$ hold. 
\item $\bm{\theta}=$100: set $flag\leftarrow fail_{\test}$ if $q_1=1$ and $u(k_1,y_1,d_1)\oplus b(k_2,y_2)\cdot b(k_3,y_3)\neq v_1$ hold.
\item $\bm{\theta}=$010: set $flag\leftarrow fail_{\test}$ if $q_2=1$ and $u(k_2,y_2,d_2)\oplus b(k_1,y_1)\cdot b(k_3,y_3)\neq v_2$ hold. 
\item
$\bm{\theta}=$001: set $flag\leftarrow fail_{\test}$ if $q_3=1$ and $u(k_3,y_3,d_3)\oplus b(k_1,y_1)\cdot b(k_2,y_2)\neq v_3
   $ hold.
   \item
   $\bm{\theta}=$111: set $flag\leftarrow fail_{\U{Hyper}}$ if one of the following holds: 
\\
   $\bm{q}=100~\U{and}~u(k_1,y_1,d_1)\neq v_1\oplus v_2\cdot v_3,$
\\
   $\bm{q}=010~\U{and}~u(k_2,y_2,d_2)\neq v_2\oplus v_1\cdot v_3,$
\\
 $\bm{q}=001~\U{and}~u(k_3,y_3,d_3)\neq v_3\oplus v_1\cdot v_2$
 \\
with $\bm{q}:=q_1q_2q_3$.
\end{enumerate}
\end{enumerate}
{\bf Completeness}.
We show in Theorem~\ref{Th:comp} that Protocol~1 satisfies the aforementioned completeness.
\begin{theorem}
\label{Th:comp}
There exists a computationally bounded quantum prover that is accepted in Protocol~1 with probability $1-\negl(\lambda)$. 
Here, $\negl(\lambda)$ is a negligible function in the security parameter $\lambda$, namely 
a function that decays faster than any inverse polynomial in $\lambda$. 
\end{theorem}
The device is accepted in Protocol~1 if all the checks in the preimage and Hadamard rounds are passed, 
whose details are given in 
Sec.~III of the  Supplemental Material~\cite{supple}. 
Here, we particularly explain the procedures for the honest device that can pass step~(e). 
Since step~(e) corresponds to the check of the generalized stabilizers, the honest device passes this check if it 
generates the entangled magic state. 
Figure~\ref{Fig:contri} shows how to generate this state. 
After returning $d_1,d_2,d_3$, the state of the honest device is close to 
a tensor product of three Pauli-$X$ basis eigenstates due to the claw-free 
property of function family $\mathcal{F}$, and hence applying the $CCZ$ gate to this state results in the entangled magic state 
up to Pauli-$Z$ operators.\\\\
{\bf Soundness}.
We next show in Theorem~\ref{Th:sound} that Protocol~1 satisfies the aforementioned soundness. 
For the purpose of self-testing, we are interested in the last round of the interaction 
[step~4 (ii)] when $\bm{\theta}=111$. 
Here, the verifier sends the measurement bases $\bm{q}\in\{0,1\}^3$ to the device and receives 
the outcomes $\bm{v}:=v_1v_2v_3\in\{0,1\}^3$. 
We can model the behavior of the device in step~4 (ii) when $\bm{\theta}=111$ by the unnormalized state 
$\sigma^{(s_1,s_2,s_3)}$ on the device's Hilbert space $\mathcal{H}$ with $s_1,s_2,s_3\in\{0,1\}$ 
and projective measurements $\{P_{\bm{q}}^{(\bm{v})}\}_{\bm{v}}$ on this state that output $\bm{v}$ given inputs $\bm{q}$ 
to the device. Here, $s_i$ is determined by bit $u(k_i,y_i,d_i)$ for $i\in\{1,2,3\}$.

The goal of Protocol~1 is to ensure that the state 
$\sigma'^{(s_1,s_2,s_3)}:=\sigma^{(s_1,s_2,s_3)}/\tr[\sigma^{(s_1,s_2,s_3)}]$ is close to the 
entangled magic state defined in Eq.~(\ref{eq:hyper}), 
which is the target state to certify, and measurements $P_{\bm{q}}^{(\bm{v})}$ are specific tensor products of Pauli measurements, 
up to an isometry and a small error. 
This error is quantified by the probabilities that the verifier rejects the prover, namely 
the verifier sets a $flag$ to $fail_{\U{Pre}}$, $fail_{\U{Test}}$ or $fail_{\U{Hyper}}$. 
We now present the soundness as follows, where $p_{\U{a}}:=\Pr\{flag\leftarrow fail_{\U{a}}\}$ with 
$\U{a}\in\{\U{Pre,}\U{Test},\U{Hyper}\}$, 
$||\cdot||_1$ being the trace norm, and $P[\ket{\cdot}]:=\ket{\cdot}\bra{\cdot}$. 

\begin{theorem}
\label{Th:sound}
Consider a device that is rejected by the verifier 
with probabilities $p_{\U{Pre}}$, $p_{\U{Test}}$ and $p_{\U{Hyper}}$, and make the LWE assumption. 
Let $\ket{\phi_{\U{H}}^{(s_1,s_2,s_3)}}$ be the target entangled magic state to certify with $s_1,s_2,s_3\in\{0,1\}$, 
state $\sigma'^{(s_1,s_2,s_3)}$ defined above, $\lambda$ the security parameter, 
$\mathcal{H}$ the device's Hilbert space, and $\mathcal{H}'$ some Hilbert space. 
Then, there exists an isometry $V:\mathcal{H}\to \mathbb{C}^8\otimes\mathcal{H}'$,  
states $\zeta^{(s_1,s_2,s_3)}_{\mathcal{H}'}$ on $\mathcal{H}'$, and a constant $r>0$ 
such that in the case of $\bm{\theta}=111$ (hypergraph case), 
\begin{align}
&\Big|\Big|V\sigma'^{(s_1,s_2,s_3)}V^{\dagger}-
\ket{\phi_{\U{H}}^{(s_1,s_2,s_3)}}\bra{\phi_{\U{H}}^{(s_1,s_2,s_3)}}\otimes\zeta^{(s_1,s_2,s_3)}_{\mathcal{H}'}\Big|\Big|^2_1\no\\
&\le O(p_{\U{Pre}}^r+p_{\U{Test}}^r+p_{\U{Hyper}}^r)+\negl(\lambda),
\label{th:state}
\end{align}
and for any $a,b,c\in\{0,1\}$ and $q_1,q_2,q_3\in\{0,1\}$,
\begin{align}
&\Big|\Big|
VP^{(abc)}_{q_1q_2q_3}
\sigma'^{(s_1,s_2,s_3)}P^{(abc)}_{q_1q_2q_3}V^{\dagger}-P[\ket{a_{q_1},b_{q_2},c_{q_3}}]
\no\\
&\ket{\phi_{\U{H}}^{(s_1,s_2,s_3)}}\bra{\phi_{\U{H}}^{(s_1,s_2,s_3)}}
P[\ket{a_{q_1},b_{q_2},c_{q_3}}]\otimes\zeta^{(s_1,s_2,s_3)}_{\mathcal{H}'}\Big|\Big|_1^2
\no\\
&
\le O(p_{\U{Pre}}^r+p_{\U{Test}}^r+p_{\U{Hyper}}^r)+\negl(\lambda).
\label{th:msn}
\end{align}  
Here, $\ket{a_{q_1}}$ with $a,q_1\in\{0,1\}$ is $\ket{a_{q_1}}:=\ket{a}$ if $q_1=0$ and 
$\ket{a_{q_1}}:=(\ket{0}+(-1)^{a}\ket{1})/\sqrt{2}$ if $q_1=1$. $\ket{b_{q_2}}$ and $\ket{c_{q_3}}$ are defined analogously.
\end{theorem}
Here, Eq.~(\ref{th:state}) guarantees how precisely the prover generates the entangled magic state under the isometry $V$, and 
Eq.~(\ref{th:msn}) how precisely it implements the specific single-qubit measurements on it 
according to the measurement bases $\bm{q}$. 
Using $V^{\dagger}V=I$, 
Eq.~(\ref{th:msn}) also reveals that the actual probability distribution of the device 
$\{\tr[P^{(abc)}_{q_1q_2q_3}\sigma'^{(s_1,s_2,s_3)}]\}_{a,b,c}$ 
is close to the ideal one obtained by measuring $\ket{\phi_{\U{H}}^{(s_1,s_2,s_3)}}$ in the Pauli-$Z$ and $X$ bases.
Note that Eqs.~(\ref{th:state}) and (\ref{th:msn}) are analogous to the statements in the traditional self-testing 
(see e.g.,~\cite{McKague_2012,CGS17,selftestreview}). 
One notable difference from the traditional self-testing is that our isometry $V$ is allowed to be a global operation acting on 
the whole device's Hilbert space $\mathcal{H}$ because we do consider the single quantum device. 
The proof of Theorem~\ref{Th:sound} is 
given in Sec.~IV of the Supplemental Material~\cite{supple}. 

{\it Applications to the proof of quantumness.}
Recently, various protocols have been invented to enable the classical verifier to certify the quantumness of the 
device~\cite{BCM+,mahadev,yamakawa,simpler,LH21,Alex21}. 
Here, the meaning of quantumness differs depending on the protocols. 
For instance, the protocols~\cite{BCM+,LH21,Alex21} verify whether the prover has a superposed state or not, 
the protocols~\cite{mahadev,yamakawa} verify whether the prover can efficiently solve ${\sf BQP}$ problems, and 
the protocol~\cite{simpler} verifies that the prover can query to an oracle in superposition. 
Importantly, if the prover is accepted by the verifier, then the prover has quantum capability.

Our C-ST protocol given as Protocol 1 can be used for the proof of 
magic under the IID scenario where the device's functionality is the same for each repetition of the protocol. 
To measure the magic, we focus on the max-relative entropy of magic~\cite{LW20}. 
We adopt this measure for simplicity, but our arguments 
can be applied to any reasonable measure of the magic.
Let $\mathfrak{D}_{\rm max}(\rho):=\log{(1+R_g(\rho))}$ be the max-relative entropy of magic of an $n$-qubit state $\rho$, where 
$R_g(\rho)$ is defined by the minimum of $t\ge0$ such that $\rho\in(1+t){\rm STAB}-t\mathcal{S}$, 
$\U{STAB}\subset\mathcal{S}$ is the convex hull of all $n$-qubit stabilizer states, and $\mathcal{S}$ is the set of $n$-qubit states.
If $\rho$ is a stabilizer state, $R_g(\rho)=0$, and hence $\mathfrak{D}_{\rm max}(\rho)=0$.
By contraposition, if $\mathfrak{D}_{\rm max}(\rho)>0$, state $\rho$ is a non-stabilizer state.
Based on above observations, 
we outline the protocol for the proof of magic as follows
\footnote{
Note that as a related work to our proof of magic, the problem of asking whether a given state is any stabilizer state 
was studied in the {\it device-dependent} scenario~\cite{Gross2021}. 
Our protocol considers its opposite problem, i.e., 
asking whether a given state is {\it not} any stabilizer state, in the {\it device-independent} scenario.
} 
(see Sec.~V of the Supplemental Material~\cite{supple}).  
\\
{\bf Protocol~2}
\begin{enumerate}
\item The verifier and prover repeat Protocol~1 
a constant number of times, and the verifier estimates the error probabilities $p_{\U{Pre}}, p_{\U{Test}}$ and $p_{\U{Hyper}}$ using 
Hoeffding's inequality from the numbers of set flags.

\item If the estimated trace norm $T_{\U{est}}$ [the square root of the right-hand side of Eq.~(\ref{th:state})] is strictly less than $1/3$, 
then the verifier accepts the prover. Otherwise, the verifier rejects the prover.
\end{enumerate}

We first show that if our protocol is passed, with a small significance level
\footnote{
Note that the significant level is defined by the maximum probability 
of passing our protocol with a state having no magic. 
}, 
which can be set to any value such as $10^{-10}$,
the verifier can guarantee that the prover generates a state having non-zero magic up to the isometry. 
If state $\rho$ has no magic, we have $\bra{\phi_{\U{H}}^{(s_1,s_2,s_3)}}\rho\ket{\phi_{\U{H}}^{(s_1,s_2,s_3)}}\le9/16$ 
because 
for any stabilizer state $\ket{\psi}$, $F:=|\langle\psi|\phi_{\U{H}}^{(s_1,s_2,s_3)}\rangle|^2\le9/16$~\cite{BBCCGH19}. 
Since $F\le9/16$ results in $||\rho-\ket{\phi_{\U{H}}^{(s_1,s_2,s_3)}}\bra{\phi_{\U{H}}^{(s_1,s_2,s_3)}}||_1\ge1/2$~\cite{nielsen10}, 
Hoeffding's inequality with precision 1/6 implies that 
$T_{\U{est}}<1/3$ holds with probability $10^{-10}$. 
Therefore, such a state $\rho$ is accepted with probability of at most $10^{-10}$.

On the other hand, 
there is a strategy that passes this protocol with probability $1-10^{-10}$. 
This is because Theorem~\ref{Th:comp} states that there exists a prover's strategy that achieves all of the error probabilities 
$p_{\U{Pre}}, p_{\U{Test}}$ and $p_{\U{Hyper}}$ being $\negl(\lambda)$, and hence from Hoeffding's inequality, 
$T_{\U{est}}\le\negl(\lambda)+1/6<1/3$ holds except for probability $10^{-10}$.

{\it Discussions.}
In this Letter, we have constructed a computational self-testing protocol for the three-qubit entangled magic state. 
To generalize~\cite{MV20} to $n$-qubit states, there are two obstacles: 
(1) The verifier chooses the state bases $\theta_1...\theta_n\in_R\{0,1\}^n$ with which the prover is requested to 
generate the state for $n$ times. 
Since 
the target state is prepared only when all the $\theta$'s are 1, it takes exponential time on average to generate the target state.  
(2) The verifier checks all the patterns of measurements, namely it checks the correctness of Pauli-$Z$ and $X$ 
measurements for each qubit, which takes $2^n$ times. 

Our construction would solve the first problem. We have shown for $n=3$ that the number of state bases is 
sufficient to be $n+2$, 
which means the target state is prepared on average by repeating the protocol $(n+2)$ times. 
We leave its rigorous analysis and the second problem as future work. 

{\it Note added.} 
Recently, we became aware of independent related works~\cite{FWZ22} and~\cite{GMP22} that extend the result~\cite{MV20}
to self-test $n$ Bell states and $n$ BB84 states, respectively. 
By exploiting these results, it could be possible to extend our result to self-test $n$ tensor products of $CCZ$ magic states 
$CCZ\ket{+}^{\otimes3}$. 

{\it Acknowledgments.}
The authors thank Tony Metger for valuable discussions on~\cite{MV20}, and Go Kato, Yasuhiro Takahashi, and Tomoyuki Morimae for 
helpful comments.
AM is supported by JST, ACT-X Grant Number JPMJAX210O, Japan. 
YT is supported by MEXT Quantum Leap Flagship Program (MEXT Q-LEAP) Grant Number JPMXS0118067394, 
JPMXS0120319794, 
the Grant-in-Aid for Scientific Research (A) No.JP22H00522 of JSPS, 
and JST [Moonshot R\&D -- MILLENNIA Program] Grant Number JPMJMS2061.
ST is partially supported by the Grant-in-Aid for Transformative Research Areas No.JP20H05966 of JSPS, 
and the Grant-in-Aid for Scientific Research (A) No.JP22H00522 of JSPS.

\clearpage
\begin{widetext}
\begin{center}
\textbf{\large Supplementary material: Computational self-testing for entangled magic states}
\end{center}
\renewcommand{\bibnumfmt}[1]{[S#1]}
\renewcommand{\citenumfont}[1]{S#1}

\section{Preliminaries}
\subsection{Notations}
We use the bold symbol $\bm{A}$ meaning $A_1A_2A_3$. 
For $i\in\{1,2,3\}$, $\bm{A}_{\bar{i}}$ denotes $\bm{A}$ except for $A_i$. 
Let the Kronecker delta be $\delta_{x,y}=0$ if $x\neq y$ and 1 if $x=y$. 
We denote by $|\mathcal{S}|$ the cardinality of set $\mathcal{S}$. 
For bit $b\in\{0,1\}$, $\bar{b}$ denotes $b\oplus1$. 
We denote $\wt(x)$ by the number of 1's in bit string $x$. 

We denote $\mathcal{H}$ by an arbitrary finite-dimensional Hilbert space. 
The set of linear operators on Hilbert space $\mathcal{H}$ is denoted by $\mathcal{L}(\mathcal{H})$. 
For $A,B\in\mathcal{L}(\mathcal{H})$, we denote the commutator by $[A,B]=AB-BA$ and the anti-commutator by $\{A,B\}=AB+BA$. 
$\Pos(\mathcal{H})$ denotes the set of positive semidefinite operators on $\mathcal{H}$, and 
we denote the set of density matrices on $\mathcal{H}$ by 
$\mathcal{D}(\mathcal{H})=\{A\in\mathcal{L}(\mathcal{H})|A\in\Pos(\mathcal{H}), \tr[A]=1\}$. 
A binary observable is defined as an observable (Hermitian operator) that 
only has eigenvalues $\in\{1,0,-1\}$. For any binary observable $O$ and $b\in\{0,1\}$, $O^{(b)}$ denotes the projector 
onto the $(-1)^b$-eigenspace of $O$. 
We denote the Pauli-$Z$ and $X$ observables by $\sigma_Z=\sum_{b=0}^1(-1)^b\ket{b}\bra{b}$ and 
$\sigma_X=\sum_{b=0}^1(-1)^b\ket{(-)^b}\bra{(-)^b}$, respectively. Here, $\ket{(-)^b}:=(\ket{0}+(-1)^b\ket{1})/\sqrt{2}$. 

Let $\negl(\lambda)$ be a negligible function in the security parameter $\lambda$, 
namely a function that decays faster than any inverse polynomial in $\lambda$. 
For a countable set $\Xcal$, $x\chosen\Xcal$ denotes that $x$ is chosen uniformly at random from $\Xcal$.

\subsection{Cryptographic Primitives}
Here, we explain the noisy trapdoor claw-free function family, which is the 
cryptographic primitive underlying our self-testing protocol described in Sec.~\ref{sec:proto}. 

\begin{define}[Hellinger Distance]
        For two probability densities $f_{1}$ and $f_{2}$ over finite set $\Xcal$,
        the Hellinger distance between $f_{1}$ and $f_{2}$ is defined as
        \begin{eqnarray*}
                H^{2}(f_{1}, f_{2}):=1-\sum_{x\in\Xcal}\sqrt{f_{1}(x)f_{2}(x)}.
        \end{eqnarray*}
\end{define}

\begin{define}[Noisy Trapdoor Claw-free Family~\cite{mahadev}]
        Let $\lambda\in\Nbb$ be a security parameter.
        Let  $\Xcal$ and  $\Ycal$ be finite sets.
        Let  $\Kcal_{\Fcal}$ be a finite set of keys.
        A family of functions
        \begin{eqnarray*}
                 \Fcal:=\{f_{k, b}: \Xcal\rightarrow \Dcal_{\Ycal}\}_{k\in\Kcal_{\Fcal}, b\in\bit}
        \end{eqnarray*}
        is called a noisy trapdoor claw-free (NTCF) family if the following conditions hold:
        \begin{itemize}
                \item Efficient Function Generation:
                        there exists an efficient probabilistic algorithm $\GEN_{\Fcal}$ 
                        that generates a key  $k\in\Kcal_{\Fcal}$ together with a trapdoor $t_{k}$, $(k, t_{k})\chosen\GEN_{\Fcal}(1^{\lambda})$.
                \item Trapdoor Injective Pair:
                         for all $k\in\Kcal_{\Fcal}$, the following conditions hold.
                         \begin{itemize}
                                 \item Trapdoor:
                                         for all $b\in\bit$ and $x\neq x'\in\Xcal$,  $\Supp(f_{k,b}(x))\cap\Supp(f_{k,b}(x'))=\emptyset$.
                                         Moreover, there exists an efficient deterministic algorithm $\INV_{\Fcal}$ such that
                                         for all  $b\in\bit$,  $x\in\Xcal$ and $y\in\Supp(f_{k, b}(x))$,
                                         $\INV_{\Fcal}(t_{k}, b, y)=x$.
                                 \item Injective Pair:
                                         there exists a perfect matching $\Rcal_{k}\subseteq \Xcal\times\Xcal$ such that
                                         $f_{k, 0}(x_{0})=f_{k, 1}(x_{1})$ if and only if  $(x_{0}, x_{1})\in\Rcal_{k}$.
                         \end{itemize}
                \item Efficient Range Superposition:
                        for all $k\in\Kcal_{\Fcal}$ and  $b\in\bit$ there exists a function  $f_{k, b}':\Xcal\rightarrow\Dcal_{\Ycal}$
                        such that
                        \begin{itemize}
                                \item For all $(x_{0}, x_{1})\in\Rcal_{k}$ and $y\in\Supp(f_{k, b}'(x_b))$,
                                        $\INV_{\Fcal}(t_{k}, b, y)=x_{b}$ and $\INV_{\Fcal}(t_{k}, b\oplus 1, y)=x_{b\oplus 1}$.
                                \item There exists an efficient deterministic procedure $\CHK_{\Fcal}$ that on input $k$, $b\in\bit$, $x\in\Xcal$,
                                        and  $y\in\Ycal$, returns $1$ if $y\in\Supp(f_{k, b}'(x))$ and $0$ otherwise.
                                        Note that $\CHK_{\Fcal}$ is not provided the trapdoor $t_{k}$.
                                \item For every $k\in\Kcal_{\Fcal}$ and $b\in\bit$,
                                        \begin{eqnarray*}
                                            \Erm_{x\chosen\Xcal}[H^{2}(f_{k, b}(x), f'_{k, b}(x))]=\negl(\lambda)
                                        \end{eqnarray*}
                                        for some negligible function $\negl(\cdot)$, where the expectation is taken over $x\chosen\Xcal$. 
                                        Here $H^{2}(\cdot, \cdot)$ is the Hellinger distance.
                                        Moreover, there exists an efficient procedure $\SAMP_{\Fcal}$ that
                                        on input $k$ and $b\in\bit$, prepares the state
                                        \begin{eqnarray*}
                                            \frac{1}{\sqrt{|\Xcal|}}\sum_{x\in\Xcal, y\in\Ycal}\sqrt{(f'_{k, b}(x))(y)}\ket{x}\ket{y}.
                                        \end{eqnarray*}
                        \end{itemize}
                 \item Adaptive Hardcore Bit:
                         for all $k\in\Kcal_{\Fcal}$ the following conditions hold 
                         for some integer $w$ that is a polynomially bounded function in $\lambda$.
                         \begin{itemize}
                                 \item For all $b\in\bit$ and  $x\in\Xcal$, 
                                         there exists a set  $G_{k, b, x}\subseteq\bit^{w}$ such that
                                         $\Pr_{d\chosen\bit^{w}}\{d\notin G_{k, b, x}\}$ is negligible in $\lambda$,
                                         and moreover there exists an efficient algorithm that checks for membership in  $G_{k, b, x}$
                                         given $k, b, x$ and the trapdoor $t_{k}$. 
                                 \item There is an efficiently computable injection $J:\Xcal\rightarrow\bit^{w}$ such that
                                         $J$ can be inverted efficiently on its range, and such that the following holds.
                                         Let 
                                         \begin{eqnarray*}
                                                \begin{aligned}
 & H_{k}:=
                                                        \left\{
                                                                \left(
                                                                        b, x_b, d, d\cdot
                                                                        \left(
                                                                                J(x_{0})\oplus J(x_{1})
                                                                        \right)
                                                                \right)|
                                                                b\in\bit, (x_{0}, x_{1})\in\Rcal_{k}, d\in G_{k,0,x_0}\cap G_{k,1,x_1}
                                                        \right\},
                                                        \\
                                                        & \overline{H}_{k}:=\{(b, x_b, d, c\oplus 1)| (b, x_b, d, c)\in H_{k}\}.\\
                                                \end{aligned}
                                         \end{eqnarray*}
                                        Then for any efficient quantum algorithm $\Acal$, 
                                        there exists a negligible function $\negl(\cdot)$ such that
                                        \begin{align}
                                                \left|
                                                \Pr_{(k, t_{k})\chosen\GEN_{\Fcal}(1^{\lambda})}\{\Acal(k)\in H_{k}\}
                                                -\Pr_{(k, t_{k})\chosen\GEN_{\Fcal}(1^{\lambda})}\{\Acal(k)\in \overline{H}_{k}\}
                                                \right|=\negl(\lambda).
\label{eq:AHBP}
                                        \end{align}
                         \end{itemize}
        \end{itemize}
\end{define}

\begin{define}[Trapdoor Injective Function Family \cite{mahadev}]
\label{def:TIFF}
Let $\lambda\in\Nbb$ be a security parameter. 
Let  $\Xcal$ and  $\Ycal$ be finite sets.
Let  $\Kcal_{\Gcal}$ be a finite set of keys.
A family of functions
\begin{eqnarray*}
        \Gcal:=\{f_{k, b}:\Xcal\rightarrow \Dcal_{\Ycal}\}_{k\in\Kcal_{\Gcal}, b\in\bit}
\end{eqnarray*}
is called a trapdoor injective function family if the following conditions hold:
\begin{itemize}
        \item Efficient Function Generation:
                There exists an efficient probabilistic algorithm $\GEN_{\Gcal}$
                which generates a key $k\in\Kcal_{\Gcal}$ together with a trapdoor $t_{k}$,
                $(k, t_{k})\chosen\GEN_{\Gcal}(1^{\lambda})$.
        \item Disjoint Trapdoor Injective Pair:
                For all $k\in\Kcal_{\Gcal}$, for all  $b, b'\in\bit$ and  $x, x'\in\Xcal$,
                if  $(b, x)\neq (b', x')$,  $\Supp(f_{k, b}(x))\cap\Supp(f_{k, b'}(x'))=\emptyset$.
                Moreover, there exists an efficient deterministic algorithm  $\INV_{\Gcal}$ such that
                for all  $b\in\bit$,  $x\in\Xcal$ and  $y\in\Supp(f_{k, b}(x))$,
                $\INV_{\Gcal}(t_{k}, y)=(b, x)$.
        \item Efficient Range Superposition:
                For all $k\in\Kcal_{\Gcal}$ and $b\in\bit$,
                \begin{enumerate}
                        \item There exists an efficient deterministic procedure $\CHK_{\Gcal}$ that
                                on input $k$, $b\in\bit$, $x\in\Xcal$, and $y\in\Ycal$, 
                                outputs $1$ if $y\in\Supp(f_{k, b}(x))$ and $0$ otherwise.
                                Note that $\CHK_{\Gcal}$ is not provided the trapdoor $t_{k}$.
                        \item There exists an efficient procedure $\SAMP_{\Gcal}$ that
                                on input $k$ and  $b\in\bit$ returns the state
                                \begin{eqnarray*}
                                         \frac{1}{\sqrt{|\Xcal|}}\sum_{x\in\Xcal, y\in\Ycal}\sqrt{(f_{k, b}(x))(y)}\ket{x}\ket{y}.
                                \end{eqnarray*}
                \end{enumerate}
\end{itemize}
\end{define}

\begin{define}\label{def:II}[Injective Invariance \cite{mahadev}]
        A NTCF family $\Fcal$ is injective invariant if 
        there exists a trapdoor injective function family $\Gcal$ such that
        \begin{itemize}
                \item The algorithm $\CHK_{\Fcal}$ and  $\SAMP_{\Fcal}$ are the same as the algorithms $\CHK_{\Gcal}$ and $\SAMP_{\Gcal}$.
                \item For all quantum polynomial-time procedures $\Acal$, 
                        there exists a negligible function $\negl(\cdot)$ such that
                        \begin{eqnarray*}
                            \left|
                            \Pr_{(k, t_{k})\chosen\GEN_{\Fcal}(1^{\lambda})}\{\Acal(k)=0\}
                            -
                            \Pr_{(k, t_{k})\chosen\GEN_{\Gcal}(1^{\lambda})}\{\Acal(k)=0\}
                            \right|
                            \leq \negl(\lambda).
                        \end{eqnarray*}
        \end{itemize}
\end{define}

\begin{define}[Extended Trapdoor Claw-free Family \cite{mahadev}]
        A NTCF family $\Fcal$ is an extended trapdoor claw-free family if
        \begin{itemize}
                \item $\Fcal$ is injective invariant.
                \item For all $k\in\Kcal_{\Fcal}$ and  $d\in\bit^{w}$, let
                        \begin{eqnarray*}
                                 H_{k, d}':=\{d\cdot(J(x_{0})\oplus J(x_{1}))| (x_{0}, x_{1})\in \mathcal{R}_{k}\}.
                        \end{eqnarray*}
                        For all quantum polynomial-time algorithms $\Acal$, there exists a negligible function $\negl(\cdot)$ such that
                        \begin{eqnarray*}
                            \left|
                            \Pr_{(k, t_{k})\chosen\GEN_{\Fcal}(1^{\lambda})}\{\Acal(k)\in H_{k, d}'\} -\frac{1}{2}
                            \right| \leq \negl(\lambda).
                        \end{eqnarray*}
                \end{itemize}
\end{define}

\begin{define}[Decoding maps for the ENTCF families~\cite{MV20}]
We define the following maps that decode the output of an ENTCF. 
\label{def:dec}
        \begin{itemize}
        \item
                        For a key $k\in\Kcal_{\Gcal}$ and $y\in\Ycal$, let $\hat{b}(k,y)$ be the bit such that 
                        $y$ is in the union of the supports of the distributions $f_{k, \hat{b}(k, y)}(x)$ over $x\in\Xcal$. 
                        This is well-defined because the function pairs in $\Gcal$ have disjoint images. 
        \item
                        For a key $k\in\Kcal_{\Fcal}$ or $\Kcal_{\Gcal}$, $y\in\Ycal$, and $b\in\{0, 1\}$,
                        let $\hat{x}_{b}(k, y)$ be the preimage of the function such that 
                        $y$ is in the support of the distribution $f_{k, b}(\hat{x}_{b}(k, y))$.
                        If $y$ is not in the support, then nothing is defined for $\hat{x}_{b}(k, y)$ 
                        (so instead we define $\hat{x}_{b}(k, y):=\bot$).
        \item
                        For a key $k\in\Kcal_{\Fcal}$, $y\in\Ycal$ and $d\in \Xcal$, 
                        we define $\hat{u}(k,y,d):=d\cdot(\hat{x}_0(k,y)\oplus\hat{x}_1(k,y))$,
                        where the preimages $\hat{x}_{0}(k,y)$ and $\hat{x}_{1}(k,y)$ can be efficiently computed by using 
the trapdoor information $t_{k}$.
                \end{itemize}
\end{define}

\subsection{Definitions}
Throughout the paper, we adopt the following definitions based on~\cite{MV20}. 
\begin{define} (Distance measures)
\label{def:distance}
\begin{enumerate}[(i)]
\item
For $A\in\mathcal{L}(\mathcal{H})$, the schatten-$p$ norm is defined by
\begin{align}
||A||_p:=[\tr(|A|^p)]^{1/p},
\no
\end{align} 
where $|A|:=\sqrt{A^{\dagger}A}$. 
Note that $||A||_1$ is called the trace norm, and $||A||_{\infty}$ is called the operator norm (largest singular value).  
\item
For $A\in\mathcal{L}(\mathcal{H})$ and $\psi\in\Pos(\mathcal{H})$, we define the state-dependent (semi) norm 
of $A$ with respect to $\psi$ as 
\begin{align}
||A||_{\psi}:=\sqrt{\tr[A^{\dagger}A\psi]}.
\no
\end{align}
\end{enumerate}
\end{define}

\begin{define} (Approximate equality)
\label{def:approx}
We use the following symbol for describing an approximate equality. 
\begin{enumerate}[(i)]
\item 
For $a,b\in\mathbb{C}$, we define
 \begin{align}
 a\approx_{\epsilon}b\Leftrightarrow |a-b|=O(\epsilon)+\negl(\lambda).
 \no
 \end{align}

\item
For $A,B\in\mathcal{L}(\mathcal{H})$, we define
 \begin{align}
 A\approx_{\epsilon}B\Leftrightarrow ||A-B||^2_1=O(\epsilon)+\negl(\lambda).
 \no
 \end{align}

\item
For $A,B\in\mathcal{L}(\mathcal{H})$ and $\psi\in\U{Pos}(\mathcal{H})$, we define
 \begin{align}
 A\approx_{\epsilon,\psi}B\Leftrightarrow ||A-B||^2_{\psi}=O(\epsilon)+\negl(\lambda).
 \no
 \end{align}
\end{enumerate}
\end{define}

\begin{define}(Computational indistinguishability) 
\label{def:compindis}
The two states $\psi, \psi'\in\mathcal{D}(\mathcal{H})$ are computationally indistinguishable up to $O(\delta)$ if 
any efficient distinguisher, which takes as input either $\psi$ or $\psi'$ and outputs the bit $b$, satisfies
\begin{align}
\Pr\{b=0|\psi\}\approx_{\delta}\Pr\{b=0|\psi'\}. 
\no
\end{align}
\end{define}
We use the notation
\begin{align}
\psi\overset{c}{\approx}_{\delta}\psi'.
\no
\end{align}

\subsection{Auxiliary Lemmas}	
We summarize auxiliary lemmas that will be frequently used in our soundness in Sec.~\ref{sec:soundness}. 
All the lemmas in this section have been derived in~\cite{MV20}. We state them here for the reader's convenience. 
\begin{lemma}
\label{l:l25}
Let $A_1$ and $A_2$ be efficient commuting binary observables. Then $A_1A_2$ is also an efficient binary observable. 
\end{lemma}

\begin{lemma}
 \label{l:L218} 
\begin{enumerate} [(i)]
\item\label{l:L218i}
Let $\psi\in\U{Pos}(\mathcal{H})$, and $A,B\in\mathcal{L}(\mathcal{H})$. For $C\in\mathcal{L}(\mathcal{H})$ such that $C^{\dagger}C\le I$ 
we have
\begin{align}
A\approx_{\epsilon,\psi}B\Rightarrow CA\approx_{\epsilon,\psi}CB.
\no
\end{align}
\item\label{l:L218ii}
Let $\psi_i\in\U{Pos}(\mathcal{H})$ for $i\in\{1,...,n\}$ with constant $n$, and $A,B\in\mathcal{L}(\mathcal{H})$. 
Define $\psi=\sum_i\psi_i$. Then, 
\begin{align}
\forall i\in\{1,...,n\}: A\approx_{\epsilon,\psi_i}B\Leftrightarrow  A\approx_{\epsilon,\psi}B.
\no
\end{align}
\end{enumerate}
\end{lemma}

\begin{lemma}
 \label{l:L219} 
 Let $\psi\in \U{Pos}(\mathcal{H})$, $\{M^{(a)}\}_{a\in\mathcal{S}}$ a projective measurement with index set $\mathcal{S}$, and $O$ denotes a 
 binary observable 
\begin{align}
O=\sum_a(-1)^{s_a}M^{(a)},
\no
\end{align}
where $s_a\in\{0,1\}$. Suppose there exists an $a'\in\mathcal{S}$ such that
\begin{align}
\tr[M^{(a')}\psi]\approx_\epsilon\tr[\psi]. 
\no
\end{align}
Then, 
\begin{align}
O\approx_{\epsilon,\psi}(-1)^{s_{a'}}I. 
\no
\end{align}
\end{lemma}

\begin{lemma}
 \label{l:L216} 
 Let $\mathcal{H}_1$, $\mathcal{H}_2$ be Hilbert spaces with $\dim(\mathcal{H}_1)\le\dim(\mathcal{H}_2)$ and $V : \mathcal{H}_1\to \mathcal{H}_2$ 
an isometry. 
Let $A$ and $B$ be binary observables on $\mathcal{H}_1$ and $\mathcal{H}_2$, respectively, 
$\psi_1\in\U{Pos}(\mathcal{H}_1)$, $\psi_2\in\U{Pos}(\mathcal{H}_2)$, and $\epsilon\ge0$. Then,
 \begin{align}
\tr\left[V^{\dagger}BVA\psi_1\right]\approx_\epsilon\tr[\psi_1]\Rightarrow V^{\dagger}BV\approx_{\epsilon,\psi_1}A,\no\\
\tr\left[VAV^{\dagger}B\psi_2\right]\approx_\epsilon\tr[\psi_2]\Rightarrow VAV^{\dagger}\approx_{\epsilon,\psi_2}B. \no
\end{align} 
 \end{lemma}

\begin{lemma}
 \label{l:L420} 
 Let $O$ be a binary observable on $\mathcal{H}$ and $\psi\in\U{Pos}(\mathcal{H})$. Then,
 \begin{align}
O\approx_{\epsilon,\psi}(-1)^bI\Rightarrow O^{(b)}\approx_{\epsilon,\psi}I~\U{and}~O^{(\overline{b})}\approx_{\epsilon,\psi}0.
 \no
 \end{align}
\end{lemma}

\begin{lemma}
 \label{l:repl} (Replacement lemma)
 \begin{enumerate} [(i)]
\item\label{l:repli}
Let $\psi\in\U{Pos}(\mathcal{H})$, and $A,B,C\in\mathcal{L}(\mathcal{H})$. If $A\approx_{\epsilon,\psi}B$ and 
$||C||_{\infty}=O(1)$, then 
 \begin{align}
& \tr[CA\psi]\approx_{\sqrt{\epsilon}}\tr[CB\psi],
\no\\
& \tr[AC\psi]\approx_{\sqrt{\epsilon}}\tr[BC\psi].
 \no
 \end{align}
 \item\label{l:replii}
Let $\psi, \psi'\in\U{Pos}(\mathcal{H})$, and $A\in\mathcal{L}(\mathcal{H})$. 
If $\psi\approx_{\epsilon}\psi'$ and $||A||_{\infty}=O(1)$, then 
 \begin{align}
& \tr[A\psi]\approx_{\sqrt{\epsilon}}\tr[A\psi'].
\no
\end{align}
  \end{enumerate}
 \end{lemma}
 
 \begin{lemma}
 \label{l:L222}
 Let $A,B\in\mathcal{L}(\mathcal{H})$ be linear operators, $C\in\mathcal{L}(\mathcal{H})$ a linear operator with constant operator norm, 
 and $\psi\in\U{Pos}(\mathcal{H})$ with $\tr[\psi]\le1$. Then, 
  \begin{align}
  A\approx_{\epsilon,\psi}B\Rightarrow A\psi C\approx_{\epsilon}B\psi C~\U{and~}C\psi A^{\dagger}\approx_{\epsilon}C\psi B^{\dagger}. 
  \no
 \end{align}
 \end{lemma}
 
 \begin{lemma} 
\label{l:L223}
 Let $\mathcal{H}_1$, $\mathcal{H}_2$ be Hilbert spaces with $\dim(\mathcal{H}_1)\le\dim(\mathcal{H}_2)$, 
$V : \mathcal{H}_1\to \mathcal{H}_2$ an isometry, and $A$ and $B$ binary observables on $\mathcal{H}_1$ and $\mathcal{H}_2$, respectively. 
Then, the following holds for any 
$\psi\in\U{Pos}(\mathcal{H}_1)$: 
\begin{align}
&VAV^{\dagger}\approx_{\epsilon,V\psi V^{\dagger}}B\Rightarrow A\approx_{\epsilon,\psi}V^{\dagger}BV,\no\\
&A\approx_{\epsilon,\psi}V^{\dagger}BV\Rightarrow VAV^{\dagger}\approx_{\sqrt{\epsilon},V\psi V^{\dagger}}B.
\no
\end{align}
\end{lemma}

\begin{lemma} 
\label{l:l224}
 Let $\mathcal{H}_1$, $\mathcal{H}_2$ be Hilbert spaces with $\dim(\mathcal{H}_1)\le\dim(\mathcal{H}_2)$ and $V : \mathcal{H}_1\to \mathcal{H}_2$ 
an isometry. 
Let $A$ and $B$ be binary observables on $\mathcal{H}_1$ and $\mathcal{H}_2$, respectively, 
$\psi\in\U{Pos}(\mathcal{H}_1)$, and $\epsilon\ge0$. Then, for any $b\in\{0,1\}$:
\begin{align}
&V^{\dagger}BV\approx_{\epsilon,\psi}A\Rightarrow V^{\dagger}B^{(b)}V\approx_{\epsilon,\psi}A^{(b)},\no\\
&B\approx_{\epsilon,V\psi V^{\dagger}}VAV^{\dagger}\Rightarrow B^{(b)}\approx_{\epsilon,V\psi V^{\dagger}}VA^{(b)}V^{\dagger}.\no
\end{align}
\end{lemma}

\begin{lemma} 
\label{l:lift}
(Lifting lemma) 
Let $\psi$, $\psi'\in \mathcal{D}(\mathcal{H})$ be computationally indistinguishable: 
$\psi\overset{c}{\approx}_{\delta}\psi'$. 
\begin{enumerate} [(i)]
\item\label{rlii}
Let $A$, $B$ be efficient binary observables on $\mathcal{H}$. Then,
\begin{align}
A\approx_{\epsilon,\psi}B\Rightarrow A\approx_{\delta+\epsilon,\psi'}B.
\no
\end{align}
\item\label{rliii}
Let $A$, $B$ be efficient binary observables on $\mathcal{H}$. Then,
\begin{align}
[A,B]\approx_{\epsilon,\psi}0\Rightarrow[A,B]\approx_{\delta+\epsilon,\psi'}0. 
\no
\end{align}
\item\label{rlvi}
 Let $\mathcal{H}'$ be another Hilbert space with $\dim(\mathcal{H}')\ge\dim(\mathcal{H})$. 
 Also, let $\psi,\psi'\in\mathcal{D}(\mathcal{H}')$ such that $\psi\overset{c}{\approx}_{\delta}\psi'$, 
$A$ be an efficient binary observable on $\mathcal{H}$, 
$B$ an efficient binary observable on $\mathcal{H}'$, and $V:\mathcal{H}\to\mathcal{H}'$ an efficient isometry. 
Then
\begin{align}
VAV^{\dagger}\approx_{\epsilon,\psi}B\Rightarrow VAV^{\dagger}\approx_{\epsilon^{1/4}+\delta,\psi'}B.
\no
\end{align}
\end{enumerate}
\end{lemma}

\section{Protocol description}
\label{sec:proto}
In this section, just for self-consistency of the Supplemental Material, we redescribe our self-testing protocol for the entangled magic state 
$CCZ\ket{+}^{\otimes3}$ presented in the main text. 
We remark that 
apart from the difference of the target state to certify, 
our protocol design differs from the one in~\cite{MV20} in the following sense. 
In our protocol, the verifier chooses the state basis $\bm{\theta}$ from $n+2=5$ candidates, 
which is linear in the number $(n)$ of qubits to certify. 
On the other hand, in~\cite{MV20}, the verifier chooses the state basis $\bm{\theta}$ from $2^n$ candidates, which 
takes exponential time on average to generate the target state. 
We reduce the number of state bases by doing checks of $Z$-basis measurement only when the verifier
chooses $\bm{\theta}=000$ and checks $X$-basis measurement using $\bm{\theta}$ with $\wt(\bm{\theta})=1$. 
\\
{\bf Protocol~1}
\begin{enumerate}
\item
The verifier chooses the state bases $\bm{\theta}:=\theta_1\theta_2\theta_3\in_R\mathcal{B}:=\{000,001,010,100,111\}$. 
The basis choices 0 and 1 correspond to the computational basis and the Hadamard basis, respectively. 
We call the basis choice 
$\bm{\theta}\in\{000,001,010,100\}$ the {\it test case} and $\bm{\theta}=111$ the {\it hypergraph case}. 
\item
The verifier samples public keys $k_1,k_2,k_3$ and trapdoors $t_{k_1},t_{k_2},t_{k_3}$ as
\begin{align}
\left\{
\begin{array}{l}
(k_i,t_{k_i})\leftarrow \GEN_{\Gcal}(1^{\lambda})~\U{if}~\theta_i=0,
\\
(k_i,t_{k_i})\leftarrow \GEN_{\Fcal}(1^{\lambda})~\U{if}~\theta_i=1.
\end{array}
\right.
\no
\end{align}
Then, the verifier sends $k_1,k_2,k_3$ to the prover but keeps trapdoors $t_{k_1},t_{k_2},t_{k_3}$ secret from the prover. 
\item
The verifier receives $y_1,y_2,y_3\in\mathcal{Y}$ from the prover. 
\item
The verifier chooses the round type from $\{\U{preimage~round,~Hadamard~round}\}$ uniformly at random and sends it to the 
prover. 
\begin{enumerate}[(i)]
\item
For a preimage~round: The verifier receives $(b_1,x_1;b_2,x_2;b_3,x_3)$ from the prover with $b_i\in\{0,1\}$ and $x_i\in\mathcal{X}$. 
The verifier sets a flag $flag\leftarrow fail_{\U{Pre}}$ except $\CHK(k_i,y_i,b_i,x_i)=1$ holds for all $i\in\{1,2,3\}$.
\item
For an Hadamard~round: 
The verifier receives $d_1,d_2,d_3\in\{0,1\}^w$ 
from the prover. Then, the verifier sends the questions $q_1,q_2,q_3\in_R\{0,1\}$ to the prover, and 
the prover returns the answers $v_1,v_2,v_3\in\{0,1\}$ to the verifier. 
Depending on the basis choice $\bm{\theta}$, the verifier executes the following checks. 
If the flag is set, the verifier rejects the prover. 
\begin{table}[h]
 \centering
  \begin{tabular}{|c|c|}
   \hline
   Basis choice $$& Verifier's check\\
   \hline
   $\bm{\theta}$=000 & Set $flag\leftarrow fail_{\test}$ if the following is true for $i\in_R\{1,2.3\}$:\\
  & ~~~~$
   q_i=0~~\wedge~~\hat{b}(k_i,y_i)\neq v_i.
   $
    \\
   $\bm{\theta}$=100 & Set $flag\leftarrow fail_{\test}$ if the following is true:\\
  & ~~~~$
   q_1=1~~\wedge~~\hat{u}(k_1,y_1,d_1)\oplus\hat{b}(k_2,y_2)\cdot\hat{b}(k_3,y_3)\neq v_1.
   $  \\
  $\bm{\theta}$=010 & Set $flag\leftarrow fail_{\test}$ if the following is true:\\
  & ~~~~$
   q_2=1~~\wedge~~\hat{u}(k_2,y_2,d_2)\oplus\hat{b}(k_1,y_1)\cdot\hat{b}(k_3,y_3)\neq v_2.
   $ \\
   $\bm{\theta}$=001 & Set $flag\leftarrow fail_{\test}$ if the following is true:\\
  & ~~~~$
   q_3=1~~\wedge~~\hat{u}(k_3,y_3,d_3)\oplus\hat{b}(k_1,y_1)\cdot\hat{b}(k_2,y_2)\neq v_3. 
   $ \\
   $\bm{\theta}$=111 & Set $flag\leftarrow fail_{\U{Hyper}}$ if one of the following is true:\\
  & ~~~~$
   \bm{q}=100~~\wedge~~\hat{u}(k_1,y_1,d_1)\neq v_1\oplus v_2\cdot v_3,
   $   \\
   & ~~~~$
   \bm{q}=010~~\wedge~~\hat{u}(k_2,y_2,d_2)\neq v_2\oplus v_1\cdot v_3,
   $   
   \\
   & ~~~~$
   \bm{q}=001~~\wedge~~\hat{u}(k_3,y_3,d_3)\neq v_3\oplus v_1\cdot v_2.
   $   \\
   \hline
  \end{tabular}
\end{table}
\end{enumerate}

\end{enumerate}

\section{protocol completeness}
\label{sec:completness}
In this section, we prove our Theorem~1 in the main text. 
Specifically, we show that there exists an honest prover's strategy, which is accepted by the verifier 
with probability negligibly close to 1. 

First, after receiving the keys $k_1, k_2, k_3$ from the verifier, the prover treats each key separately and prepares the following state for 
$i\in\{1,2,3\}$: 
\begin{align}
\frac{1}{\sqrt{2|\mathcal{X}|}}\sum_{b=0}^1\sum_{x\in\mathcal{X}, y\in\mathcal{Y}}\sqrt{(f_{k_i,b}(x))(y)}\ket{b}\ket{x}\ket{y}.
\no
\end{align}
The preparation of this state can be efficiently done up to negligible error using the procedures from the definition of ENTCF families 
(definition~4.2 in~\cite{mahadev}). 
Then, the prover measures the $y$-register and returns the outcomes $y_1,y_2,y_3\in\mathcal{Y}$ to the verifier. 
At this point, the post-measurement state for each $i\in\{1,2,3\}$ is written as
\begin{eqnarray}
  \begin{cases}
\ket{\hat{b}(k_i,y_i)}\ket{\hat{x}(k_i,y_i)}&~~\U{if~}k_i\in \mathcal{K}_{\mathcal{G}},\no\\
\frac{1}{\sqrt{2}}(\ket{0}\ket{\hat{x}_0(k_i,y_i)}+\ket{1}\ket{\hat{x}_1(k_i,y_i)})&~~\U{if~}k_i\in \mathcal{K}_{\mathcal{F}}.
\no
\end{cases}
\end{eqnarray}
Note that bit $\hat{b}(k,y)$ for $k\in \mathcal{K}_{\mathcal{G}}$ and $y\in\mathcal{Y}$, and preimage 
$\hat{x}_b(k,y)$ with $b\in\{0,1\}$, $k\in \mathcal{K}_{\mathcal{G}}\cup\mathcal{K}_{\mathcal{F}}$ and $y\in\mathcal{Y}$
are defined in Definition~\ref{def:dec}. For simplicity of notation, we define $\hat{x}(k,y):=\hat{x}_{\hat{b}(k,y)}(k,y)$. 
If the verifier chooses the preimage round, it is easy to figure out that the prover is accepted by the verifier with probability negligibly 
close to 1. 

If the verifier chooses the Hadamard round, the prover measures the $x$-register in the Hadamard basis, obtains the outcomes 
$d_1,d_2,d_3\in\{0,1\}^w$ and returns these to the verifier. 
At this point, the prover's state for each $i\in\{1,2,3\}$ is given by
\begin{eqnarray}
  \begin{cases}
\ket{\hat{b}(k_i, y_i)}&~~\U{if~}k_i\in \mathcal{K}_{\mathcal{G}},\no\\
\ket{(-)^{\hat{u}(k_i,y_i,d_i)}}&~~\U{if~}k_i\in \mathcal{K}_{\mathcal{F}}.
\no
\end{cases}
\end{eqnarray}
Here, we define $\hat{u}(k_i,y_i,d_i):=d_i\cdot(\hat{x}_0(k_i,y_i)\oplus \hat{x}_1(k_i,y_i))$. 
Now, the prover performs the $CCZ$ gate among the three qubits and obtains
\begin{eqnarray}
  \begin{cases}
    \ket{\hat{b}(k_1, y_1)}\ket{\hat{b}(k_2, y_2)}\ket{\hat{b}(k_3, y_3)}~~(\U{if}~k_1, k_2, k_3\in \mathcal{K}_{\mathcal{G}}), \\
    \ket{(-)^{\hat{u}(k_i,y_i,d_i)\oplus\hat{b}(k_j, y_j)\cdot \hat{b}(k_l, y_l)}}\ket{\hat{b}(k_j, y_j)}\ket{\hat{b}(k_l, y_l)}
    ~~(\U{if~}k_i\in \mathcal{K}_{\mathcal{F}}, k_j, k_l\in \mathcal{K}_{\mathcal{G}}),\\ 
    \ket{\phi_{\U{H}}^{(\hat{u}(k_1,y_1,d_1),\hat{u}(k_2,y_2,d_2),\hat{u}(k_3,y_3,d_3))}}~~(\U{if}~k_1, k_2, k_3\in \mathcal{K}_{\mathcal{F}}),
  \end{cases}
  \label{comple:CCZ}
\end{eqnarray}
where we define 
$\ket{\phi_{\U{H}}^{(a,b,c)}}:=(\sigma_Z^a\otimes\sigma_Z^b\otimes\sigma_Z^c)CCZ\ket{+}^{\otimes3}$.
It is easy to find that in the first two cases of Eq.~(\ref{comple:CCZ}), the prover's answer is accepted by the verifier. 
For the third case of Eq.~(\ref{comple:CCZ}), by rewriting $\ket{\phi_{\U{H}}^{(u_1,u_2,u_3)}}$ 
(we use the simplified notation: $u_i=\hat{u}(k_i,y_i,d_i)$) depending on $\bm{q}$ as 
\begin{align}
\bm{q}=100:~&\frac{
\ket{(-)^{u_1}}\otimes\left[\sqrt{2}\ket{0,(-)^{u_3}}+
(-1)^{u_2}\ket{1,0}\right]
+(-1)^{u_2+u_3}\ket{(-)^{u_1\oplus1}}\otimes\ket{1,1}
}{2},
\no\\
\bm{q}=010:~&\frac{\left[\sqrt{2}\ket{0,(-)^{u_2},(-)^{u_3}}+(-1)^{u_1}\ket{1,(-)^{u_2},0}\right]
+(-1)^{u_1+u_3}\ket{1,(-)^{u_2\oplus1},1}}{2},
\no\\
\bm{q}=001:~&\frac{
\left[\sqrt{2}\ket{0,(-)^{u_2}}+
(-1)^{u_1}\ket{1,0}\right]\otimes\ket{(-)^{u_3}}
+(-1)^{u_1+u_2}\ket{1,1}\otimes\ket{(-)^{u_3\oplus1}}
}{2},
\no
\end{align}
and if the honest prover measures the qubits in the Pauli-$Z$ or $X$ basis depending on $q_i=0$ or $q_i=1$, 
by returning the measurement outcome as the answer $v_i$, 
it is straightforward to figure out that the prover is accepted by the verifier. \sq

\section{protocol soundness}
\label{sec:soundness}
In this section, we provide the proof of our Theorem~2 presented in the main text. 
\subsection{Modeling Devices in the Protocol}
\label{sec:De}
\begin{define}(Device)
The behavior of an arbitrary prover can be modeled by a device $D:=(S,\Pi,M,P)$, which are specified as follows~\cite{MV20}. 
\begin{enumerate}
\item
(State just after returning images $\bm{y}$) 
We define set of states $S:=\{\psi^{(\bm{\theta})}\}_{\bm{\theta}\in\{0,1\}^3}$ as
\begin{align}
\psi^{(\bm{\theta})}=\sum_{\bm{y}\in\mathcal{Y}^3}\psi^{(\bm{\theta})}_{\bm{y}}\otimes\ket{\bm{y}}\bra{\bm{y}}_Y\in 
\mathcal{D}(\mathcal{H}_D\otimes\mathcal{H}_Y). 
\label{eq:psi}
\end{align}
Note that $\psi^{(\bm{\theta})}$ with $\bm{\theta}\in\mathcal{B}$
represents the state of the prover just after step~3 of Protocol~1, namely the state just after returning 
images $\bm{y}$ to the verifier. 
The state $\psi^{(\bm{\theta})}$ is  implicitly averaged over the keys $(k_1,k_2,k_3)$ chosen by the verifier, and 
all the statements we make in terms of the device $D$ hold on overage over the keys.

\item
(Measurement in the preimage round)  
A projective measurement on systems $\mathcal{H}_D\otimes\mathcal{H}_Y$ performed in the preimage round is defined as
\begin{align}
\Pi=\left\{\Pi^{(\bm{b},\bm{x})}=\sum_{\bm{y}\in\mathcal{Y}^3}
\Pi^{(\bm{b},\bm{x})}_{\bm{y}}\otimes\ket{\bm{y}}\bra{\bm{y}}_Y\right\}_{\bm{b},\bm{x}}. 
\no
\end{align}
Here, $\Pi^{(\bm{b},\bm{x})}_{\bm{y}}$ represents the projective measurement to obtain outcomes $\bm{b}\in\{0,1\}^3$ and $\bm{x}\in\mathcal{X}^3$ given $\bm{y}\in\mathcal{Y}^3$. 
\item (Measurement and post-measurement states in the Hadamard round)  
A projective measurement on systems $\mathcal{H}_D\otimes\mathcal{H}_Y$ performed in the Hadamard round to obtain 
$\bm{d}\in\{0,1\}^{3w}$ is defined as
\begin{align}
M=\left\{M^{(\bm{d})}=\sum_{\bm{y}\in\mathcal{Y}^3}M^{(\bm{d})}_{\bm{y}}\otimes\ket{\bm{y}}\bra{\bm{y}}_Y\right\}_{\bm{d}}. 
\no
\end{align}
For any $\bm{\theta}\in\{0,1\}^3$, 
the post-measurement normalized state after measurement $M$ is written as 
\begin{align}
\rho^{(\bm{\theta})}=\sum_{\bm{y}\in\mathcal{Y}^3,\bm{d}\in\{0,1\}^{3w}}
\sigma^{(\bm{\theta})}_{\bm{y},\bm{d}}\otimes\ket{\bm{y},\bm{d}}\bra{\bm{y},\bm{d}}_{YR},
\label{eq:rho}
\end{align}
where $\sigma^{(\bm{\theta})}_{\bm{y},\bm{d}}:=M^{(\bm{d})}_{\bm{y}}\psi^{(\bm{\theta})}_{\bm{y}}M^{(\bm{d})}_{\bm{y}}$. 
\item (Measurement after receiving questions $\bm{q}$ in the Hadamard round) 
Given the verifier's questions $\bm{q}\in\{0,1\}^3$, $P_{\bm{q}}$ denotes the projective measurement on systems 
$\mathcal{H}_D\otimes\mathcal{H}_Y\otimes\mathcal{H}_R$:
\begin{align}
P_{\bm{q}}=\left\{P_{\bm{q}}^{(\bm{v})}=\sum_{\bm{y}\in\mathcal{Y}^3,\bm{d}\in\{0,1\}^{3w}}P_{\bm{q},\bm{y},\bm{d}}^{(\bm{v})}
\otimes\ket{\bm{y},\bm{d}}\bra{\bm{y},\bm{d}}_{YR}\right\}_{\bm{v}\in\{0,1\}^3}.
\no
\end{align}
By performing this measurement, the prover obtains the outcomes $\bm{v}\in\{0,1\}^3$ that are returned to the verifier.
\end{enumerate}
\end{define}

\begin{define}
\label{def:MO}
For a device $D=(S,\Pi,M,P)$, we define a set of binary observables with projective measurement $P_{\bm{q}}$:
\begin{align}
\left\{A_{i,\bm{q}}:=\sum_{\bm{v}\in\{0,1\}^3}(-1)^{v_i}P^{(\bm{v})}_{\bm{q}}\right\}_{\bm{q}\in\{0,1\}^3}.
\label{def:Marginal}
\end{align}
\end{define}
We call $\{A_{i,\bm{q}=000}\}_{i=1}^3$ and $\{A_{i,\bm{q}=111}\}_{i=1}^3$ {\it non-tilde observables}. 
Any other binary observables $\{A_{i,\bm{q}}\}_{i,\bm{q}}$ are called {\it tilde observables}. 
Note that all the $A_{i,\bm{q}}$ act on the same Hilbert space regardless of $i$ and $\bm{q}$. 
The difference lies in classical post-processing of the answers $\bm{v}$, where 
$A_{i,\bm{q}}$ focuses only on 
the outcome $v_i$ with the other outcomes being marginalized. 
If two binary observables $A_{i,\bm{q}}$ and $A_{j,\bm{q}}$ have the same input $\bm{q}$, 
the only difference is classical post-processing of the measurement outcomes. 
As classical post-processing obviously commute, 
\begin{align}
[A_{i,\bm{q}},A_{j,\bm{q}}]=0
\label{L420Z}
\end{align}
holds for any $i,j\in\{1,2,3\}$.

\begin{define}(Efficient device)
A device $D$ is called efficient if state preparations for $\psi^{(\bm{\theta})}$ and measurements $\Pi, M, P_{\bm{q}}$ can be performed efficiently. 
\end{define}

For any efficient device, from the injective invariance property (Definition~\ref{def:II}), 
post-measurement states $\rho^{(\bm{\theta})}$ in Eq.~(\ref{eq:rho}) are shown to be computationally indistinguishable.  
\begin{lemma}
\label{l:L412}
Let $D$ be an efficient device and $\rho^{(\thetabf)}$ be a post-measurement state defined in Eq.~(\ref{eq:rho}). 
Then, for any $\thetabf, \thetabf'\in\bit^{3}$ and quantum polynomial-time algorithm $\Dcal$,
there exists a negligible function $\negl(\cdot)$ such that
        \begin{align}
            \left|
            \Pr\{\Dcal(\rho^{(\thetabf)})=0\}-\Pr\{\Dcal(\rho^{(\thetabf')})=0\}
            \right|\leq\negl(\lambda).
\label{Cineq}
        \end{align}
The same statement holds for states $\psi^{(\bm{\theta})}$ in Eq.~(\ref{eq:psi}) 
because the following proof is valid also for $\psi^{(\bm{\theta})}$. 
\end{lemma}
(Proof) It suffices to show Eq.~(\ref{Cineq}) for $\thetabf, \thetabf'\in\bit^{3}$ with $\wt(\thetabf\oplus\thetabf')=1$ and 
$\theta_{i}\neq\theta_{i}'$. 
This is because once this is in hand, we can lift it to any $\bm{\theta}, \bm{\theta}'$. 
For instance, when $\bm{\theta}=000$ and $\bm{\theta}'=111$, 
by considering the three inequalities of Eq.~(\ref{Cineq}) with 
$(\bm{\theta},\bm{\theta}')=(000,001), (001,011)$ and $(011,111)$ such that 
the weight of each pair becomes 1 and by using the triangle inequality, 
we obtain Eq.~(\ref{Cineq}) for $\bm{\theta}=000$ and $\bm{\theta}'=111$. 
To prove Eq.~(\ref{Cineq}) for $\thetabf, \thetabf'\in\bit^{3}$ with $\wt(\thetabf\oplus\thetabf')=1$, 
we use the algorithm $\Dcal$ to construct an algorithm $\Acal$ for the injective invariance of $\Fcal$. 
Algorithm $\Acal$ is given a key $k$ of $\Fcal$ or $\Gcal$, sets $k_i:=k$ 
and computes the other keys $k_j$ from $\GEN_{\Gcal}(1^{\lambda})$ if $\theta_j=0$ ($\GEN_{\Fcal}(1^{\lambda})$ if $\theta_j=1$). 
                    $\Acal$ prepares a post-measurement state $\rho^{(\bm{\theta})}$ with $k_{1}, k_{2}$ and $k_{3}$, which is input to $\Dcal$, and 
$\Dcal$ outputs bit $b\in\{0,1\}$. 
Then, we have
    \begin{eqnarray*}
            \begin{aligned}
                & \Pr\{\Dcal(\rho^{(\thetabf)})=0\}=\Pr_{(k, t_{k})\chosen\GEN_{\Gcal}(1^{\lambda})}\{\Acal(k)=0\},\\
                & \Pr\{\Dcal(\rho^{(\thetabf')})=0\}=\Pr_{(k, t_{k})\chosen\GEN_{\Fcal}(1^{\lambda})}\{\Acal(k)=0\},
            \end{aligned}
    \end{eqnarray*}
and substituting these RHS to the LHS of Eq.~(\ref{Cineq}), Eq.~(\ref{Cineq}) holds from the injective invariance of $\Fcal$.
\sq

\subsection{Success Probabilities of a Device}
If the prover's answer is incorrect in the protocol, the verifier sets a flag. In this section, we relate the probabilities that the prover passes these 
checks to the states and measurements in Sec.~\ref{sec:De}. 
Note that Lemmas~\ref{l:pr}, \ref{lgammaTest} and \ref{l:hgc} correspond to Lemmas 4.10 (i), (ii) and (iii) in~\cite{MV20}, respectively. 
\begin{lemma} (Preimage check)
\label{l:pr}
Let $D=(S,\Pi,M,P)$ be a device. 
The probability of passing $i^{\U{th}}$ preimage check (namely $\CHK(k_i,y_i,b_i,x_i)=1$) 
conditioned on basis choice $\bm{\theta}\in\mathcal{B}$ and the preimage round is written as
\begin{align}
&\Pr\{\U{Prover~passes~the}~i^{\U{th}}\U{~preimage~check}|\bm{\theta}, \U{preimage~round}\}\no\\
=&\delta_{\theta_i,0}\sum_{\bm{y},\bm{x}_{\bar{i}},\bm{b}_{\bar{i}}}
\tr\left[\Pi^{(\hat{b}(k_i,y_i),\hat{x}(k_i,y_i);\bm{b}_{\bar{i}},\bm{x}_{\bar{i}})}_{\bm{y}}
\psi^{(\bm{\theta})}_{\bm{y}}\right]
+\delta_{\theta_i,1}\sum_{\bm{y},\bm{x}_{\bar{i}},\bm{b}_{\bar{i}},b}
\tr\left[\Pi^{(b,\hat{x}_b(k_i,y_i);\bm{b}_{\bar{i}},\bm{x}_{\bar{i}})}_{\bm{y}}
\psi^{(\bm{\theta})}_{\bm{y}}\right].
\label{gammaPre2}
\end{align}
Let $p_{\min}$ denote the minimum probability of Eq.~(\ref{gammaPre2}) over $i\in\{1,2,3\}$ and $\bm{\theta}\in\mathcal{B}$, 
and we define
\begin{align}
\gamma_P(D):=1-p_{\min}. 
\label{gammaPre}
\end{align}
Then, the upper bound on $\gamma_P(D)$ is obtained as
\begin{align}
\gamma_P(D)\le 15\cdot\Pr\{flag=fail_{\U{Pre}}|\U{preimage~round}\}.
\label{eq:upper_gammaP}
\end{align}
\end{lemma}
Note that $\Pr\{flag=fail_{\U{Pre}}|\U{preimage~round}\}$ can be estimated through repeating the self-testing protocol. 
\\
(Proof) 
The way of checks in the preimage round is exactly 
the same as that in~\cite{MV20}, and hence by the same argument done in the proof of Lemma 4.10 (i), 
we obtain Eq.~(\ref{eq:upper_gammaP}). 
Note that 15 in Eq.~(\ref{eq:upper_gammaP}) comes from $n\times|\mathcal{B}|=3\times5$ with $n$ being the 
number of qubits to certify, where $n=2$ and $|\mathcal{B}|=4$ in~\cite{MV20}. 
\sq

\begin{lemma} (Test case) 
\label{lgammaTest}
Let $D=(S,\Pi,M,P)$ be a device. 
We define
\begin{align}
\gamma_T(D):=1-\min\left\{\sum_{\bm{v}\in\{0,1\}^3}
\tr\left[A^{(v_i)}_{i,\bm{q}|q_i=0}\sigma^{(0,v_1; 0,v_2;0,v_3)}\right],
\sum_{\bm{v}\in\{0,1\}^3}\tr\left[A^{(v_i)}_{i,\bm{q}|q_i=1}\sigma^{(\theta_1,v_1; \theta_2,v_2;\theta_3,v_3)}\right]\right\}_{i,\bm{q}_{\bar{i}}}
\label{gammaTest}
\end{align}
with
\begin{align}
\sigma^{(0,v_1; 0,v_2;0,v_3)}:=&\sum_{\substack{
\bm{y},\bm{d}:\\ \hat{b}(k_1,y_1)=v_1, \hat{b}(k_2,y_2)=v_2,\hat{b}(k_3,y_3)=v_3}}
\sigma^{(000)}_{\bm{y},\bm{d}}\otimes\ket{\bm{y},\bm{d}}\bra{\bm{y},\bm{d}},
\label{def:sigma000}\\
\sigma^{(\theta_1,v_1; \theta_2,v_2;\theta_3,v_3)}:=
&\sum_{\substack{\bm{y},\bm{d}:\\\hat{u}(k_i,y_i,d_i)\oplus\prod_{j\neq i}\hat{b}(k_j,y_j)=v_i, 
\hat{b}(k_j,y_j)=v_j,\hat{b}(k_l,y_l)=v_l}}
\sigma^{(\bm{\theta}~\U{s.t.}~\wt(\bm{\theta})=1, \theta_i=1)}_{\bm{y},\bm{d}}\otimes\ket{\bm{y},\bm{d}}\bra{\bm{y},\bm{d}}.
\label{def:sigma001}
\end{align}
Then, the upper bound on $\gamma_T(D)$ is given by
\begin{align}
\gamma_T(D)\le 96\cdot\Pr\{flag=fail_{\test}|\U{Test}, \U{Hadamard~round}\}.
\label{eq:upper_gammaT}
\end{align}
\end{lemma}
Note that $\Pr\{flag=fail_{\test}|\U{Test}, \U{Hadamard~round}\}$ can be estimated through repeating the self-testing protocol. \\
(Proof) 
Since the way of checks in the Hadamard round differs from that in~\cite{MV20}, 
we provide the complete proof. 
The probability of obtaining $fail_{\test}$ in the Hadamard round (HR) with the test case is written as 
\begin{align}
\Pr\{flag=fail_{\test}|\U{Test}, \U{HR}\}
=\Pr\{\bm{\theta}=000,  flag=fail_{\test}|\U{Test}, \U{HR}\}
+\Pr\{\wt(\bm{\theta})=1,  flag=fail_{\test}|\U{Test}, \U{HR}\}.
\label{mainbi}
\end{align}
We calculate the first and second terms in turn. First, we focus on the first one: 
\begin{align}
&\Pr\{\bm{\theta}=000,  flag=fail_{\test}|\U{Test}, \U{HR}\}
=\Pr\{\bm{\theta}=000|\U{Test}, \U{HR}\}\cdot\Pr\{flag=fail_{\test}|\bm{\theta}=000, \U{HR}\}
\no\\
=&\frac{1}{12}
\sum_{i=1}^3\sum_{b=0}^1\sum_{y_i}\Pr\{q_i=0,\hat{b}(k_i,y_i)=b,  v_i=\bar{b}|\bm{\theta}=000, \U{HR},i\}
\no\\
=&\frac{1}{8}
-\frac{1}{96}
\sum_{i=1}^3\sum_{b=0}^1\sum_{y_i}\sum_{\bm{q}_{\bar{i}}\in\{0,1\}^2}
\Pr\{\hat{b}(k_i,y_i)=v_i=b|\bm{\theta}=000, \U{HR},i,q_i=0,\bm{q}_{\bar{i}}\}.
\label{main0}
\end{align}
Here, $\sum_{b=0}^1\sum_{y_i}\Pr\{\hat{b}(k_i,y_i)=v_i=b|\bm{\theta}=000, \U{HR},i,q_i=0,\bm{q}_{\bar{i}}\}$ represents the 
probability that the prover's answer $v_i$ is accepted by the verifier conditioned on measuring state $\rho^{(000)}$ when the input to the 
device is $\bm{q}$ with $q_i=0$. 
This probability can be rewritten by using the expressions of the states and measurements as
\begin{align}
\sum_{b=0}^1\tr\left[A^{(v_i=b)}_{i,\bm{q}|q_i=0}\sum_{y_i:\hat{b}(k_i,y_i)=b}
\left(\ket{y_i}\bra{y_i}\rho^{(000)}\ket{y_i}\bra{y_i}\right)
\right]
=\sum_{b=0}^1\tr\left(A^{(v_i=b)}_{i,\bm{q}|q_i=0}
\sum_{\bm{y},\bm{d}: \hat{b}(k_i,y_i)=b}
\sigma_{\bm{y},\bm{d}}^{(000)}\otimes\ket{\bm{y},\bm{d}}\bra{\bm{y},\bm{d}}\right).
\label{main}
\end{align}
By using the definition of Eq.~(\ref{def:sigma000}), Eq.~(\ref{main}) is rewritten as 
\begin{align}
\sum_{\bm{v}\in\{0,1\}^3}\tr(A^{(v_i)}_{i,\bm{q}|q_i=0}\sigma^{(0,v_1; 0,v_2;0,v_3)}).
\label{eq:HR1}
\end{align}
Next, we calculate the second term of Eq.~(\ref{mainbi}): 
\begin{align}
&\Pr\{\wt(\bm{\theta})=1, flag=fail_{\test}|\U{Test}, \U{HR}\}
=\frac{1}{4}\sum_{i=1}^3\Pr\{flag=fail_{\test}|\U{Test}, \U{HR}, \wt(\bm{\theta})=1, \theta_i=1\}
\no\\
=&\frac{1}{4}\sum_{i=1}^3\Pr\{q_i=1, \hat{u}(k_i,y_i,d_i)\oplus\prod_{j\neq i}\hat{b}(k_j,y_j)\neq v_i
|\U{Test}, \U{HR}, \wt(\bm{\theta})=1, \theta_i=1\}
\no\\
=&\frac{1}{32}
\Big[\sum_{i=1}^3\sum_{\bm{q}_{\bar{i}}}1
-\sum_{i=1}^3\sum_{b=0}^1\sum_{\bm{y},d_i}\sum_{\bm{q}_{\bar{i}}}\Pr\{\hat{u}(k_i,y_i,d_i)\oplus\prod_{j\neq i}\hat{b}(k_j,y_j)=v_i=b
|\U{Test}, \U{HR}, \wt(\bm{\theta})=1, \theta_i=1,q_i=1,\bm{q}_{\bar{i}}\}\Big].
\label{TESTLINT}
\end{align}
Here, $\sum_{b=0}^1\sum_{\bm{y},d_i}
\Pr\{\hat{u}(k_i,y_i,d_i)\oplus\prod_{j\neq i}\hat{b}(k_j,y_j)=v_i=b|\U{Test}, \U{HR}, \wt(\bm{\theta})=1, 
\theta_i=1,q_i=1,\bm{q}_{\bar{i}}\}$ expresses the 
probability that the prover's answer $v_i$ is accepted by the verifier conditioned on measuring state $\rho^{(\bm{\theta})}$ 
with $\wt(\bm{\theta})=1$ and $\theta_i=1$ when $\bm{q}$ with $q_i=1$ is input to the device. 
This probability can be written by using the expressions of the states and measurements as
\begin{align}
\sum_{b=0}^1\tr\left(A^{(v_i=b)}_{i,\bm{q}|q_i=1}
\sum_{\bm{y},\bm{d}: \hat{u}(k_i,y_i,d_i)\oplus\prod_{j\neq i}\hat{b}(k_j,y_j)=b}
\sigma_{\bm{y},\bm{d}}^{(\bm{\theta}~\U{s.t.}~\wt(\bm{\theta})=1, \theta_i=1)}\otimes\ket{\bm{y},\bm{d}}\bra{\bm{y},\bm{d}}\right).
\label{main2}
\end{align}
By using the definition of Eq.~(\ref{def:sigma001}), 
for $\bm{\theta}$ such that $\wt(\bm{\theta})=1$ and $\theta_i=1$, Eq.~(\ref{main2}) is rewritten as 
\begin{align}
\sum_{\bm{v}}\tr\left(A^{(v_i)}_{i,\bm{q}|q_i=1}\sigma^{(\theta_1,v_1; \theta_2,v_2;\theta_ 3,v_3)}\right).
\label{eq:HR2}
\end{align}
By substituting Eq.~(\ref{eq:HR1}) to Eq.~(\ref{main0}) and Eq.~(\ref{eq:HR2}) to Eq.~(\ref{TESTLINT}), Eq.~(\ref{mainbi}) results in 
\begin{align}
&\Pr\{flag=fail_{\test}|\U{Test}, \U{HR}\}
=\frac{1}{2}-\frac{1}{96}\sum_{i=1}^3\sum_{\bm{q}_{\bar{i}}}
\left(\tr\left[\sum_{\bm{v}}A^{(v_i)}_{i,\bm{q}|q_i=0}
\sigma^{(0,v_1; 0,v_2;0,v_3)}\right]
+3\tr\left[\sum_{\bm{v}}A^{(v_i)}_{i,\bm{q}|q_i=1}
\sigma^{(\theta_1,v_1; \theta_2,v_2;\theta_3,v_3)}\right]\right).
\no
\end{align}
The RHS has $3\times2^3$ trace terms, and its minimum term is 
$1-\gamma_T(D)$ as defined in Eq.~(\ref{gammaTest}). 
To take a lower bound on the RHS, we replace the $(3\times2^3-1)$ trace terms by 1 and only one term by $1-\gamma_T(D)$. 
By doing so, we have 
\begin{align}
\Pr\{flag=fail_{\test}|\U{Test}, \U{Hadamard~round}\}
&\ge \frac{1}{2}-\frac{1}{96}
\left\{[(3\times2^2-1)\times 1+1\times (1-\gamma_T(D))]+(3\times3\times2^2)\times 1\right\}
\no\\
&=\frac{\gamma_T(D)}{96},
\no
\end{align}
which results in Eq.~(\ref{eq:upper_gammaT}).
\sq

\begin{lemma} (Hypergraph case) 
\label{l:hgc}
Let $D=(S,\Pi,M,P)$ be a device. 
We define
\begin{align}
\gamma_H(D):=1-r_{\min}
\label{gammaH}
\end{align}
with
\begin{align}
r_{\min}:=
\min\Big\{
&\sum_{\bm{s}\in\{0,1\}^3}\tr\left[\left(A_{1,\bm{q}=100}A_{2,\bm{q}=100}^{(0)}
+A_{1,\bm{q}=100}A_{2,\bm{q}=100}^{(1)}A_{3,\bm{q}=100}\right)^{(s_1)}\sigma^{(1,s_1;1,s_2;1,s_3)}\right],\no\\
&\sum_{\bm{s}\in\{0,1\}^3}\tr\left[\left(A_{1,\bm{q}=010}^{(0)}A_{2,\bm{q}=010}
+A_{1,\bm{q}=010}^{(1)}A_{2,\bm{q}=010}A_{3,\bm{q}=010}\right)^{(s_2)}\sigma^{(1,s_1;1,s_2;1,s_3)}\right],\no\\
&\sum_{\bm{s}\in\{0,1\}^3}\tr\left[\left(A_{1,\bm{q}=001}^{(0)}A_{3,\bm{q}=001}
+A_{1,\bm{q}=001}^{(1)}A_{2,\bm{q}=001}A_{3,\bm{q}=001}\right)^{(s_3)}\sigma^{(1,s_1;1,s_2;1,s_3)}\right]
\Big\},
\no
\end{align}
and
\begin{align}
\sigma^{(1,s_1;1,s_2;1,s_3)}:=\sum_{\substack{\bm{y},\bm{d}:\\ \hat{u}(k_1,y_1,d_1)=s_1, \hat{u}(k_2,y_2,d_2)=s_2, \hat{u}(k_3,y_3,d_3)=s_3}}
\sigma^{(111)}_{\bm{y},\bm{d}}\otimes\ket{\bm{y},\bm{d}}\bra{\bm{y},\bm{d}}.
\label{def:sigma111}
\end{align}
Then, the upper bound on $\gamma_H(D)$ is given by
\begin{align}
\gamma_H(D)\le 8\cdot\Pr\{flag=fail_{\U{Hyper}}|\bm{\theta}=111, \U{Hadamard~round}\}.
\label{eq:upper_gammaH}
\end{align}
\end{lemma}
Note that $\Pr\{flag=fail_{\U{Hyper}}|\bm{\theta}=111, \U{Hadamard~round}\}$ can be estimated through repeating the protocol. \\
(Proof) The probability of obtaining $fail_{\U{Hyper}}$ conditioned on choosing the hypergraph case ($\bm{\theta}=111$)
and the Hadamard round (HR) is calculated as 
\begin{align}
&\Pr\{flag=fail_{\U{Hyper}}|\bm{\theta}=111, \U{HR}\}
=\sum_{i=1}^3\Pr\{q_i=1,\bm{q}_{\bar{i}}=00, \hat{u}(k_i,y_i,d_i)\neq v_i\oplus\prod_{j\neq i}v_j|\bm{\theta}=111, \U{HR}\}\no\\
=&\frac{1}{8}\sum_{b=0}^1\sum_{i=1}^3
\sum_{y_i,d_i}
\Pr\{\hat{u}(k_i,y_i,d_i)=b, v_i\oplus\prod_{j\neq i}v_j=\bar{b}|\bm{\theta}=111, \U{HR},q_i=1,\bm{q}_{\bar{i}}=00\}\no\\
=&\frac{3}{8}-\frac{1}{8}\sum_{b=0}^1\sum_{i=1}^3
\sum_{y_i,d_i}\Pr\{\hat{u}(k_i,y_i,d_i)=v_i\oplus\prod_{j\neq i}v_j=b|\bm{\theta}=111, \U{HR},q_i=1,\bm{q}_{\bar{i}}=00\}.
\label{mainbiH}
\end{align}
Here, in the case of $i=1$, 
\begin{align}
\sum_{b=0}^1
\sum_{y_i,d_i}\Pr\{\hat{u}(k_i,y_i,d_i)=v_i\oplus\prod_{j\neq i}v_j=b|\bm{\theta}=111, \U{HR},q_i=1,\bm{q}_{\bar{i}}=00\}
\label{eq:HPR}
\end{align}
represents the 
probability that the prover's answer $v_1\oplus v_2\cdot v_3$ is accepted by the verifier conditioned on measuring state 
$\rho^{(111)}$ when $\bm{q}=100$ is input to the device. 
This probability can be rewritten by using the expressions of the states and measurements as
\begin{align}
&\sum_{b=0}^1
\tr\left[(A_{1,\bm{q}=100}A_{2,\bm{q}=100}^{(0)}
+A_{1,\bm{q}=100}A_{2,\bm{q}=100}^{(1)}A_{3,\bm{q}=100})^{(b)}
\sum_{\substack{y_1,d_1:\\\hat{u}(k_1,y_1,d_1)=b}}\left(\ket{y_1,d_1}\bra{y_1,d_1}\rho^{(111)}\ket{y_1,d_1}\bra{y_1,d_1}
\right)\right]
\no\\
=&\sum_{\bm{s}\in\{0,1\}^3}\tr\left[(A_{1,\bm{q}=100}A_{2,\bm{q}=100}^{(0)}
+A_{1,\bm{q}=100}A_{2,\bm{q}=100}^{(1)}A_{3,\bm{q}=100})^{(s_1)}
\sigma^{(1,s_1;1,s_2;1,s_3)}\right],
\label{HC1}
\end{align}
where $\sigma^{(1,s_1;1,s_2;1,s_3)}$ is defined in Eq.~(\ref{def:sigma111}). 
Also, we obtain analogous expressions for $i=2,3$ in Eq.~(\ref{eq:HPR}). 
By substituting these three expressions to Eq.~(\ref{mainbiH}) 
and using the definition in Eq.~(\ref{gammaH}), we obtain its lower bound as 
$
\Pr\{flag=fail_{\U{Hyper}}|\bm{\theta}=111, \U{HR}\}
\ge3/8-(2+r_{\min})/8=\gamma_H(D)/8$, which results in Eq.~(\ref{eq:upper_gammaH}).
\sq
\\

Next, we introduce a {\it perfect device}, whose $\gamma_P(D)$ in Eq.~(\ref{gammaPre}) is negligible. 
This means that the perfect device can pass the preimage round of our protocol with probability $1-\negl(\lambda)$. 
\begin{define} (Perfect device). We call a device $D=(S,\Pi,M,P)$ perfect if 
$\gamma_P(D) = \negl(\lambda).
$
\end{define}
The following lemma claims that for any efficient device $D$, we can efficiently construct another efficient perfect device $D'$, 
which uses the same measurements as $D$, and whose initial state is close to the one of $D$. 
Lemma~\ref{lper} implies that the efficient device can be replaced with the corresponding perfect one by adding 
an approximation error of $O(\sqrt{\gamma_P(D)})$, 
it suffices to show the soundness proof to the efficient perfect device. 
We omit the proof of Lemma~\ref{lper} as it is essentially the same as that of Lemma 4.13 in~\cite{MV20}.
\begin{lemma}
\label{lper}
Let $D=(S,\Pi, M, P)$ be an efficient device with $S=\{\psi^{(\bm{\theta})}\}_{\bm{\theta}\in\mathcal{B}}$ and 
$\gamma_P(D)<1-1/\U{poly}(\lambda)$. 
Then there exists an efficient perfect device $D'=(S',\Pi, M, P)$, which uses the same measurements $\Pi, M, P$ and 
whose states $S'=\{\psi'^{(\bm{\theta})}\}_{\bm{\theta}\in\mathcal{B}}$
satisfy the following for any $\bm{\theta}\in\mathcal{B}$: 
\begin{align}
||\psi^{(\bm{\theta})}-\psi'^{(\bm{\theta})}||_1\approx_{\sqrt{\gamma_P(D)}}0.
\label{tracePSI}
\end{align}
\label{perfectdevice}
\end{lemma}

At the end of this section, we describe Lemma~\ref{L47} and Corollary~\ref{C48} 
that are frequently used in the rest of our soundness proof. 
We omit these proofs since these are essentially the same as those of Lemma~4.8 and Corollary~4.9~\cite{MV20}. 
\begin{lemma}
\label{L47}
Let $D=(S,\Pi,M,P)$ be a device. For any binary observable $O$, $\bm{\theta}\in\mathcal{B}$, $i\in\{1,2,3\}$ and $\epsilon\ge0$, 
\begin{align}
\sum_{\bm{v}\in\{0,1\}^3}\tr
(O^{(v_i)}\sigma^{(\theta_1,v_1; \theta_2,v_2; \theta_3,v_3)})\approx_{\epsilon}1
\Rightarrow 
\forall\bm{v}\in\{0,1\}^3: 
\tr(\sigma^{(\theta_1,v_1; \theta_2,v_2; \theta_3,v_3)})\approx_{\epsilon}
\tr(O^{(v_i)}\sigma^{(\theta_1,v_1; \theta_2,v_2; \theta_3,v_3)}),
\no
\end{align}
where the definitions of $\sigma^{(\theta_1,v_1; \theta_2,v_2; \theta_3,v_3)}$ are given in Eqs.~(\ref{def:sigma000}), (\ref{def:sigma001}) and 
(\ref{def:sigma111}). 
\end{lemma}

\begin{coro}
\label{C48}
Let $D=(S,\Pi,M,P)$ be a device. For any binary observable $O$, $\bm{\theta}\in\mathcal{B}$, $i\in\{1,2,3\}$ and $\epsilon\ge0$, 
\begin{align}
\sum_{\bm{v}\in\{0,1\}^3}\tr
(O^{(v_i)}\sigma^{(\theta_1,v_1; \theta_2,v_2; \theta_3,v_3)})\approx_{\epsilon}1
\Rightarrow 
\forall\bm{v}\in\{0,1\}^3: 
O\approx_{\epsilon,\sigma^{(\theta_1,v_1; \theta_2,v_2; \theta_3,v_3)}}(-1)^{v_i}I,
\no
\end{align}
where the definitions of states $\sigma^{(\theta_1,v_1; \theta_2,v_2; \theta_3,v_3)}$ are given in Eqs.~(\ref{def:sigma000}), (\ref{def:sigma001}) and 
(\ref{def:sigma111}). 
\end{coro}

In the following Secs.~\ref{sec:anti} and \ref{ApNon-tilde}, 
we only discuss the non-tilde binary observables, namely 
$\{A_{i,\bm{q}=000}\}_{i=1}^3$ and $\{A_{i,\bm{q}=111}\}_{i=1}^3$ in Definition~\ref{def:MO}. 
For simplicity, we use the notations:
\begin{align}
A_{i,\bm{0}}:=A_{i,\bm{q}=000},~~A_{i,\bm{1}}:=A_{i,\bm{q}=111}.
\no
\end{align}

\subsection{Anti-Commutation and Commutation Relations of Non-Tilde Observables}
\label{sec:anti}
If the prover is honest, binary observables 
$A_{i,\bm{0}}$ and $A_{i,\bm{1}}$ are equal to the Pauli ones, which satisfy
the exact commutation and anti-commutation relations. 
In this section, 
we show in Proposition~\ref{Pro:antic} and Lemma~\ref{LACR}
that these relations hold approximately for a general prover modeled by Definition~\ref{sec:De}. 
\begin{pro} (Anti-commutation relation)
\label{Pro:antic}
For any efficient perfect device $D=(S,\Pi,M,P)$, the following approximate anti-commutation relation 
holds for any $i\in\{1,2,3\}$ and $\bm{\theta}\in\mathcal{B}$: 
\begin{align}
\{A_{i,\bm{0}},A_{i,\bm{1}}\}\approx_{\sqrt{\gamma_T(D)},\rho^{(\bm{\theta})}}0. 
\no
\end{align}
\end{pro}
(Proof) 
Once we have the lemmas derived so far, the proof is obtained by following the same argument in Sec.~4.5 of~\cite{MV20}. 

Next, we turn to the commutation relation. 
This corresponds to Proposition 4.24 in~\cite{MV20}, but due to the difference of the protocol design 
mentioned in Sec.~\ref{sec:proto}, we cannot obtain Lemma~\ref{LACR} just by applying the proof in~\cite{MV20}.
We solve this problem in Lemma~\ref{newlemma} by proving 
that the statistics of the measurement outcome $v_i$ given $\bm{q}=000$ for $\bm{\theta}=000$
is close to that for other state basis $\bm{\theta}$. 
\begin{lemma}
\label{newlemma}
For any efficient device $D=(S,\Pi,M,P)$, the following approximate relation holds for any 
$\bm{\theta}\in\mathcal{B}$ with $\theta_j=1$ and $\theta_i=0$ for any $i\neq j$:
\begin{align}
\tr\left[\sum_{\bm{v}\in\{0,1\}^3}A^{(v_i)}_{i,\bm{0}}\sigma^{(\theta_1,v_1;\theta_2,v_2;\theta_3,v_3)}\right]
\approx_0
\tr\left[\sum_{\bm{v}\in\{0,1\}^3}A^{(v_i)}_{i,\bm{0}}\sigma^{(0,v_1;0,v_2; 0,v_3)}\right].
\label{newle}
\end{align}
\end{lemma}
(Proof) 
The RHS [LHS] represents the probability that the prover's answer $v_i$ is accepted by the verifier 
conditioned on measuring state $\rho^{(000)}$ [
$\rho^{(\bm{\theta})}$ (with $\theta_j=1$ and $\theta_i=0$ for any $i\neq j$)] when the input to the device is $\bm{q}=000$. 
We prove Eq.~(\ref{newle}) by contradiction, namely if there exists a non-negligible difference between both 
sides of Eq.~(\ref{newle}), 
we can construct an adversary $\mathcal{A}$ that distinguishes $\rho^{(000)}$ and 
$\rho^{(\bm{\theta})}$ with non-negligible advantage, which contradicts Lemma~\ref{l:L412}. 
The construction of $\mathcal{A}$ is as follows. 

First, adversary $\mathcal{A}$ receives the $j^{\U{th}}$ key $k_j$ from the verifier and samples 
the other keys and trapdoors from the distribution $(k_i,t_{k_i})\leftarrow \GEN_{\Gcal}(1^{\lambda})$ for $i\in\{1,2,3\}\setminus\{j\}$.  
Note that $\mathcal{A}$ does not know whether $k_j\in \mathcal{K}_{\mathcal{G}}$ or $k_j\in \mathcal{K}_{\mathcal{F}}$ that indicates 
$\theta_j=0$ or $\theta_j=1$, respectively. 
Then  $\mathcal{A}$ prepares the state $\psi^{(\bm{\theta})}$ and measures the state to obtain $\bm{y}\in\mathcal{Y}^3$. 
After that  $\mathcal{A}$ performs measurement $M$ and obtains $\bm{d}\in\{0,1\}^{3w}$. 
Next, by using binary observable $\{A_{i,\bm{0}}^{(v_i)}\}_{v_i}$, $\mathcal{A}$ performs 
measurement to know whether his outcome $v_i$ is accepted by the verifier, that is $v_i=\hat{b}(k_i,y_i)$ holds, or not. 
If the outcome $v_i$ is accepted, $\mathcal{A}$ outputs $b=0$. 
The reason why $\mathcal{A}$ can judge whether $v_i$ is accepted or not is that $\mathcal{A}$ knows the $i^{\U{th}}$ trapdoor. 
With the negation of Eq.~(\ref{newle}), we have
\begin{align}
|\Pr\{b=0|\rho^{(\bm{\theta})}
\}-\Pr\{b=0|\rho^{(000)}\}|\ge\mu(\lambda). 
\no
\end{align}
This breaks the computational indistinguishability of $\rho^{(\bm{\theta})}$ stated in Lemma~\ref{l:L412}. 
Note that the proof of Lemma~\ref{l:L412} reveals that 
this lemma also holds even when an efficient adversary $\mathcal{A}$ uses the $l^{\U{th}}$ trapdoor, 
where $l$ indicates the common $\theta$ in $\bm{\theta}$ and $\bm{\theta}'$ with $\wt(\bm{\theta}\oplus\bm{\theta}')\in\{1,2\}$. 
\sq

\begin{lemma} (Commutation relation)
\label{LACR}
For any efficient device $D=(S,\Pi,M,P)$, the approximate commutation relation holds for any $\bm{\theta}\in\mathcal{B}$ and 
any $i, j$ of $i\neq j$: 
\begin{align}
[A_{i,\bm{0}},A_{j,\bm{1}}]\approx_{\gamma_T(D),\rho^{(\bm{\theta})}}0.
\label{L421}
\end{align}
\end{lemma}
(Proof) 
From Lemma~\ref{l:lift} (\ref{rliii}) and computational indistinguishability of $\rho^{(\bm{\theta})}$ stated in Lemma~\ref{l:L412}, 
it suffices to show Eq.~(\ref{L421}) for a specific $\bm{\theta}$. 
We here fix $\bm{\theta}$ to be $\theta_j=1$ and $\theta_i=0$ for any $i\neq j$. 
By the definition of $\gamma_T(D)$ in Eq.~(\ref{gammaTest}), we have
\begin{align}
\tr\left(\sum_{\bm{v}}A^{(v_i)}_{i,\bm{0}}\sigma^{(0,v_1; 0,v_2; 0,v_3)}\right)\approx_{\gamma_T(D)}1,~~~
\tr\left(\sum_{\bm{v}}A^{(v_j)}_{j,\bm{1}}\sigma^{(\theta_1,v_1; \theta_2,v_2; \theta_3,v_3)}\right)\approx_{\gamma_T(D)}1.
\label{L13in1}
\end{align}
Since our protocol checks the $Z$-basis measurement outcome only when 
$\bm{\theta}=000$, we cannot obtain the first equation of 
Eq.~(\ref{L13in1}) for $\sigma^{(\theta_1,v_1; \theta_2,v_2; \theta_3,v_3)}$. 
To make $\sigma$'s in the above two equations to be the same, we apply Lemma~\ref{newlemma} to the first equation of Eq.~(\ref{L13in1}), 
which results in
\begin{align}
\tr\left(\sum_{\bm{v}}A^{(v_i)}_{i,\bm{0}}\sigma^{(\theta_1,v_1; \theta_2,v_2; \theta_3,v_3)}\right)\approx_{\gamma_T(D)}1.
\label{L13in2}
\end{align}
By using Corollary~\ref{C48}, the second equation of Eq.~(\ref{L13in1}) and (\ref{L13in2}) respectively lead to
$A_{j,\bm{1}}\approx_{\gamma_T(D),\sigma^{(\theta_1,v_1; \theta_2,v_2; \theta_3,v_3)}}(-1)^{v_j}I$ and 
$A_{i,\bm{0}}\approx_{\gamma_T(D),\sigma^{(\theta_1,v_1; \theta_2,v_2; \theta_3,v_3)}}(-1)^{v_i}I$. 
Finally, Lemma~\ref{l:L218} (\ref{l:L218i}) implies
\begin{align}
A_{i,\bm{0}}A_{j,\bm{1}}&\approx_{\gamma_T(D),\sigma^{(\theta_1,v_1; \theta_2,v_2; \theta_3,v_3)}}(-1)^{v_j}A_{i,\bm{0}}
\approx_{\gamma_T(D),\sigma^{(\theta_1,v_1; \theta_2,v_2; \theta_3,v_3)}}(-1)^{v_i+v_j}I\no\\
&\approx_{\gamma_T(D),\sigma^{(\theta_1,v_1; \theta_2,v_2; \theta_3,v_3)}}A_{j,\bm{1}}\cdot(-1)^{v_i}I
\approx_{\gamma_T(D),\sigma^{(\theta_1,v_1; \theta_2,v_2; \theta_3,v_3)}}A_{j,\bm{1}}A_{i,\bm{0}},\no
\end{align}
which ends the proof. \sq

\subsection{Approximate Relations of Non-Tilde Observables and Pauli Observables}
\label{ApNon-tilde}
In this section, we introduce {\it swap isometry}. 
This isometry is a completely positive and trace preserving (CPTP) 
map that adds three-qubit Hilbert space $\mathbb{C}^8$ to prover's Hilbert space $\mathcal{H}$ and 
swaps the three-qubit space in $\mathcal{H}$ to $\mathbb{C}^8$. 
This isometry is an extension of the one in Definition 4.27~\cite{MV20} to the three-qubit case. 
\begin{define} (Swap isometry) 
Given a device $D=(S,\Pi,M,P)$ with Hilbert space $\mathcal{H}$, 
we define swap isometry $V_S: \mathcal{H}\to\mathbb{C}^8\otimes \mathcal{H}$ using non-tilde observables 
introduced in Eq.~(\ref{def:Marginal}) as
\begin{align}
V_S=\sum_{a,b,c\in\{0,1\}}\ket{a,b,c}\otimes
\left[A^c_{3,\bm{1}}A^{(c)}_{3,\bm{0}}A^b_{2,\bm{1}}A^{(b)}_{2,\bm{0}}A^a_{1,\bm{1}}A^{(a)}_{1,\bm{0}}\right].
\label{defV}
\end{align}
Here, superscript $a$ in $A^a$ indicates the exponent, and $A^{(a)}$ is the projector onto $(-1)^{a}$-eigenspace of $A$. 
\end{define}
The goal of this section is to prove Lemmas~\ref{conjPauliequal}, \ref{1stXPauli}, \ref{Lemma430}, 
and \ref{NewLemma430}, which state 
that non-tilde observables $A_{i,\bm{0}}$ and $A_{i,\bm{1}}$ are close to the Pauli observables under isometry $V_S$. 
\begin{lemma}
\label{conjPauliequal}
Conjugating Pauli observables by swap isometry $V_S$ gives the following. 
\begin{align}
V_S^{\dagger}\sigma_{Z,1}V_S&=A_{1,\bm{0}}
\label{eq:i}\\
V_S^{\dagger}\sigma_{Z,2}V_S&=\sum_{a=0}^1A^{(a)}_{1,\bm{0}}A^a_{1,\bm{1}}A_{2,\bm{0}}
A^a_{1,\bm{1}}A^{(a)}_{1,\bm{0}}
\label{eq:ii}\\
V_S^{\dagger}\sigma_{Z,3}V_S
&=\sum_{a,b=0}^1A^{(a)}_{1,\bm{0}}A^a_{1,\bm{1}}
A^{(b)}_{2,\bm{0}}A_{2,\bm{1}}^bA_{3,\bm{0}}A_{2,\bm{1}}^bA^{(b)}_{2,\bm{0}}A_{1,\bm{1}}^aA^{(a)}_{1,\bm{0}}
\label{eq:iii}
\\
V_S^{\dagger}\sigma_{X,1}V_S&=\sum_{a=0}^1A^{(\overline{a})}_{1,\bm{0}}A_{1,\bm{1}}
A^{(a)}_{1,\bm{0}}
\label{eq:iv}
\\
V_S^{\dagger}\sigma_{X,2}V_S
&=\sum_{a,b=0}^1A^{(a)}_{1,\bm{0}}A^a_{1,\bm{1}}
A^{(b)}_{2,\bm{0}}A_{2,\bm{1}}A^{(\overline{b})}_{2,\bm{0}}A_{1,\bm{1}}^aA^{(a)}_{1,\bm{0}}
\label{eq:v}\\
V_S^{\dagger}\sigma_{X,3}V_S&=
\sum_{a,b,c=0}^1A^{(a)}_{1,\bm{0}}A^a_{1,\bm{1}}
A^{(b)}_{2,\bm{0}}A^b_{2,\bm{1}}A^{(\overline{c})}_{3,\bm{0}}A_{3,\bm{1}}A^{(c)}_{3,\bm{0}}
A_{2,\bm{1}}^bA^{(b)}_{2,\bm{0}}A^a_{1,\bm{1}}A^{(a)}_{1,\bm{0}}
\label{eq:vi}
\end{align}
Here, $\sigma_{Z,i}$ and $\sigma_{X,i}$ denote $\sigma_Z$ and $\sigma_X$ acting on the $i^{\U{th}}$ qubit, respectively.
\end{lemma}
(Proof) These can be proven by inserting Eq.~(\ref{defV}). \sq

Next, we show that under isometry $V_S$, the binary observable $A_{1,\bm{1}}$ is approximately equal to $\sigma_X$. 
\begin{lemma}
\label{1stXPauli}
For any efficient perfect device $D=(S,\Pi,M,P)$, we have for any $\bm{\theta}\in\mathcal{B}$,
\begin{align}
V_S^{\dagger}\sigma_{X,1}V_S\approx_{\sqrt{\gamma_T(D)},\rho^{(\bm{\theta})}}A_{1,\bm{1}}.
\label{L427re}
\end{align}
\label{Lemma427}
\end{lemma}
(Proof) 
The proof follows from that in Lemma 4.30~\cite{MV20} using Eq.~(\ref{eq:iv}), 
Proposition~\ref{Pro:antic} and Lemmas~\ref{l:L216} and \ref{l:repl} (\ref{l:repli}).\sq

From Eqs.~(\ref{eq:i}) and (\ref{L427re}), $A_{1,\bm{q}=\bm{0}}$ and $A_{1,\bm{q}=\bm{1}}$ 
are shown to be close to Pauli-$Z$ and $X$ observables, respectively. 
Using these equations, we partially characterize the prover's states. Specifically, we show that the prover's states 
can be written as product states where the first qubit is the eigenstate of either $\sigma_Z$ or $\sigma_X$ depending on $\theta_1\in\{0,1\}$. 
\begin{lemma}
\label{L428S}
Let $D=(S,\Pi,M,P)$ be an efficient perfect device. For any $\bm{v}\in\{0,1\}^3$, there exists positive matrices 
$\alpha^{(0,v_1;0,v_2;0,v_3)}, \alpha^{(0,v_1;0,v_2;1,v_3)}, \alpha^{(0,v_1;1,v_2;0,v_3)}$ and $\alpha^{(1,v_1;0,v_2;0,v_3)}$ 
such that the following holds:
\begin{align}
&V_S\sigma^{(0,v_1;0,v_2;0,v_3)}V_S^{\dagger}\approx_{\gamma_T(D)^{1/4}}\ket{v_1}\bra{v_1}\otimes\alpha^{(0,v_1;0,v_2;0,v_3)},
\label{L428first}\\
&V_S\sigma^{(0,v_1;0,v_2;1,v_3)}V_S^{\dagger}\approx_{\gamma_T(D)^{1/4}}\ket{v_1}\bra{v_1}\otimes\alpha^{(0,v_1;0,v_2;1,v_3)},
\label{L428second}\\
&V_S\sigma^{(0,v_1;1,v_2;0,v_3)}V_S^{\dagger}\approx_{\gamma_T(D)^{1/4}}\ket{v_1}\bra{v_1}\otimes\alpha^{(0,v_1;1,v_2;0,v_3)},
\label{L428third}\\
&V_S\sigma^{(1,v_1;0,v_2;0,v_3)}V_S^{\dagger}\approx_{\gamma_T(D)^{1/4}}\ket{(-)^{v_1}}\bra{(-)^{v_1}}\otimes\alpha^{(1,v_1;0,v_2;0,v_3)}.
\label{L428last}
\end{align}
\end{lemma}
There are four approximate relations since there are four states corresponding to $\bm{\theta}=000, 001,010$ and 100 
in the test case of our protocol. The proof is similar to Lemma 4.31~\cite{MV20}, but we give the full proof for completeness.
\\
(Proof) We first prove~Eq.~(\ref{L428last}). 
From Lemma~\ref{Lemma427}, we have
\begin{align}
A_{1,\bm{1}}\approx_{\sqrt{\gamma_T(D)},\rho^{(100)}}V_S^{\dagger}\sigma_{X,1}V_S,
\label{PL428F}
\end{align}
and applying Lemmas~\ref{l:L218} (\ref{l:L218ii}), \ref{l:l224} and \ref{l:repl} (\ref{l:repli}) in this order results in
\begin{align}
\sum_{\bm{v}}\tr\left[V_S^{\dagger}
(\ket{(-)^{v_1}}\bra{(-)^{v_1}}\otimes I^{\otimes2}_2\otimes I)V_S\sigma^{(1,v_1;0,v_2;0,v_3)}\right]
\approx_{\gamma_T(D)^{1/4}}\sum_{\bm{v}}\tr[A^{(v_1)}_{1,\bm{1}}\sigma^{(1,v_1;0,v_2;0,v_3)}].
\end{align}
From the definition of $\gamma_T(D)$ in Eq.~(\ref{gammaTest}), the RHS is approximately equal as $\approx_{\gamma_T(D)}1$,  
and hence the LHS results in
\begin{align}
\sum_{\bm{v}}\tr\left[V_S^{\dagger}
(\ket{(-)^{v_1}}\bra{(-)^{v_1}}\otimes I^{\otimes2}_2\otimes I)V_S\sigma^{(1,v_1;0,v_2;0,v_3)}\right]
\approx_{\gamma_T(D)^{1/4}}1. 
\no
\end{align}
From Lemma~\ref{l:L420} and Corollary~\ref{C48}, this leads to
$\ket{(-)^{v_1}}\bra{(-)^{v_1}}\otimes I^{\otimes2}_2\otimes I
\approx_{\gamma_T(D)^{1/4},V_S\sigma^{(1,v_1;0,v_2;0,v_3)}V_S^{\dagger}}I$. 
Finally, using Lemma~\ref{l:L222} implies
\begin{align}
V_S\sigma^{(1,v_1;0,v_2;0,v_3)}V_S^{\dagger}\approx_{\gamma_T(D)^{1/4}}
(\ket{(-)^{v_1}}\bra{(-)^{v_1}}\otimes I^{\otimes2}_2\otimes I)V_S\sigma^{(1,v_1;0,v_2;0,v_3)}V_S^{\dagger}
(\ket{(-)^{v_1}}\bra{(-)^{v_1}}\otimes I^{\otimes2}_2\otimes I).
\no
\end{align}
By defining
\begin{align}
\alpha^{(1,v_1;0,v_2;0,v_3)}:=(\bra{(-)^{v_1}}\otimes I^{\otimes2}_2\otimes I)V_S\sigma^{(1,v_1;0,v_2;0,v_3)}V_S^{\dagger}
(\ket{(-)^{v_1}}\otimes I^{\otimes2}_2\otimes I),
\no
\end{align}
we obtain the desired relation of Eq.~(\ref{L428last}). 
The other relations Eq.~(\ref{L428first})-(\ref{L428third}) 
can be proven in the same way just by replacing Eq.~(\ref{PL428F}) in the above proof with Eq.~(\ref{eq:i}). 
\sq
\\

In the following discussions, we use the simplified notation $A\approx_{R,\psi} B$ 
if there exists a constant $c>0$ such that 
$A\approx_{\gamma_T(D)^c,\psi} B$. 

We next show that the prover's auxiliary states are computationally indistinguishable to the prover. 
We omit the proof because once Lemma~\ref{L428S} is in hand, 
the proof is exactly the same argument with that of Lemma 4.32~\cite{MV20}. 
\begin{lemma}
\label{Lemma429S}
Let $D=(S,\Pi,M,P)$ be an efficient perfect device. There exists a normalized state $\alpha$ 
such that the following holds for any $v_1\in\{0,1\}$:
\begin{align}
&\sum_{v_2,v_3}V_S\sigma^{(0,v_1;0,v_2;0,v_3)}V_S^{\dagger}\overset{c}{\approx}_R
\frac{\ket{v_1}\bra{v_1}}{2}\otimes\alpha,~~
\sum_{v_2,v_3}V_S\sigma^{(0,v_1;0,v_2;1,v_3)}V_S^{\dagger}\overset{c}{\approx}_R
\frac{\ket{v_1}\bra{v_1}}{2}\otimes\alpha,
\no
\\
&\sum_{v_2,v_3}V_S\sigma^{(0,v_1;1,v_2;0,v_3)}V_S^{\dagger}\overset{c}{\approx}_R
\frac{\ket{v_1}\bra{v_1}}{2}\otimes\alpha,~~
\sum_{v_2,v_3}V_S\sigma^{(1,v_1;0,v_2;0,v_3)}V_S^{\dagger}\overset{c}{\approx}_R
\frac{\ket{(-)^{v_1}}\bra{(-)^{v_1}}}{2}\otimes\alpha.
\no
\end{align}
\end{lemma}

So far, we have shown that the state of the first register is approximately equal to the eigenstate of the Pauli observable. 
Using Lemmas~\ref{l:L412}	, \ref{LACR}, \ref{1stXPauli} and \ref{Lemma429S} and Proposition~\ref{Pro:antic}, 
the second observables are shown to be close to the Pauli observables under isometry $V_S$. 
We omit the poof since it is essentially the same as Lemma 4.33 in~\cite{MV20}. 
\begin{lemma}
\label{Lemma430}
For any efficient perfect device $D=(S,\Pi,M,P)$, we have for any $\bm{\theta}\in\mathcal{B}$, 
\begin{align}
&V_S^{\dagger}\sigma_{Z,2}V_S\approx_{R,\rho^{(\bm{\theta})}}A_{2,\bm{0}},
\label{L430Z2}\\
&V_S^{\dagger}\sigma_{X,2}V_S\approx_{R,\rho^{(\bm{\theta})}}A_{2,\bm{1}}.
\label{L430X2}
\end{align}
\end{lemma}
Using this result, we will characterize the state of the first and the second registers, which is an extension of Lemma~\ref{L428S}. 
Specifically, we show that the prover's states are approximately equal to the product states 
where the first and the second qubits are the eigenstates of the Pauli observables depending on $\theta_1$ and $\theta_2$. 
\begin{lemma} (Extension of Lemma~\ref{L428S})
\label{Lemma434} 
For any efficient and perfect device 
$D=(S,\Pi,M,P)$, we have for any $\bm{v}\in\{0,1\}^3$, there exists positive matrices 
$\tilde{\alpha}^{(0,v_1;0,v_2;0,v_3)}$, $\tilde{\alpha}^{(0,v_1;0,v_2;1,v_3)}$, $\tilde{\alpha}^{(0,v_1;1,v_2;0,v_3)}$ and 
$\tilde{\alpha}^{(1,v_1;0,v_2;0,v_3)}$ such that the 
following holds with $P[\cdot]:=\ket{\cdot}\bra{\cdot}$.
\begin{align}
&V_S\sigma^{(0,v_1; 0,v_2; 0,v_3)}V_S^{\dagger}\approx_{R}P[\ket{v_1,v_2}]\otimes \tilde{\alpha}^{(0,v_1;0,v_2;0,v_3)}
\label{L481}\\
&V_S\sigma^{(0,v_1; 0,v_2; 1,v_3)}V_S^{\dagger}\approx_{R}P[\ket{v_1,v_2}]\otimes \tilde{\alpha}^{(0,v_1;0,v_2;1,v_3)}
\label{L482}\\
&V_S\sigma^{(0,v_1; 1,v_2; 0,v_3)}V_S^{\dagger}\approx_{R}P[\ket{v_1,(-)^{v_2}}]\otimes 
\tilde{\alpha}^{(0,v_1;1,v_2;0,v_3)}
\label{L483}\\
&V_S\sigma^{(1,v_1; 0,v_2; 0,v_3)}V_S^{\dagger}\approx_{R}P[\ket{(-)^{v_1},v_2}]\otimes 
\tilde{\alpha}^{(1,v_1;0,v_2;0,v_3)}
\label{L484}
\end{align}
\end{lemma}
(Proof) 
The proof is similar to Lemma~\ref{L428S} and Lemma 4.37 in~\cite{MV20}, so we sketch it for Eq.~(\ref{L484}). 
Applying Lemmas~\ref{Lemma430}, \ref{l:L218} (\ref{l:L218ii}), Lemma~\ref{l:l224} and Lemma~\ref{l:repl} (\ref{l:repli}) in this order, we have
\begin{align}
\sum_{\bm{v}}\tr\left[A_{2,\bm{0}}^{(v_2)}\sigma^{(1,v_1;0,v_2;0,v_3)}\right]
\approx_R
\sum_{\bm{v}}\tr\left[V_S^{\dagger}(I_2\otimes\ket{v_2}\bra{v_2}\otimes I_2\otimes I)V_S\sigma^{(1,v_1;0,v_2;0,v_3)}\right].
\label{L434inProof1}
\end{align}
From Lemma~\ref{newlemma},  
\begin{align}
\sum_{\bm{v}}\tr[A_{2,\bm{0}}^{(v_2)}\sigma^{(1,v_1;0,v_2;0,v_3)}]
\approx_0
\sum_{\bm{v}}\tr[A_{2,\bm{0}}^{(v_2)}\sigma^{(0,v_1;0,v_2;0,v_3)}]
\no
\end{align}
holds, and by the definition of $\gamma_T(D)$ in Eq.~(\ref{gammaTest}), the RHS is approximately equal to 1. 
This means that the RHS of Eq.~(\ref{L434inProof1}) is also approximately equal to 1. 
Combining this fact, Corollary~\ref{C48} and Lemma~\ref{l:L420} results in 
$I_2\otimes\ket{v_2}\bra{v_2}\otimes I_2\otimes I\approx_{R,V_S\sigma^{(1,v_1;0,v_2;0,v_3)}V_S^{\dagger}}I$. 
From Lemma~\ref{L428S}, the state in the subscript of $\approx$ is close to $\ket{(-)^{v_1}}\bra{(-)^{v_1}}\otimes\alpha^{(1,v_1;0,v_2;0,v_3)}$, and 
Lemma~\ref{l:repl} (\ref{l:replii}) enables us to replace the state as
$I_2\otimes\ket{v_2}\bra{v_2}\otimes I_2\otimes I\approx_{R,\ket{(-)^{v_1}}\bra{(-)^{v_1}}\otimes\alpha^{(1,v_1;0,v_2;0,v_3)}}I$. 
Finally, from Eq.~(\ref{L428last}) and Lemma~\ref{l:L222}, we have 
\begin{align}
V_S\sigma^{(1,v_1;0,v_2;0,v_3)}V_S^{\dagger}\approx_R
P[\ket{(-)^{v_1},v_2}]\otimes\tilde{\alpha}^{(1,v_1;0,v_2;0,v_3)}.
\no
\end{align}
With $\tilde{\alpha}^{(1,v_1;0,v_2;0,v_3)}:=(\bra{v_2}\otimes I)\alpha^{(1,v_1;0,v_2;0,v_3)}(\ket{v_2}\otimes I)$, we obtain the desired relation. 
\sq

Using Lemma~\ref{Lemma434}, we next show an extension of Lemma~\ref{Lemma429S}, which states 
that prover's auxiliary states with the first and second registers are computationally indistinguishable to the prover. 
\begin{lemma} 
\label{Lemma435} 
(Extension of Lemma~\ref{Lemma429S})
Let $D=(S,\Pi,M,P)$ be an efficient perfect device. There exists a normalized state $\tilde{\alpha}$ such that 
the following holds for any $v_1,v_2\in\{0,1\}$:
\begin{align}
&\sum_{v_3}V_S\sigma^{(0,v_1; 0,v_2; 0,v_3)}V_S^{\dagger}\overset{c}{\approx}_{R}\frac{P[\ket{v_1,v_2}]}{4}\otimes \tilde{\alpha},~~
\sum_{v_3}V_S\sigma^{(0,v_1; 0,v_2; 1,v_3)}V_S^{\dagger}\overset{c}{\approx}_{R}\frac{P[\ket{v_1,v_2}]}{4}\otimes \tilde{\alpha},
\no\\
&\sum_{v_3}V\sigma^{(0,v_1; 1,v_2; 0,v_3)}V_S^{\dagger}\overset{c}{\approx}_{R}\frac{P[\ket{v_1,(-)^{v_2}}]
}{4}\otimes \tilde{\alpha},~~
\sum_{v_3}V_S\sigma^{(1,v_1; 0,v_2; 0,v_3)}V_S^{\dagger}\overset{c}{\approx}_{R}\frac{P[\ket{(-)^{v_1},v_2}]}{4}\otimes \tilde{\alpha}.
\no
\end{align}
\end{lemma}
(Proof) 
By repeating exactly the same argument done in the proof of Lemma~\ref{Lemma429S} and Lemma 4.39 in~\cite{MV20}, we have this lemma. 
\sq

With Lemma~\ref{Lemma435} in hand, we obtain a simple corollary describing an approximate relation of state $\rho^{(\bm{\theta})}$. 
\begin{coro} 
\label{Corollary436} 
Let $D=(S,\Pi,M,P)$ be an efficient perfect device. 
There exists a normalized state $\tilde{\alpha}$ such that $\forall\bm{\theta}\in\mathcal{B}$,
\begin{align}
V_S\rho^{(\bm{\theta})}V_S^{\dagger}\overset{c}{\approx}_R\frac{I_2\otimes I_2}{4}\otimes\tilde{\alpha}. 
\end{align}
\end{coro}
(Proof) The proof follows from the one in Corollary 4.40~\cite{MV20}. 
Taking the sum of the equations in 
Lemma~\ref{Lemma435} over $v_1$ and $v_2$ yields the the statement for $\bm{\theta}$ of the test case. 
We can lift up the statement for any $\bm{\theta}\in\mathcal{B}$ thanks to Lemma~\ref{l:L412}. \sq

So far, we have shown that the prover's states are approximately equal to the product states 
where the first and the second registers are in the qubit states. 
Using Lemmas~\ref{LACR} and \ref{Lemma430}, Proposition~\ref{Pro:antic} and Corollary~\ref{Corollary436}, we can prove that 
$A_{3,\bm{0}}$ and $A_{3,\bm{1}}$ are approximately equal to the Pauli observables under isometry $V_S$. 
\begin{lemma} 
\label{NewLemma430} 
For any efficient perfect device $D=(S,\Pi,M,P)$, we have for any $\bm{\theta}\in\mathcal{B}$, 
\begin{align}
&V_S^{\dagger}\sigma_{Z,3}V_S\approx_{R,\rho^{(\bm{\theta})}}A_{3,\bm{0}},
\label{newL430Z}\\
&V_S^{\dagger}\sigma_{X,3}V_S\approx_{R,\rho^{(\bm{\theta})}}A_{3,\bm{1}}.
\label{newL430X}
\end{align}
\end{lemma}

\subsection{Approximate Relations of Tilde Observables and Pauli Observables}
\label{subsec:tilde}
In this section, we prove in Corollary~\ref{Corollary433} that 
the tilde observables in Eq.~(\ref{def:Marginal}) are also close to the Pauli observables under isometry $V_S$. 
In so doing, we first prove that the tilde observables are close to the non-tilde ones. 
\begin{lemma}
\label{Lemma432} 
For any efficient perfect device $D=(S,\Pi,M,P)$, we have for any $\bm{\theta}\in\mathcal{B}$ and $i\in\{1,2,3\}$,
\begin{align}
&
A_{i,\bm{q}}\approx_{\gamma_T(D),\rho^{(\bm{\theta})}}A_{0,\bm{0}}~~\U{for}~q_i=0~\U{and}~\bm{q}_{\bar{i}}\in\{0,1\}^2\setminus\{00\},
\no\\
&
A_{i,\bm{q}}\approx_{\gamma_T(D),\rho^{(\bm{\theta})}}A_{1,\bm{1}}~~\U{for}~q_i=1~\U{and}~\bm{q}_{\bar{i}}\in\{0,1\}^2\setminus\{11\}.
\no
\end{align}
\end{lemma}
(Proof) The proof follows from that of Lemma 4.35 in~\cite{MV20}.  
We prove $A_{1,001}\approx_{\gamma_T(D),\rho^{(\bm{\theta})}}A_{1,\bm{0}}$, and the others can be proven analogously. 
Once we prove 
\begin{align}
A_{1,001}\approx_{\gamma_T(D),\rho^{(000)}}A_{1,\bm{0}},
\label{L431Z}
\end{align}
Lemma~\ref{l:lift} (\ref{rlii}) implies $A_{1,001}\approx_{\gamma_T(D),\rho^{(\bm{\theta})}}A_{1,\bm{0}}$ for any $\bm{\theta}\in\mathcal{B}$. 
The conditions of Lemma~\ref{l:lift} (\ref{rlii}) are guaranteed by 
computational indistinguishability of $\rho^{(\bm{\theta})}$ in Lemma~\ref{l:L412} and by the fact that 
$A_{1,001}$ and $A_{1,\bm{0}}$ are efficient binary observables. 
Therefore, it suffices to show Eq.~(\ref{L431Z}). 
Moreover, thanks to Lemma~\ref{l:L218} (\ref{l:L218ii}), the proof is reduced to showing 
\begin{align}
A_{1,001}\approx_{\gamma_T(D),\sigma^{(0,v_1;0,v_2;0,v_3)}}A_{1,\bm{0}}.
\label{A1001}
\end{align}
From the definition of $\gamma_T(D)$ in Eq.~(\ref{gammaTest}) and Corollary~\ref{C48}, we have
$A_{1,001}\approx_{\gamma_T(D),\sigma^{(0,v_1;0,v_2;0,v_3)}}(-1)^{v_1}I$ and 
$A_{1,\bm{0}}\approx_{\gamma_T(D),\sigma^{(0,v_1;0,v_2;0,v_3)}}(-1)^{v_1}I$ for any $\bm{v}\in\{0,1\}^3$. 
Hence, the triangle inequality of the state dependent norm results in Eq.(\ref{A1001}).\sq

\begin{coro}
\label{Corollary433}
For any efficient perfect device $D=(S,\Pi,M,P)$, we have the following for any $\bm{\theta}\in\mathcal{B}$ and $i\in\{1,2,3\}$, 
\begin{align} 
&A_{i,\bm{q}}\approx_{R,\rho^{(\bm{\theta})}}V_S^{\dagger}\sigma_{Z,i}V_S~~\U{for}~q_i=0~\U{and}~\bm{q}_{\bar{i}}\in\{0,1\}^2\setminus\{00\},
\label{C433Z1}\\
&A_{i,\bm{q}}\approx_{R,\rho^{(\bm{\theta})}}V_S^{\dagger}\sigma_{X,i}V_S~~\U{for}~q_i=1~\U{and}~\bm{q}_{\bar{i}}\in\{0,1\}^2\setminus\{11\}.
\label{C433X3}
\end{align}
\end{coro}
(Proof) We follow the proof of Corollary 4.36 in~\cite{MV20}. 
Eqs.~(\ref{C433Z1}) and (\ref{C433X3}) can be proven by combining Lemma~\ref{Lemma432} 
with Eqs.~(\ref{eq:i}), (\ref{L427re}), (\ref{L430Z2}), (\ref{L430X2}), (\ref{newL430Z}) and (\ref{newL430X}). 
\sq

\subsection{Approximate Relations of Joint Observables and Products of Pauli Observables}
\label{subsec:joint}
In this section, we prove that the 
joint observables are close to the products of Pauli ones. 
Specifically, any two and three joint observables are close to the products of Pauli observables in Lemmas~\ref{Lemma437} and 
\ref{NewLemma437}, and observables of the generalized stabilizers are close to the ideal ones 
in Lemma~\ref{AltLemma437}. 
These lemmas are the crux of proving our main result, Theorem~\ref{Theorem438}. 
To derive these relations, we first prepare the extended statements of Lemmas~\ref{Lemma434} and \ref{Lemma435} 
and Corollary~\ref{Corollary436} in Lemmas~\ref{NewLemma434} and \ref{NewLemma435} and Corollary~\ref{NewCorollary436}, respectively.

\begin{lemma}
\label{NewLemma434} 
(Extension of Lemma~\ref{Lemma434})
For any efficient and perfect device $D=(S,\Pi,M,P)$, we have for any $\bm{v}\in\{0,1\}^3$, there exists positive matrices 
$\tilde{\tilde{\alpha}}^{(0,v_1;0,v_2;0,v_3)}$, $\tilde{\tilde{\alpha}}^{(0,v_1;0,v_2;1,v_3)}$, $\tilde{\tilde{\alpha}}^{(0,v_1;1,v_2;0,v_3)}$ and 
$\tilde{\tilde{\alpha}}^{(1,v_1;0,v_2;0,v_3)}$ such that the following holds:
\begin{align}
&V_S\sigma^{(0,v_1; 0,v_2; 0,v_3)}V_S^{\dagger}\approx_{R}P[\ket{v_1,v_2,v_3}]\otimes \tilde{\tilde{\alpha}}^{(0,v_1;0,v_2;0,v_3)},
\label{L160_O}\\
&V_S\sigma^{(0,v_1; 0,v_2; 1,v_3)}V_S^{\dagger}\approx_{R}
P[\ket{v_1,v_2,(-)^{v_3}}]\otimes \tilde{\tilde{\alpha}}^{(0,v_1;0,v_2;1,v_3)},
\label{L161_O}\\
&V_S\sigma^{(0,v_1; 1,v_2; 0,v_3)}V_S^{\dagger}\approx_{R}P[\ket{v_1,(-)^{v_2},v_3}]\otimes \tilde{\tilde{\alpha}}^{(0,v_1;1,v_2;0,v_3)},\no\\
&V_S\sigma^{(1,v_1; 0,v_2; 0,v_3)}V_S^{\dagger}\approx_{R}P[\ket{(-)^{v_1},v_2,v_3}]\otimes \tilde{\tilde{\alpha}}^{(1,v_1;0,v_2;0,v_3)}.
\label{L163_O}
\end{align}
\end{lemma}
(Proof) The proof is similar to Lemma~\ref{Lemma434}, so we sketch it for Eq.~(\ref{L163_O}). 
Applying Lemmas~\ref{NewLemma430}, \ref{l:L218} (\ref{l:L218ii}), \ref{l:l224} and \ref{l:repl} (\ref{l:repli}) in this order, we have
\begin{align}
\sum_{\bm{v}}\tr\left[A^{(v_3)}_{3,\bm{0}}\sigma^{(1,v_1;0,v_2;0,v_3)}\right]\approx_R
\sum_{\bm{v}}\tr\left[V_S^{\dagger}(I^{\otimes2}_2\otimes\ket{v_3}\bra{v_3}\otimes I)V_S\sigma^{(1,v_1;0,v_2;0,v_3)}\right].
\label{L434main}
\end{align}
The LHS is approximately equal to 1 from Lemma~\ref{newlemma} and the definition of $\gamma_T(D)$ 
in Eq.~(\ref{gammaTest}), and hence the RHS of Eq.~(\ref{L434main}) is also approximately equal to 1. 
Applying Corollary~\ref{C48} and Lemma~\ref{l:L420} implies
$I^{\otimes2}_2\otimes\ket{v_3}\bra{v_3}\otimes I\approx_{R,V_S\sigma^{(1,v_1;0,v_2;0,v_3)}V_S^{\dagger}}I$. 
From Lemma~\ref{Lemma434}, $V_S\sigma^{(1,v_1;0,v_2;0,v_3)}V_S^{\dagger}$ is close to 
$P[\ket{(-)^{v_1},v_2}]\otimes\tilde{\alpha}^{(1,v_1;0,v_2;0,v_3)}$, and Lemma~\ref{l:repl} (\ref{l:replii}) 
enables us to replace these states. 
By combining the replaced equation, Eq.~(\ref{L484}) and  Lemma~\ref{l:L222}, we finally obtain
\begin{align}
V_S\sigma^{(1,v_1;0,v_2;0,v_3)}V_S^{\dagger}
&\approx_RP[\ket{(-)^{v_1},v_2,v_3}]\otimes\tilde{\tilde{\alpha}}^{(1,v_1;0,v_2;0,v_3)},
\no
\end{align}
where we define $\tilde{\tilde{\alpha}}^{(1,v_1;0,v_2;0,v_3)}:=(\bra{v_3}\otimes I)\tilde{\alpha}^{(1,v_1;0,v_2;0,v_3)}(\ket{v_3}\otimes I)$. \sq

\begin{lemma}
\label{NewLemma435} 
(Extension of Lemma~\ref{Lemma435}) 
Let $D=(S,\Pi,M,P)$ be an efficient perfect device. There exists a normalized state $\tilde{\tilde{\alpha}}$ such that 
the following holds for any $\bm{v}\in\{0,1\}^3$:
\begin{align}
&V_S\sigma^{(0,v_1; 0,v_2; 0,v_3)}V_S^{\dagger}\overset{c}{\approx}_{R}\frac{P[\ket{v_1,v_2,v_3}]}{8}\otimes \tilde{\tilde{\alpha}},
\label{L160}
\\
&V_S\sigma^{(0,v_1; 0,v_2; 1,v_3)}V_S^{\dagger}\overset{c}{\approx}_{R}\frac{P[\ket{v_1,v_2,(-)^{v_3}}]}{8}\otimes \tilde{\tilde{\alpha}},
\label{L161}
\\
&V_S\sigma^{(0,v_1; 1,v_2; 0,v_3)}V_S^{\dagger}\overset{c}{\approx}_{R}\frac{P[\ket{v_1,(-)^{v_2},v_3}]}{8}\otimes \tilde{\tilde{\alpha}},
\label{eq3:Lemma56}\\
&V_S\sigma^{(1,v_1; 0,v_2; 0,v_3)}V_S^{\dagger}\overset{c}{\approx}_{R}\frac{P[\ket{(-)^{v_1},v_2,v_3}]}{8}\otimes \tilde{\tilde{\alpha}}.
\label{L163}
\end{align}
\end{lemma}
(Proof) 
The proof is similar to Lemma~\ref{Lemma435}, but we spell out all the details for completeness. 
We first prove Eq.~(\ref{eq3:Lemma56}), and by using Eq.~(\ref{eq3:Lemma56}), we prove the rest of the equations. 
In so doing, we need to show that 
$\{\tiltila^{(0,v_1;1,v_2;0,v_3)}\}_{\bm{v}}$ are computationally indistinguishable. 
For this, we already have shown in Lemma~\ref{Lemma435} that 
$\{\sum_{v_3}\tilde{\alpha}^{(0,v_1;1,v_2;0,v_3)}\}_{v_2}$ are computationally indistinguishable for any $v_1$, and 
by considering Lemmas~\ref{Lemma434} and \ref{NewLemma434}, this implies that 
$\{\sum_{v_3}\tiltila^{(0,v_1;1,v_2;0,v_3)}\}_{v_2}$ are also computationally indistinguishable for any $v_1$. 
Therefore, the remaining task is to prove that 
$\{\tiltila^{(0,v_1;1,v_2;0,v_3)}\}_{v_3}$ are computationally indistinguishable for any fixed $v_1$ and $v_2$. 
In the following discussions, we fix $v_1$ and $v_2$.  
From Lemma~\ref{NewLemma434}, there exists a $d>0$ such that for any $v_1,v_2,v_3$, 
\begin{align}
&\left|\left|
V_S\sigma^{(0,v_1; 1,v_2; 0,v_3)}V_S^{\dagger}-P[\ket{v_1,(-)^{v_2},v_3}]
\otimes\tiltila^{(0,v_1;1,v_2;0,v_3)}
\right|\right|_1^2\le\epsilon,
\label{eq010}\\
&\left|\left|
V_S\sigma^{(0,v_1; 0,v_2; 1,v_3)}V_S^{\dagger}-P[\ket{v_1,v_2,(-)^{v_3}}]
\otimes\tiltila^{(0,v_1;0,v_2;1,v_3)}
\right|\right|_1^2\le\epsilon
\label{eq001}
\end{align}
hold with $\epsilon:=O(\gamma_T(D)^d)$. 
From Lemmas~\ref{Lemma434} and \ref{NewLemma434}, we have that 
\begin{align}
\left|
\tr\left[M_0\sum_{v_3'}\tiltila^{(0,v_1;1,v_2=0;0,v_3')}\right]-
\tr\left[M_0\sum_{v_3'}\tiltila^{(0,v_1;1,v_2=1;0,v_3')}\right]
\right|\le 2\sqrt{\epsilon}
\label{bunkatuHT}
\end{align}
holds for any $v_1$ and efficient measurement $M:=\{M_0,M_1\}$. 
For the sake of contradiction, we assume that there exists a POVM $\Lambda:=\{\Lambda_0,\Lambda_1\}$ with 
$\Lambda_0+\Lambda_1=I$ such that
\begin{align}
\left|
\tr\left[\Lambda_0\tiltila^{(0,v_1;1,v_2;0,v_3=0)}\right]
-\tr\left[\Lambda_0\tiltila^{(0,v_1;1,v_2;0,v_3=1)}\right]\right|
\ge 2\mu(\lambda)+42\sqrt{\epsilon}
\label{katei_b}
\end{align}
holds with a non-negligible function $\mu(\lambda)$. 
Under the existence of this POVM, we can construct an adversary $\mathcal{A}$ that 
breaks the injective invariance property in Definition~\ref{def:II} 
using an efficient measurement $\{\Gamma,I-\Gamma\}$ with
\begin{align}
\Gamma:=V_S^{\dagger}(P[\ket{v_1,(-)^{v_2},0}]\otimes\Lambda_0)V_S.
\no
\end{align}
Below, we describe the procedure of $\mathcal{A}$ that breaks the injective invariance property. 

$\mathcal{A}$ is given keys $(k_2,k_3)$, and the task is to distinguish whether the input is $(\theta_2,\theta_3)=(0,1)$ or 
$(\theta_2,\theta_3)=(1,0)$. For this, 
$\mathcal{A}$ samples the other key and a trapdoor 
$(k_1,t_{k_1})\leftarrow\GEN_{\Gcal}(1^{\lambda})$, prepares the state $\psi^{(\bm{\theta})}$ by performing the same operations as the device 
$D$, measures the $Y$-register to obtain $\bm{y}$, followed by measuring the $R$-register to obtain $\bm{d}$. 
At this moment, $\mathcal{A}$ prepares the state $\rho^{(\bm{\theta})}$. 
Finally, $\mathcal{A}$ performs the measurement $\{\Gamma,I-\Gamma\}$. 
This procedure is efficient because the device $D$ and the POVM $\{\Gamma,I-\Gamma\}$ are efficient. 
In this procedure, we calculate the distinguishing advantage $\U{Adv}=|\tr[\Gamma(\rho^{(010)}-\rho^{(001)})]|$
in obtaining the outcome 
corresponding to $\Gamma$ for the states $\rho^{(010)}$ and $\rho^{(001)}$. 
Once we show that this advantage is non-negligible under Eq.~(\ref{katei_b}), 
this contradicts the injective invariance property.  
Hence, by taking a contraposition,  
we obtain the negation of Eq.~(\ref{katei_b}), which is the required statement in the proof. 
By this discussion, we only need to prove that the advantage $\U{Adv}$ is non-negligible from Eq.~(\ref{katei_b}). 
First, by the definitions of $\Gamma$ and $\rho^{(\bm{\theta})}$, we have
\begin{align}
\U{Adv}=\left|
\tr\left[
(P[\ket{v_1,(-)^{v_2},0}]\otimes\Lambda_0)
\left(\sum_{\bm{v}'}V_S\sigma^{(0,v_1';1,v_2';0,v_3')}V_S^\dagger-\sum_{\bm{v}'}V_S\sigma^{(0,v_1';0,v_2';1,v_3')}V_S^\dagger\right)
\right]
\right|.
\label{defadv}
\end{align}
To obtain its lower bound, we exploit the relation from H\"{o}lder's inequality:
\begin{align}
\left|\tr[A(\varphi-\varphi')]\right|\le||A||_{\infty}\cdot||\varphi-\varphi'||_1\le
\tau
\label{star2}
\end{align}
for positive operators $\varphi$ and $\varphi'$ satisfying $||\varphi-\varphi'||_1\le\tau$ and a linear operator $A$ satisfying 
$||A||_{\infty}\le1$. 
By applying Eq.~(\ref{star2}) with Eqs.~(\ref{eq010}) and (\ref{eq001}), Adv in Eq.~(\ref{defadv}) is lower-bounded by 
\begin{align}
&\Big|\sum_{\bm{v}'\in\{0,1\}^3}
\tr\Big[(P[\ket{v_1,(-)^{v_2},0}\otimes\Lambda_0)
\Big(P[\ket{v_1',(-)^{v_2'},v_3'}]\otimes\tiltila^{(0,v_1';1,v_2';0,v_3')}-
P[\ket{v_1',v_2',(-)^{v_3'}}]\otimes\tiltila^{(0,v_1';0,v_2';1,v_3')}\Big)\Big]
\Big|-16\sqrt{\epsilon}\no\\
=&
\left|\tr\left[\Lambda_0\tiltila^{(0,v_1;1,v_2;0,0)}\right]-
\frac{1}{4}\sum_{v_2',v_3'}\tr\left[\Lambda_0\tiltila^{(0,v_1;0,v_2';1,v_3')}\right]
\right|-16\sqrt{\epsilon}
\no\\
=&\left|\frac{1}{2}\tr\left[\Lambda_0(\tiltila^{(0,v_1;1,v_2;0,0)}-\tiltila^{(0,v_1;1,v_2;0,1)})\right]
+\frac{1}{2}\sum_{v_3'}\tr[\Lambda_0\tiltila^{(0,v_1;1,v_2;0,v_3')}]-
\frac{1}{4}\sum_{v_2',v_3'}\tr\left[\Lambda_0\tiltila^{(0,v_1;0,v_2';1,v_3')}\right]	\right|-16\sqrt{\epsilon}\no\\
\ge&
\mu(\lambda)+5\sqrt{\epsilon}
-\frac{1}{2}
\left|\sum_{v_3'}\tr[\Lambda_0\tiltila^{(0,v_1;1,v_2;0,v_3')}]
-\frac{1}{2}\sum_{v_2',v_3'}\tr\left[\Lambda_0\tiltila^{(0,v_1;0,v_2';1,v_3')}\right]\right|
\no\\
\ge&\mu(\lambda)+5\sqrt{\epsilon}
-\frac{1}{2}
\left|\sum_{v_3'}\tr\left[\Lambda_0\left(\tiltila^{(0,v_1;1,v_2;0,v_3')}
-\sum_{v_2'}\frac{\tiltila^{(0,v_1;1,v_2';0,v_3')}}{2}\right)\right]\right|-
\frac{1}{4}\left|\sum_{v_2',v_3'}\tr\left[\Lambda_0\left(\tiltila^{(0,v_1;1,v_2';0,v_3')}-\tiltila^{(0,v_1;0,v_2';1,v_3')}\right)\right]\right|
\no\\
\ge&\mu(\lambda)+4\sqrt{\epsilon}
-\frac{1}{4}\left|
\sum_{v_2',v_3'}\tr\left[\Lambda_0\left(\tiltila^{(0,v_1;1,v_2';0,v_3')}-\tiltila^{(0,v_1;0,v_2';1,v_3')}\right)\right]
\right|
\no\\
=&\mu(\lambda)+4\sqrt{\epsilon}
-\frac{1}{4}\left| 
\sum_{v_2',v_3'}\tr\left[\Lambda_0\left\{\ket{v_1}\bra{v_1}\otimes
\left(P[\ket{(-)^{v_2'},v_3'}]
\otimes
\tiltila^{(0,v_1;1,v_2';0,v_3')}
-P[\ket{v_2,(-)^{v_3'}}]\otimes\tilde{\alpha}^{(0,v_1;0,v_2';1,v_3')}\right\}\right)\right]
\right|,
\no
\end{align}
where we use the triangle inequality and Eq.~(\ref{katei_b}) in the first inequality, 
the second one follows from the triangle inequality, and the third one comes from Eq.~(\ref{bunkatuHT}). 
Again, by applying Eq.~(\ref{star2}) with Eqs.~(\ref{eq010}) and (\ref{eq001}), we obtain
\begin{align}
\U{Adv}\ge&\mu(\lambda)
-\frac{1}{4}\left|
\tr\left[\Lambda_0\left(\sum_{v_2',v_3'}V_S\sigma^{(0,v_1;1,v_2';0,v_3')}V_S^{\dagger}
-\sum_{v_2',v_3'}V_S\sigma^{(0,v_1;0,v_2';1,v_3')}V_S^{\dagger}
\right)\right]
\right|.
\no
\end{align}
Finally, by Eq.~ (\ref{def:sigma001}) and 
$\rho^{(\bm{\theta})}=\sum_{\bm{v}}\sigma^{(\theta_1,v_1;\theta_2,v_2;\theta_3,v_3)}$, 
the RHS is equal to 
$\mu(\lambda)-\frac{1}{4}\left|
\tr\left[W
\left(
\rho^{(010)}-\rho^{(001)}\right)\right]
\right|$ 
with 
$W:=\sum_{y_1:\hat{b}(k_1,y_1)=v_1}\ket{y_1}\bra{y_1}V_S^{\dagger}\Lambda_0V_S\sum_{y_1:\hat{b}(k_1,y_1)=v_1}\ket{y_1}\bra{y_1}.
$ 
The measurement $\{W,I-W\}$ is efficient since $\mathcal{A}$ has the information of the trapdoor $t_{k_1}$ and 
computing $\hat{b}(k_1,y_1)$ is efficient. 
Hence, the computational indistinguishability in Lemma~\ref{l:L412} reveals that the second term is $\negl(\lambda)$, namely
\begin{align}
\U{Adv}=|\tr[\Gamma(\rho^{(010)}-\rho^{(001)})]|\ge\mu(\lambda)-\negl(\lambda).
\no
\end{align}
This contradicts Lemma~\ref{l:L412} and completes the proof of Eq.~(\ref{eq3:Lemma56}). 

Next, we prove Eqs.~(\ref{L160}), (\ref{L161}) and (\ref{L163}) using Eq.~(\ref{eq3:Lemma56}). 
First, from Eq.~(\ref{eq3:Lemma56}) and using the fact that $\rho^{(\bm{\theta})}$ are computationally indistinguishable from Lemma~\ref{l:L412} 
and isometry $V_S$ is efficient, we have
\begin{align}
V_S\rho^{(\bm{\theta})}V_S^\dagger=
\sum_{\bm{v}}V_S\sigma^{(\theta_1,v_1; \theta_2, v_2; \theta_3,v_3)} V_S^{\dagger}\overset{c}{\approx}_R \frac{I_2^{\otimes3}}{8}\otimes \tiltila
\label{Vs010}
\end{align}
for $\bm{\theta}\in\{000,001,010,100\}$. 
Combining Eq.~(\ref{Vs010})
with Eqs.~(\ref{L160_O}), (\ref{L161_O}) and (\ref{L163_O}), we respectively obtain 
\begin{align}
&\sum_{\bm{v}}P[\ket{v_1,v_2,v_3}]\otimes\tiltila^{(0,v_1; 0,v_2;0,v_3)}
\overset{c}{\approx}_R I_2^{\otimes3}\otimes \frac{\tiltila}{8},
\no\\
&\sum_{\bm{v}}P[\ket{v_1,v_2,(-)^{v_3}}]\otimes\tiltila^{(0,v_1; 0,v_2;1,v_3)}
\overset{c}{\approx}_R I_2^{\otimes3}\otimes \frac{\tiltila}{8},
\no
\\
&\sum_{\bm{v}}P[\ket{(-)^{v_1},v_2,v_3}]\otimes\tiltila^{(1,v_1; 0,v_2;0,v_3)}
\overset{c}{\approx}_R I_2^{\otimes3}\otimes \frac{\tiltila}{8}.
\no
\end{align}
Since these approximate relations hold for any efficient prover, these relations hold when the first, second and third 
registers are measured in the Pauli bases. By considering such a prover, we have 
\begin{align}
\tiltila^{(0,v_1; 0,v_2;0,v_3)}\overset{c}{\approx}_R\frac{\tiltila}{8},~~
\tiltila^{(0,v_1; 0,v_2;1,v_3)}\overset{c}{\approx}_R\frac{\tiltila}{8},~~
\tiltila^{(1,v_1; 0,v_2;0,v_3)}\overset{c}{\approx}_R\frac{\tiltila}{8}.\no
\end{align}
By applying these three approximate relations to Eqs.~(\ref{L160_O}), (\ref{L161_O}) and (\ref{L163_O}), we respectively obtain 
Eqs.~(\ref{L160}), (\ref{L161}) and (\ref{L163}), which completes the proof. 
\sq

\begin{coro}
\label{NewCorollary436} (Extension of Corollary~\ref{Corollary436}) 
Let $D=(S,\Pi,M,P)$ be an efficient perfect device. 
There exists a normalized state $\tilde{\tilde{\alpha}}$ such that for any $\bm{\theta}\in\mathcal{B}$,
\begin{align}
V_S\rho^{(\bm{\theta})}V_S^{\dagger}\overset{c}{\approx}_R\frac{I_2^{\otimes3}}{8}\otimes\tilde{\tilde{\alpha}}. 
\end{align}
\end{coro}
(Proof) Taking the sum of the equations in 
Lemma~\ref{NewLemma435} over $\bm{v}$ yields the statement for $\bm{\theta}$ of the test case. 
We can lift up the statement for any $\bm{\theta}\in\mathcal{B}$ thanks to Lemma~\ref{l:L412}. 
\sq

Below, we present crucial Lemmas~\ref{Lemma437}, \ref{NewLemma437} and \ref{AltLemma437} 
for proving our main result, Theorem~\ref{Theorem438}. 
Lemma~\ref{Lemma437} shows that any two joint observables are close to the products of the Pauli observables. 
\begin{lemma}
\label{Lemma437} 
For any efficient perfect device $D=(S,\Pi,M,P)$, we have the following for any $\bm{\theta}\in\mathcal{B}$, 
$\bm{q}\in\{0,1\}^3$ and $i\neq i'\in\{1,2,3\}$: 
\begin{align}
V_S(A_{i,\bm{q}}A_{i',\bm{q}})V_S^{\dagger}\approx_{R,V_S\rho^{(\bm{\theta})}V_S^{\dagger}}\sigma_{W,i}\otimes\sigma_{W',i'},
\label{eq50}
\end{align}
where $W=Z$ ($W=X$) if $q_i=0~(1)$ and likewise for $W'$. 
\end{lemma}
Note that Lemma~\ref{Lemma437} will be used to prove Theorem~\ref{Theorem438} (ii). 
\\
(Proof) We follow the proof of Lemma 4.41 in~\cite{MV20}. 
We prove Eq.~(\ref{eq50}) for $(i,i',\bm{q})=(1,2,000)$, and the others can be shown analogously. 
As $[A_{1,\bm{0}},A_{2,\bm{0}}]=0$ from Eq.~(\ref{L420Z}), and 
$A_{1,\bm{0}}$ and $A_{2,\bm{0}}$ are efficient binary observables, 
Lemma~\ref{l:l25} implies that $A_{1,\bm{0}}A_{2,\bm{0}}$ is an efficient binary observable. 
Since $V_S$, $A_{1,\bm{0}}A_{2,\bm{0}}$ and $\sigma_Z$ are all efficient, Corollary~\ref{Corollary436} and Lemma~\ref{l:lift} (\ref{rlvi}) 
reduced the proof to showing 
$V_S(A_{1,\bm{0}}A_{2,\bm{0}})V_S^{\dagger}\approx_{R,\kappa}\sigma_{Z,1}\otimes\sigma_{Z,2}$ with 
$\kappa:=I_2\otimes I_2/4\otimes\tilde{\alpha}$. 
From Lemma~\ref{l:L216}, it suffices to show
\begin{align}
\tr\left[(\sigma_{Z,1}\otimes\sigma_{Z,2})(V_SA_{1,\bm{0}}A_{2,\bm{0}}V_S^{\dagger})\kappa\right]\approx_R1. 
\label{eq:L437in}
\end{align}
Since 
$V_SA_{1,\bm{0}}V_S^{\dagger}\approx_{R,V_S\rho^{(\bm{\theta})}V_S^{\dagger}}\sigma_{Z,1}$ 
holds from Eq.~(\ref{eq:i}) and Lemma~\ref{l:L223}, and 
$V_SA_{2,\bm{0}}V_S^{\dagger}\approx_{R,V_S\rho^{(\bm{\theta})}V_S^{\dagger}}\sigma_{Z,2}$ does from 
Lemmas~\ref{l:L223} and \ref{Lemma430},
and using Corollary~\ref{Corollary436} and Lemma~\ref{l:lift} (\ref{rlvi}), these imply
$V_SA_{i,\bm{0}}V_S^{\dagger}\approx_{R,\kappa}\sigma_{Z,i}$ for $i=1,2$. 
Hence, using this and $[\sigma_{Z,2},\kappa]=0$, 
the LHS of Eq.~(\ref{eq:L437in}) is equal to
\begin{align}
&\tr\left[(\sigma_{Z,1}\otimes\sigma_{Z,2})(V_SA_{1,\bm{0}}V_S^{\dagger})(V_SA_{2,\bm{0}}V_S^{\dagger})
\kappa
\right]
\approx_R
\tr\left[\sigma_{Z,1}(V_SA_{1,\bm{0}}V_S^{\dagger})\kappa\right]
\approx_R
\tr\left[\sigma_{Z,1}\sigma_{Z,1}\kappa
\right]
=1,
\label{similarc}
\end{align}
which ends the proof.\sq

We next prove that any three joint observables are approximately equal to the products of the Pauli observables. 
\begin{lemma}
\label{NewLemma437}
For any efficient perfect device $D=(S,\Pi,M,P)$, we have the following for any $\bm{\theta}\in\mathcal{B}$ and $\bm{q}\in\{0,1\}^3$,
\begin{align}
V_S(A_{1,\bm{q}}A_{2,\bm{q}}A_{3,\bm{q}})V_S^{\dagger}\approx_{R,V_S\rho^{(\bm{\theta})}V_S^{\dagger}}
\sigma_W\otimes\sigma_{W'}\otimes\sigma_{W''},
\label{eq80}
\end{align}
where $W,W'$ and $W''$ are defined in the same way as in Lemma~\ref{Lemma437}. 
\end{lemma}
Note that Lemma~\ref{NewLemma437} will be used to prove Theorem~\ref{Theorem438} (ii). 
\\
(Proof) 
We prove Eq.~(\ref{eq80}) with $\bm{q}=000$, and the others can be shown analogously. 
Since $[A_{3,\bm{0}},A_{1,\bm{0}}A_{2,\bm{0}}]=0$ from Eq.~(\ref{L420Z}) and 
$A_{3,\bm{0}}$ and $A_{1,\bm{0}}A_{2,\bm{0}}$ are efficient binary observables as explained in the proof of Lemma~\ref{Lemma437}, 
Lemma~\ref{l:l25} implies that $A_{1,\bm{0}}A_{2,\bm{0}}A_{3,\bm{0}}$ is also an efficient binary observable. 
As $V_S$, $A_{1,\bm{0}}A_{2,\bm{0}}A_{3,\bm{0}}$ and $\sigma_Z$ are all efficient, from Lemma~\ref{l:lift} (\ref{rlvi}) 
and Corollary~\ref{NewCorollary436}, the proof is reduced to showing
$V_S(A_{1,\bm{0}}A_{2,\bm{0}}A_{3,\bm{0}})V_S^{\dagger}\approx_{R,\phi}\sigma_Z\otimes\sigma_Z\otimes\sigma_Z$ 
with $\phi:=\frac{1}{8}I_2^{\otimes3}\otimes\tilde{\tilde{\alpha}}$. 
From Lemma~\ref{l:L216} and $V_S^{\dagger}V_S=I$, it suffices to show
\begin{align}
\tr\left[\sigma_Z^{\otimes 3}(V_SA_{1,\bm{0}}V_S^{\dagger})(V_SA_{2,\bm{0}}V_S^{\dagger})(V_SA_{3,\bm{0}}V_S^{\dagger})\phi\right]
\approx_{R}1.
\label{NL437main}
\end{align}
Since $V_SA_{i,\bm{0}}V_S^{\dagger}\approx_{R,V_S\rho^{(\bm{\theta})}V_S^{\dagger}}\sigma_{Z,i}$ 
holds from Lemmas~\ref{l:L223}, \ref{Lemma430} and \ref{NewLemma430} and Eq.~(\ref{eq:i}), 
Corollary~\ref{NewCorollary436} and Lemma~\ref{l:lift} (\ref{rlvi}) imply $V_SA_{i,\bm{0}}V_S^{\dagger}\approx_{R,\phi}\sigma_{Z,i}$
for any $i\in\{1,2,3\}$. 
Using this and Lemma~\ref{l:repl} (\ref{l:repli}), Eq.~(\ref{NL437main}) is verified by a direct calculation with 
a similar argument to Eq.~(\ref{similarc}).\sq

By exploiting Corollary~\ref{NewCorollary436}, we show that the prover's measurements 
to obtain the outcomes 
$v_3\oplus v_1v_2$, $v_2\oplus v_1v_3$ and $v_1\oplus v_2v_3$ 
at step~(e) of the protocol in Sec.~\ref{sec:proto} 
are close to the generalized stabilizers of $CCZ\ket{+}^{\otimes3}$. 
\begin{lemma}
\label{AltLemma437} 
For any efficient perfect device $D=(S,\Pi,M,P)$, the following holds for any $\bm{\theta}\in\mathcal{B}$:
\begin{align}
&V_S(A^{(0)}_{1,001}A_{3,001}+A^{(1)}_{1,001}A_{2,001}A_{3,001})V_S^{\dagger}\approx_{R,V_S\rho^{(\bm{\theta})}V_S^{\dagger}}
\sigma_Z^{(0)}\otimes I_2\otimes\sigma_X+\sigma_Z^{(1)}\otimes \sigma_Z\otimes\sigma_X,
\label{ALT4165}\\
&V_S(A^{(0)}_{1,010}A_{2,010}+A^{(1)}_{1,010}A_{2,010}A_{3,010})V_S^{\dagger}\approx_{R,V_S\rho^{(\bm{\theta})}V_S^{\dagger}}
\sigma_Z^{(0)}\otimes\sigma_X\otimes I_2+\sigma_Z^{(1)}\otimes \sigma_X\otimes\sigma_Z,
\label{ALT4165-1}\\
&V_S(A_{1,100}A^{(0)}_{2,100}+A_{1,100}A^{(1)}_{2,100}A_{3,100})V_S^{\dagger}\approx_{R,V_S\rho^{(\bm{\theta})}V_S^{\dagger}}
\sigma_X\otimes\sigma_Z^{(0)}\otimes I_2+\sigma_X\otimes\sigma_Z^{(1)}\otimes \sigma_Z.
\label{ALT4165-2}
\end{align}
\end{lemma}
Note that Lemma~\ref{AltLemma437} will be used to prove Theorem~\ref{Theorem438} (i). \\
(Proof) We prove Eq.~(\ref{ALT4165}), and the others can be proven in the same way. 
Note that $(A^{(0)}_{1,001}A_{3,001}+A^{(1)}_{1,001}A_{2,001}A_{3,001})$ is an efficient 
binary observable that determines bit $v_3\oplus v_1v_2$. 
Using Corollary~\ref{NewCorollary436} and Lemma~\ref{l:lift} (\ref{rlvi}) reduces the proof to showing 
$V_S(A^{(0)}_{1,001}A_{3,001}+A^{(1)}_{1,001}A_{2,001}A_{3,001})V_S^{\dagger}\approx_{R,\phi}
(\sigma_Z^{(0)}\otimes I_2\otimes\sigma_X+\sigma_Z^{(1)}\otimes \sigma_Z\otimes\sigma_X)$ with 
$\phi:=\frac{I_2^{\otimes3}}{8}\otimes\tilde{\tilde{\alpha}}$. 
By using Lemma~\ref{l:L216}, it suffices to prove
\begin{align}
\tr\left[(\sigma_Z^{(0)}\otimes I_2\otimes\sigma_X+\sigma_Z^{(1)}\otimes \sigma_Z\otimes\sigma_X)
V_S(A^{(0)}_{1,001}A_{3,001}+A^{(1)}_{1,001}A_{2,001}A_{3,001})
V_S^{\dagger}\phi\right]\approx_R
\tr(\phi)=1.
\label{th438MAIN}
\end{align}
Inserting $V^\dagger_SV_S=I$, the LHS is equal to
\begin{align}
\tr\left[
\underbrace{
(\sigma_Z^{(0)}\otimes I_2\otimes\sigma_X+\sigma_Z^{(1)}\otimes \sigma_Z\otimes\sigma_X)
[(V_SA^{(0)}_{1,001}V_S^{\dagger})+(V_SA^{(1)}_{1,001}V_S^{\dagger})(V_SA_{2,001}V_S^{\dagger})}_{=:C}](V_SA_{3,001}V_S^{\dagger})\phi\right].
\label{thLHS}
\end{align}
From Eq.~(\ref{C433X3}) and Lemma~\ref{l:L223}, we have 
$V_SA_{3,001}V_S^{\dagger}\approx_{R,V_S\rho^{(\bm{\theta})}V_S^{\dagger}}\sigma_{X,3}$, 
and state $V_S\rho^{(\bm{\theta})}V_S^{\dagger}$ can be replaced with $\phi$ thanks to 
Lemmas~\ref{l:lift} (\ref{rlvi}) and Corollary~\ref{NewCorollary436}. 
Since the operator norm of $C$ defined in Eq.~(\ref{thLHS}) is constant, from Lemma~\ref{l:repl} (\ref{l:repli})
we have the approximate equation of Eq.~(\ref{thLHS}) as
\begin{align}
&\tr\left[(\sigma_Z^{(0)}\otimes I_2\otimes\sigma_X+\sigma_Z^{(1)}\otimes \sigma_Z\otimes\sigma_X)
[(V_SA^{(0)}_{1,001}V_S^{\dagger})+(V_SA^{(1)}_{1,001}V_S^{\dagger})(V_SA_{2,001}V_S^{\dagger})]\sigma_{X,3}\phi\right]\no\\
=&\tr\left[(\sigma_Z^{(0)}\otimes I^{\otimes2}_2+\sigma_Z^{(1)}\otimes \sigma_Z\otimes I_2)
(V_SA^{(0)}_{1,001}V_S^{\dagger})\phi\right]
+\tr\left[(\sigma_Z^{(0)}\otimes I^{\otimes2}_2+\sigma_Z^{(1)}\otimes \sigma_Z\otimes I_2)
(V_SA^{(1)}_{1,001}V_S^{\dagger})(V_SA_{2,001}V_S^{\dagger})\phi\right],
\label{th4381st2nd}
\end{align}
where we used the commutation relation $[\sigma_{X,3},\phi]=0$ in the equation. 
Since the projector $A^{(b)}_{1,001}$ is written as 
$A^{(b)}_{1,001}=(I+(-1)^bA_{1,001})/2$ using binary observable $A_{1,001}$,
calculations of both terms in Eq.~(\ref{th4381st2nd}) can be done by applying a similar argument done in 
Lemma~\ref{NewLemma437}. It is straightforward to show that the first and second terms of Eq.~(\ref{th4381st2nd}) are 
approximately equal to
$\tr[(\ket{0}\bra{0}\otimes I)\phi]$ and 
$\tr[(\ket{1}\bra{1}\otimes I)\phi]$, respectively. 
Therefore, the LHS of Eq.~(\ref{th438MAIN}), which is the main target of computation in the proof, is approximately equal to 1, 
which ends the proof. \sq

\subsection{Certifying Entangled Magic States}
\label{subsec:th}
\begin{theorem}
\label{Theorem438}
We define $Z$-rotated entangled magic states as
\begin{align}
\ket{\phi_{\U{H}}^{(a,b,c)}}:=(\sigma_Z^a\otimes\sigma_Z^b\otimes\sigma_Z^c)CCZ\ket{+}^{\otimes3}.
\no
\end{align} 
For $b\in\{0,1\}$, we use the notation
\begin{align}
\ket{b_0}:=\ket{b},~~\ket{b_1}:=\ket{(-)^b}.
\no
\end{align} 
Let $D=(S,\Pi,M,P)$ be an efficient device, device's Hilbert space be $\mathcal{H}$, 
$\sigma^{(1,s_1;1,s_2;1,s_3)}$ be defined in Eq.~(\ref{def:sigma111}), and $\mathcal{H}'$ be some Hilbert space.
Then, there exists 
an isometry $V:\mathcal{H}\to\mathbb{C}^8\otimes\mathcal{H}'$, and 
a constant $d>0$ such that there are states 
$\zeta^{(s_1,s_2,s_3)}\in\mathcal{D}(\mathcal{H}')$ for $s_1,s_2,s_3\in\{0,1\}$ satisfying the following. 
In the description, $\gamma_P(D), \gamma_T(D)$ and $\gamma_H(D)$ are 
defined in Lemmas~\ref{l:pr}, \ref{lgammaTest}, and \ref{l:hgc}, respectively
\footnote{
Note that 
$\gamma_P(D), \gamma_T(D)$ and $\gamma_H(D)$ are upper-bounded by the probabilities of obtaining a flag in 
the preimage round, Hadamard round with the 
test case and Hadamard round with the hypergraph one, which are shown in 
Eqs.~(\ref{eq:upper_gammaP}), (\ref{eq:upper_gammaT}) and (\ref{eq:upper_gammaH}), respectively. 
}.
\\
(i) The unnormalized state in an Hadamard round is close to the entangled magic state up to isometry $V$:
\begin{align}
V\sigma^{(1,s_1;1,s_2;1,s_3)}V^{\dagger}\approx_{\gamma_P(D)^d+\gamma_T(D)^d+\gamma_H(D)^d}\frac{1}{8}
P[\ket{\phi_{\U{H}}^{(s_1,s_2,s_3)}}]\otimes\zeta^{(s_1,s_2,s_3)},
\no
\end{align} 
where $P[\ket{\cdot}]:=\ket{\cdot}\bra{\cdot}$ and the different $\zeta^{(s_1,s_2,s_3)}$ are computationally indistinguishable. \\
(ii) Under isometry $V$, measurements $\{P^{(abc)}_{q_1q_2q_3}\}_{a,b,c}$ acting on prover's state $\sigma^{(1,s_1;1,s_2;1,s_3)}$ are 
close to the Pauli-$Z$ and $X$ measurements 	acting on the entangled magic state: 
\begin{align}
VP^{(abc)}_{q_1q_2q_3}
\sigma^{(1,s_1;1,s_2;1,s_3)}P^{(abc)}_{q_1q_2q_3}V^{\dagger}
\approx_{\gamma_P(D)^d+\gamma_T(D)^d+\gamma_H(D)^d}\frac{1}{8}
\left|\expect{a_{q_1},b_{q_2},c_{q_3}\Big{|}\phi_{\U{H}}^{(s_1,s_2,s_3)}}\right|^2
P[\ket{a_{q_1},b_{q_2},c_{q_3}}]\otimes\zeta^{(s_1,s_2,s_3)}.
\no
\end{align} 
\end{theorem}
In both proofs of (i) and (ii), 
by Lemma~\ref{perfectdevice}, up to an additional error $O(\sqrt{\gamma_P(D)})$, we can assume that device $D$ is perfect. 
In these proofs, we take isometry $V$ as swap isometry $V_S$ defined in Eq.~(\ref{defV}). 

\subsubsection{Proof of (i)}
First, using Eq.~(\ref{ALT4165}) and Lemma~\ref{l:L223} leads to
\begin{align}
A^{(0)}_{1,001}A_{3,001}+A^{(1)}_{1,001}A_{2,001}A_{3,001}
\approx_{R,\rho^{(\bm{\theta})}}V_S^{\dagger}
\left(\sum_{i=0}^1\ket{i}\bra{i}\otimes\sigma_Z^i\otimes\sigma_X\right)V_S.
\label{ReplacePth1}
\end{align}
By applying Lemma~\ref{l:L218} (\ref{l:L218ii}) to this, we have that $\rho^{(\bm{\theta})}$ in Eq.~(\ref{ReplacePth1}) 
can be replaced with $\sigma^{(1,s_1;1,s_2;1,s_3)}$. 
By noting that the LHS and $(\sum_{i=0}^1\ket{i}\bra{i}\otimes\sigma_Z^i\otimes\sigma_X)$ are binary observables, 
Lemma~\ref{l:l224} implies
\begin{align}
(A^{(0)}_{1,001}A_{3,001}+A^{(1)}_{1,001}A_{2,001}A_{3,001})
^{(a)}\approx_{R,\sigma^{(1,s_1;1,s_2;1,s_3)}}
V_S^{\dagger}\left(\sum_{i=0}^1\ket{i}\bra{i}\otimes\sigma_Z^i\otimes\sigma_X\right)^{(a)}V_S.
\no
\end{align}
Using this and Lemma~\ref{l:repl} (\ref{l:repli}) leads to
\begin{align}
\tr
\left[\left(\sum_{i=0}^1\ket{i}\bra{i}\otimes\sigma_Z^i\otimes\sigma_X\right)^{(a)}\varphi^{(s_1,s_2,s_3)}\right]
\approx_R
\tr\left[(A^{(0)}_{1,001}A_{3,001}+A^{(1)}_{1,001}A_{2,001}A_{3,001})^{(a)}\sigma^{(1,s_1;1,s_2;1,s_3)}\right],
\label{mixGP}
\end{align}
where 
$\varphi^{(s_1,s_2,s_3)}:= V_S\sigma^{(1,s_1;1,s_2;1,s_3)}V_S^{\dagger}$. 
By the definition of $\gamma_H(D)$ given in Eq.~(\ref{gammaH}), we have
$\sum_{\bm{s}}\tr\left[(A^{(0)}_{1,001}A_{3,001}+A^{(1)}_{1,001}A_{2,001}A_{3,001})^{(s_3)}\sigma^{(1,s_1;1,s_2;1,s_3)}\right]
\approx_{\gamma_H(D)}1$, 
and by using Lemma~\ref{L47} and $V_S^{\dagger}V_S=I$, this leads to
\begin{align}
\tr[(A^{(0)}_{1,001}A_{3,001}+A^{(1)}_{1,001}A_{2,001}A_{3,001})^{(s_3)}\sigma^{(1,s_1;1,s_2;1,s_3)}]
\approx_{\gamma_H(D)}
\tr(\varphi^{(s_1,s_2,s_3)}). 
\label{ALEQ4167}
\end{align}
Combining Eqs.~(\ref{mixGP}), (\ref{ALEQ4167}) and the triangle inequality results in
\begin{align}
\tr\left[\left(\sum_{i=0}^1\ket{i}\bra{i}\otimes\sigma_Z^i\otimes\sigma_X\right)^{(s_3)}\varphi^{(s_1,s_2,s_3)}\right]
\approx_R\tr(\varphi^{(s_1,s_2,s_3)}).
\label{ALEQ4169}
\end{align}
Using Lemma~\ref{l:L219}, Eq.~(\ref{ALEQ4169}) and Lemma~\ref{l:L420} lead to
\begin{align}
O_1^{(a)}:=\left(\sum_{i=0}^1\ket{i}\bra{i}\otimes\sigma_Z^i\otimes\sigma_X\right)^{(a)}
\approx_{R,\varphi^{(s_1,s_2,s_3)}}\delta_{a,s_3}I.
\label{th438A}
\end{align}
By replacing Eq.~(\ref{ReplacePth1}) with 
Eqs.~(\ref{ALT4165-1}) and (\ref{ALT4165-2}) and applying the same arguments so far result in
\begin{align}
O_2^{(b)}&:=\left(\sum_{i=0}^1\ket{i}\bra{i}\otimes\sigma_X\otimes\sigma_Z^i\right)^{(b)}
\approx_{R,\varphi^{(s_1,s_2,s_3)}}\delta_{b,s_2}I,
\label{th438B}
\\
O_3^{(c)}&:=\left(\sigma_X\otimes\sum_{i=0}^1\ket{i}\bra{i}\otimes\sigma_Z^i\right)^{(c)}
\approx_{R,\varphi^{(s_1,s_2,s_3)}}\delta_{c,s_1}I.
\label{th438C}
\end{align}
Once Eqs.~(\ref{th438A})-(\ref{th438C}) are in hand, we can prove
$\varphi^{(s_1,s_2,s_3)}\approx_RO_3^{(s_1)}O_2^{(s_2)}O_1^{(s_3)}
\left(\sum_{\bm{s}}\varphi^{(s_1,s_2,s_3)}\right)O_1^{(s_3)}O_2^{(s_2)}O_3^{(s_1)}$, 
which is equivalent to
\begin{align}
V_S\sigma^{(1,s_1;1,s_2;1,s_3)}V_S^{\dagger}
\approx_RO_3^{(s_1)}O_2^{(s_2)}O_1^{(s_3)}
(V_S\rho^{(111)}V_S^{\dagger})
O_1^{(s_3)}O_2^{(s_2)}O_3^{(s_1)}.
\label{ALEQ4173}
\end{align}
Its proof can be done by showing the following equations 
and using the triangle inequality of the trace distance:
\begin{align}
&\varphi^{(s_1,s_2,s_3)}\approx_RO_3^{(s_1)}O_2^{(s_2)}O_1^{(s_3)}\varphi^{(s_1,s_2,s_3)}O_1^{(s_3)}O_2^{(s_2)}O_3^{(s_1)},
\label{TH438T1}\\
&0\approx_RO_3^{(t_1)}O_2^{(t_2)}O_1^{(t_3)}\varphi^{(s_1,s_2,s_3)}O_1^{(s_3)}O_2^{(s_2)}O_3^{(s_1)}
\label{TH438T8}
\end{align}
for any $t_1t_2t_3\neq s_1s_2s_3$. 
The proofs of Eqs.~(\ref{TH438T1}) and (\ref{TH438T8}) are as follows. 
Using Lemma~\ref{l:L218} (\ref{l:L218i}) and Eq.~(\ref{th438A}) gives 
$O_2^{(b)}O_1^{(a)}\approx_{R,\varphi^{(s_1,s_2,s_3)}}\delta_{a,s_3}O_2^{(b)}$, and according to $s_3$, 
this yields the following two approximate relations due to Eq.~(\ref{th438B}):  
\begin{align}
&O_2^{(b)}O_1^{(s_3)}\approx_{R,\varphi^{(s_1,s_2,s_3)}}O_2^{(b)}\approx_{R,\varphi^{(s_1,s_2,s_3)}}\delta_{b,s_2}I,
\label{O2O3}\\
&O_2^{(b)}O_1^{(\overline{s_3})}\approx_{R,\varphi^{(s_1,s_2,s_3)}}0.
\label{O2barO3}
\end{align}
By employing Eq.~(\ref{th438C}) and Lemma~\ref{l:L218} (\ref{l:L218i}), Eqs.~(\ref{O2O3}) and (\ref{O2barO3}) respectively lead to
\begin{align}
&O_2^{(s_2)}O_1^{(s_3)}\approx_{R,\varphi^{(s_1,s_2,s_3)}}I\Rightarrow
\left\{
\begin{array}{l}
O_3^{(s_1)}O_2^{(s_2)}O_1^{(s_3)}\approx_{R,\varphi^{(s_1,s_2,s_3)}}I
\\
O_3^{(\overline{s_1})}O_2^{(s_2)}O_1^{(s_3)}\approx_{R,\varphi^{(s_1,s_2,s_3)}}0,
\label{T438B12}
\end{array}
\right.
\\
&O_2^{(\overline{s_2})}O_1^{(s_3)}\approx_{R,\varphi^{(s_1,s_2,s_3)}}0\Rightarrow
O_3^{(c)}O_2^{(\overline{s_2})}O_1^{(s_3)}\approx_{R,\varphi^{(s_1,s_2,s_3)}}0
\label{T438B3}
\end{align}
and
\begin{align}
O_3^{(c)}O_2^{(b)}O_1^{(\overline{s_3})}\approx_{R,\varphi^{(s_1,s_2,s_3)}}0.
\label{T438B4}
\end{align}
There are eight approximate relations in Eqs.~(\ref{T438B12}), (\ref{T438B3}) and (\ref{T438B4}), and 
combining each approximate relation with Lemma~\ref{l:L222} derives Eqs.~(\ref{TH438T1})-(\ref{TH438T8}). 

Now, we have Eq.~(\ref{ALEQ4173}), and $O_3^{(s_1)}O_2^{(s_2)}O_1^{(s_3)}$ in Eq.~(\ref{ALEQ4173}) is equal to as
\begin{align}
P[\ket{\phi^{(s_1,s_2,s_3)}_{\U{H}}}]=O_3^{(s_1)}O_2^{(s_2)}O_1^{(s_3)},
\label{StabilizerHyper}
\end{align}
whose proof is as follows. We define 
$U:=CCZ(H\otimes H\otimes H)$ with $H$ denoting the Hadamard operator, and 
$P[\ket{\phi^{(0,0,0)}_{\U{H}}}]$ is written as
$P[U\ket{0,0,0}]
=\prod_{i=1}^3(I_2^{\otimes3}+U\sigma_{Z,i}U^{\dagger})/2$. 
Since a direct calculation leads to 
$
U\sigma_{Z,1}U^{\dagger}=O_3, U\sigma_{Z,2}U^{\dagger}=O_2$ and 
$U\sigma_{Z,3}U^{\dagger}=O_1$, we have
$P[\ket{\phi^{(0,0,0)}_{\U{H}}}]=O_3^{(0)}O_2^{(0)}O_1^{(0)}.$
Next, we define unitary operator $W:=(\sigma_Z^{s_1}\otimes \sigma_Z^{s_2}\otimes\sigma_Z^{s_3})$, and a direct calculation leads to
\begin{align}
P[\ket{\phi^{(s_1,s_2,s_3)}_{\U{H}}}]=P[W\ket{\phi^{(0,0,0)}_\U{H}}]
=W(O_3^{(0)}O_2^{(0)}O_1^{(0)})W^{\dagger}=(WO_3^{(0)}W^{\dagger})(WO_2^{(0)}W^{\dagger})(WO_1^{(0)}W^{\dagger})
=O_3^{(s_1)}O_2^{(s_2)}O_1^{(s_3)}.
\no
\end{align}
Finally, substituting Eq.~(\ref{StabilizerHyper}) into Eq.~(\ref{ALEQ4173}) and using Corollary~\ref{NewCorollary436} results in 
\begin{align}
V_S\sigma^{(1,s_1;1,s_2;1,s_3)}V_S^{\dagger}
&\approx_RP[\ket{\phi^{(s_1,s_2,s_3)}_{\U{H}}}]
V_S\rho^{(111)}V_S^{\dagger}
P[\ket{\phi^{(s_1,s_2,s_3)}_{\U{H}}}]
\label{Thlast1}\\
&\overset{c}{\approx}_R\frac{1}{8}
P[\ket{\phi^{(s_1,s_2,s_3)}_{\U{H}}}]\otimes\tilde{\tilde{\alpha}}_{H}.
\label{Thlast2}
\end{align}
By defining $\zeta^{(s_1,s_2,s_3)}$ to be the renormalized state of 
$\xi^{(s_1,s_2,s_3)}:=\bra{\phi^{(s_1,s_2,s_3)}_H}V_S\rho^{(111)}V_S^{\dagger}\ket{\phi^{(s_1,s_2,s_3)}_{\U{H}}}$, 
the RHS of Eq.~(\ref{Thlast1}) equals to $P[\ket{\phi^{(s_1,s_2,s_3)}_{\U{H}}}]\otimes
\tr[\xi^{(s_1,s_2,s_3)}]\zeta^{(s_1,s_2,s_3)}$. Then, using Eq.~(\ref{Thlast1}) results in 
\begin{align}
V_S\sigma^{(1,s_1;1,s_2;1,s_3)}V_S^{\dagger}{\approx}_R\frac{1}{8}
P[\ket{\phi^{(s_1,s_2,s_3)}_{\U{H}}}]\otimes\zeta^{(s_1,s_2,s_3)},
\label{th1:18}
\end{align}
which shows the desired relation. \sq

\subsubsection{Proof of (ii)}
We prove (ii) in the case of $\bm{q}=000$. The other cases can be shown analogously. 
First, we have 
\begin{align}
V_SP^{(abc)}_{000}V_S^{\dagger}&=V_SA^{(a)}_{1,\bm{0}}A^{(b)}_{2,\bm{0}}A^{(c)}_{3,\bm{0}}V_S^{\dagger}
=V_S\left(\prod_{i=1}^3\frac{I+(-1)^{a_i}A_{i,\bm{0}}}{2}\right)V_S^{\dagger},
\label{rev1}
\end{align}
where $a_1:=a, a_2:=b, a_3:=c$ and the first equation comes from using Eq.~(\ref{def:Marginal}). 
Below, we prove
\begin{align}
V_SP^{(abc)}_{000}V_S^{\dagger}&\approx_{R,V_S\rho^{(\bm{\theta})}V_S^{\dagger}}
\bigotimes_{i=1}^3
\frac{I_2+(-1)^{a_i}\sigma_{Z,i}}{2}=P[\ket{a,b,c}].
\label{Pth438(ii)in1}
\end{align}
By expanding the terms in the parenthesis in Eq.~(\ref{rev1}), 
we find that once we have the following eight approximate relations, the triangle inequality of the state dependent norm implies 
Eq.~(\ref{Pth438(ii)in1}). 
\begin{align}
&V_SV_S^{\dagger}\approx_{R,V_S\rho^{(\bm{\theta})}V_S^{\dagger}}I,~~~
V_SA_{i,\bm{0}}V_S^{\dagger}\approx_{R,V_S\rho^{(\bm{\theta})}V_S^{\dagger}}\sigma_{Z,i},~~~
V_SA_{i,\bm{0}}A_{j,\bm{0}}V_S^{\dagger}\approx_{R,V_S\rho^{(\bm{\theta})}V_S^{\dagger}}\sigma_{Z,i}\otimes\sigma_{Z,j},\no\\
&V_SA_{1,\bm{0}}A_{2,\bm{0}}A_{3,\bm{0}}V_S^{\dagger}\approx_{R,V_S\rho^{(\bm{\theta})}V_S^{\dagger}}\sigma_{Z,1}\otimes\sigma_{Z,2}\otimes
\sigma_{Z,3}
\no
\end{align}
We can prove the first equation from a direct calculation using the definition of the state dependent norm. 
The other equations have already been proven in Eq.~(\ref{eq:i}), Lemma~\ref{Lemma430}, Lemma~\ref{NewLemma430}, 
Lemma~\ref{Lemma437}, Lemma~\ref{Lemma437}, Lemma~\ref{Lemma437} and 
Lemma~\ref{NewLemma437}. 

Now we have Eq.~(\ref{Pth438(ii)in1}), and using Lemma~\ref{l:L218} (\ref{l:L218ii}) implies 
$V_SP^{(abc)}_{000}V_S^{\dagger}\approx_{R,V_S\sigma^{(1,s_1;1s_2;1,s_3)}V_S^{\dagger}}
P[\ket{a,b,c}]$. 
Hence, Lemma~\ref{l:L222} leads to
\begin{align}
V_SP^{(abc)}_{000}\sigma^{(1,s_1;1s_2;1,s_3)}P^{(abc)}_{000}V_S^{\dagger}
&=(V_SP^{(abc)}_{000}V_S^{\dagger})(V_S\sigma^{(1,s_1;1s_2;1,s_3)}V_S^{\dagger})(V_SP^{(abc)}_{000}V_S^{\dagger})
\no\\
&\approx_{R}
(P[\ket{a,b,c}\otimes I_{\mathcal{H}})
(V_S\sigma^{(1,s_1;1s_2;1,s_3)}V_S^{\dagger})
(P[\ket{a,b,c}\otimes I_{\mathcal{H}}).
\label{th2:unnormal}
\end{align}
Since acting projector does not increase trace distance, 
using Theorem~\ref{Theorem438} (i) enables us to replace $V_S\sigma^{(1,s_1;1s_2;1,s_3)}V_S^{\dagger}$ with 
$\frac{1}{8}P[\ket{\phi^{(s_1,s_2,s_3)}_{\U{H}}}]\otimes\zeta^{(s_1,s_2,s_3)}$, 
which is the desired relation. 
\sq

\section{Protocol for proof of magic}
\label{sec:appli}
In this section, we show the details of our proof of quantumness, Protocol~2 in the main text. 
In particular, we explain why this protocol works by running Protocol~1 a constant number of times at step~1 in Protocol~2. 
Protocol~2 exploits Theorem~\ref{Theorem438} (i), 
which states that there exists a positive constant $c'$ and a negligible function $\negl_1(\lambda)$ satisfying
\begin{align}
\left|\left|
V\sigma'^{(s_1;s_2;s_3)}V^\dag-
P[\ket{\phi^{(s_1,s_2,s_3)}_{\U{H}}}]\otimes\zeta^{(s_1,s_2,s_3)}_{\mathcal{H}'}
\right|\right|_1^2
\le\left[\sqrt{c'}\left(\sqrt{p^r_{\U{Pre}}}+\sqrt{p^r_{\U{Test}}}+\sqrt{p^r_{\U{Hyper}}}\right)+\sqrt{\negl_1(\lambda)}\right]^2
=:T^2.
\label{eq:T}
\end{align}
For simplicity of notations, we define $c:=\sqrt{c'}$ and $\negl_2(\lambda):=\sqrt{\negl_1(\lambda)}$. 
Since the exact value of $T$ cannot be obtained by repeating Protocol~1 a finite number of times, 
we need to estimate it from the number of set flags. 
Specifically, our goal is to derive the estimated value $T_{\U{est}}$ of $T$ satisfying
\begin{align}
\Pr\left[
|T-T_{\U{est}}|\le\epsilon\right]\ge1-\delta
\no
\end{align}
for any $\epsilon>0$ and $\delta>0$. 
Below, we show that when these $\epsilon$ and $\delta$ are constant, the number of times repeating 
Protocol~1 at step~1 in Protocol~2 is also constant. 

From the numbers of set flags obtained at step~1 in Protocol~2, 
we have the estimated value $p_a'$ of $p_a$ for each $\U{a}\in\{\U{Pre},\U{Test},\U{Hyper}\}$ by employing Hoeffding's inequality as
\begin{align}
\Pr\left[
|p_a-p'_a|\le\epsilon'\right]\ge1-\delta'
\label{est_pa}
\end{align}
for any $0<\epsilon'<1$ and $0<\delta'<1$. 
This relation can be obtained by repeating 
$$
N_{\epsilon',\delta'}:=O\left(\frac{1}{\epsilon'^2}\ln\frac{1}{\delta'}\right)
$$
times of Protocol~1 on average. 
Using these estimated probabilities $p'_{\U{Pre}},p'_{\U{Test}}$ and $p'_{\U{Hyper}}$, we define the estimated 
value of the trace norm $T_{\U{est}}$ as 
\begin{align}
T_{\U{est}}:=c\left(\sqrt{p'^r_{\U{Pre}}}+\sqrt{p'^r_{\U{Test}}}+\sqrt{p'^r_{\U{Hyper}}}\right)+\negl_2(\lambda).
\label{eq:Test}
\end{align}
Then, 
by substituting the definitions in Eqs.~(\ref{eq:T}) and (\ref{eq:Test}) to $|T-T_{\U{est}}|$, we have
\begin{align}
|T-T_{\U{est}}|
\le c\sum_{\U{a}\in\{\U{Pre},\U{Test},\U{Hyper}\}}\left|\sqrt{p^r_{\U{a}}}-\sqrt{p'^r_{\U{a}}}\right|.
\label{eq:cases}
\end{align}
Using Eq.~(\ref{est_pa}), we obtain the following with probability at least $1-\delta'$: 
\begin{align}
\left|\sqrt{p^r_{\U{a}}}-\sqrt{p'^r_{\U{a}}}\right|\le\left\{
\begin{array}{l}
\sqrt{(p_{\U{a}}+\epsilon')^r}-\sqrt{p^r_{\U{a}}}~~~(\U{if}~p_a'\ge p_a)\\
\sqrt{p^r_{\U{a}}}-\sqrt{(p_{\U{a}}-\epsilon')^r}~~~(\U{if}~p_a'< p_a~\U{and}~p_a\ge\epsilon')\\
\sqrt{\epsilon'^r}~~~(\U{if}~p_a'< p_a<\epsilon').
\end{array}
\right.
\label{eq:cases2}
\end{align}
In the first case of $p_a'\ge p_a$, by a simple calculation, 
it is easy to find that $\sqrt{(p_{\U{a}}+\epsilon')^r}$ is upper-bounded by
\begin{align}
\sqrt{(p_{\U{a}}+\epsilon')^r}\le\left\{
\begin{array}{l}
\sqrt{p^r_{\U{a}}}+\sqrt{\epsilon'^{r}}~~~~~~~~~~~~~~(0<\frac{r}{2}<1)\\
\sqrt{p^r_{\U{a}}}+(\sqrt{2^r}-1)\epsilon'~~~~~~(\frac{r}{2}\in\mathbb{N})\\
\sqrt{p^r_{\U{a}}}+(2\sqrt{2^r}-1)\epsilon'^x~~~(1\le\frac{r}{2}, \frac{r}{2}\notin\mathbb{N}),
\end{array}
\right.
\no
\end{align}
where in the third case, we express $r/2$ as $x+n$ with 
$x$ ($0<x<1$) being the decimal number and $n$ being the integer. 
Hence, $\sqrt{(p_{\U{a}}+\epsilon')^r}\le\sqrt{p^r_{\U{a}}}+O(\epsilon'^t)$ holds 
with $t$ being a non-zero constant value.

In the second case of $p_a'<p_a$ and $p_a\ge\epsilon'$ in Eq.~(\ref{eq:cases2}), by a simple calculation, 
it is easy to find that $\sqrt{(p_{\U{a}}-\epsilon')^r}$ is lower-bounded by
\begin{align}
\sqrt{(p_{\U{a}}-\epsilon')^r}\ge\left\{
\begin{array}{l}
\sqrt{p^r_{\U{a}}}-\sqrt{\epsilon'^{r}}~~~~~~~~~~~~~~(0<\frac{r}{2}<1)\\
\sqrt{p^r_{\U{a}}}-(\sqrt{2^r}-1)\epsilon'~~~~~(\frac{r}{2}\in\mathbb{N})\\
\sqrt{p^r_{\U{a}}}-\sqrt{2^r}\epsilon'^x~~~~~~~~~~~(1\le\frac{r}{2}, \frac{r}{2}\notin\mathbb{N}),
\end{array}
\right.
\no
\end{align}
where in the third case, we express $r/2$ as $x+n$ with 
$x$ ($0<x<1$) being the decimal number and $n$ being the integer. 
Hence, $\sqrt{(p_{\U{a}}-\epsilon')^r}\ge\sqrt{p^r_{\U{a}}}-O(\epsilon'^t)$ holds 
with $t$ being a non-zero constant value.

By combining the arguments so far and considering the fact that $c$ is a constant value, we finally obtain 
\begin{align}
\Pr\left[|T-T_{\U{est}}|\le O(\epsilon'^t)\right]\ge1-3\delta'.
\no
\end{align}
By setting $\delta=3\delta'$ and $\epsilon=O(\epsilon'^{t})$, 
if $\delta$ and $\epsilon$ are constant, then 
the number $N_{\epsilon',\delta'}$ of times repeating Protocol~1 at step~1 in Protocol~2 results in the constant number. 
\\

In the explanations of the main text, we have set $\epsilon=1/6$ and $\delta=10^{-10}$ for simplicity of the arguments, but 
for any $\epsilon>0$ and $\delta>0$, the number of times repeating Protocol~1 at step~1 in Protocol~2	becomes constant.

\end{widetext}
\end{document}